\documentclass[
reprint,
superscriptaddress,
nofootinbib,
amsmath,
amssymb,
aps,
prd
]{revtex4-2}

\usepackage{times}
\usepackage{graphicx}
\usepackage{dcolumn}
\usepackage{physics}
\usepackage{bm}
\usepackage{xcolor}
\usepackage{url}
\usepackage{xurl}
\usepackage[hidelinks]{hyperref}

\begin{document}
\title{Three-flavor neutrino Boltzmann transport and flavor instabilities \\
in long-lived  neutron-star merger remnants}
\author{Harry Ho-Yin Ng}
\email{hng@caltech.edu}
\affiliation{TAPIR, Mailcode 350-17, California Institute of Technology, Pasadena, CA 91125, USA}

\author{Jiabao Liu}
\affiliation{Department of Physics and Applied Physics, School of Advanced
Science \& Engineering, Waseda University, Tokyo 169-8555, Japan}

\author{Arthur Offermans}
\affiliation{Department of Physics and Astronomy, KU Leuven, B-3001 Leuven, Belgium}
\affiliation{Leuven Gravity Institute, KU Leuven, Celestijnenlaan 200D box 2415, 3001 Leuven, Belgium}

\author{Shoichi Yamada}
\affiliation{Department of Physics and Applied Physics, School of Advanced Science \& Engineering, Waseda University, Tokyo 169-8555, Japan}

\author{Elias R. Most}
\affiliation{TAPIR, Mailcode 350-17, California Institute of Technology, Pasadena, CA 91125, USA}
\affiliation{Walter Burke Institute for Theoretical Physics, California Institute of Technology, Pasadena, CA 91125, USA}

\author{Luciano Rezzolla}
\affiliation{Institut f\"ur Theoretische Physik, Goethe Universit\"at,
 Max-von-Laue-Str. 1, 60438 Frankfurt am Main, Germany}
\affiliation{School of Mathematics, Trinity College, Dublin 2, Ireland}
\affiliation{Frankfurt Institute for Advanced Studies,  Ruth-Moufang-Str. 1, 60438 Frankfurt am Main, Germany}

\begin{abstract}
Binary neutron-star (BNS) mergers are important multimessenger sources
whose kilonova emission and heavy-element nucleosynthesis depend
sensitively on neutrino microphysics and flavor conversion. Most BNS
merger simulations assume $\mu$-$\tau$ neutrino degeneracy in a
simplified two-flavor treatment. We investigate the fast flavor
instability (FFI) and the collisional flavor instability (CFI) in
long-lived BNS merger remnants using axisymmetric general-relativistic
three-flavor Boltzmann neutrino transport with a comprehensive treatment of muons and muonic interactions, following a handoff from 3D grey moment-based simulations.
Our simulations largely reproduce the properties of the parent 3D
simulations, recovering trapped weak equilibrium in the core, muonic
non-equilibrium in the semitransparent outer disk due to prolonged
demuonization, and cold weak equilibrium in transparent regions. Linear
stability analysis reveals strong and widespread CFI across all three flavor
sectors, including narrow resonance-like structures. Muonic interactions
fundamentally reshape the distribution of FFI, shifting the $e\tau$-sector instability
toward denser disk regions compared with the extended FFI outside the
disk in the two-flavor model. At the same time, disk demuonization
produces a strong $\nu_\mu$ excess and predominantly one-sided
muon-lepton-number angular distributions, suppressing FFI in the $e\mu$
and $\mu\tau$ sectors. We perform local three-flavor quantum kinetic
equation (QKE) calculations at selected points, revealing 
nonlinear coupling among flavor sectors that excites coherence even in an initially linearly stable sector.
Diagonal collisional repopulation plays a decisive role in determining the nonlinear evolution and asymptotic flavor populations, while neglecting it can overestimate flavor equilibration. 
These findings expose the limitations of simplified muonless, two-flavor approximations in describing neutrino flavor instabilities and their nonlinear evolution. 
We establish the necessity of three-flavor transport from both classical and quantum perspectives and underscore the critical importance of consistently coupling nonlinear flavor conversion with diagonal collisional repopulation in future CFI subgrid models.

\end{abstract}
\maketitle

\section{Introduction}
\label{sec:introduction}
Binary neutron-star (BNS) mergers are among the most important
multi-messenger sources in the
Universe~\cite{Lattimer74,Eichler89,Metzger:2010}, producing signals such
as gravitational waves~\cite{Baiotti2016,Burns2020,Radice2020}, short
gamma-ray bursts~\cite{Rezzolla:2011,Ruiz2021rev,Kiuchi2024}, and
kilonovae caused by the production of heavy elements via $r$-process
nucleosynthesis~\cite{Hotokezaka2013,Kasen2017,Metzger2017}. The first
multi-messenger observation of a BNS merger,
GW170817/GRB170817A/AT2017gfo, provided a new avenue for probing and
constraining a broad range of
physics~\cite{Abbott2017_etal,Abbott2017b,FermiLat2017}. BNS mergers
involve physics across widely separated scales, thus requiring
numerical-relativity simulations for quantitative modeling (see, e.g.,
the reviews in Refs.~\cite{Baiotti2016,ShibataRev19}).

Among the major uncertainties in BNS merger modeling, neutrinos play one
of the most important roles. Neutrinos carry away thermal energy and
lepton number from the hot remnant and disk, and they also regulate
cooling and heating. Neutrino microphysics plays a decisive role in
determining the electron fraction of the disk and ejecta, largely
affecting $r$-process nucleosynthesis, lanthanide production, and
electromagnetic counterparts, particularly kilonova light
curves~\cite{Foucart2023,Zappa2023,Tanaka2020_opacity,Qiu2025b}.
Modeling these effects requires solving neutrino transport together with
the hydrodynamics and a large set of sophisticated weak
interactions. Considerable progress has been made in recent years using
moment-based, Monte Carlo, lattice-Boltzmann, and Boltzmann-transport
methods, together with increasingly complete neutrino
microphysics~\cite{Radice2013,Miller2019a,Sumiyoshi2020,Weih2020c,Radice2022,Foucart2023,Cheong2023,Musolino2023,Kawaguchi2025,Olsen2025,Alford2026,Foucart2026,Rath2026}.
Nevertheless, these approaches describe neutrino transport only within a
classical picture.

A crucial missing piece of neutrino physics is the quantum-kinetic behavior of neutrinos, in particular their quantum flavor coherence.
In addition to vacuum oscillations, neutrino--neutrino self-interactions and collisions in a dense neutrino gas can trigger collective flavor instabilities.
Depending on the flavor and momentum distributions of neutrinos and the collisional environment, the quantum kinetic equations (QKEs) can have flavor instabilities in which small flavor coherence grows exponentially, initiating these flavor conversion processes~\cite{Pantaleone1992,Sigl1993,Duan2010,Richers2019,Tamborra2020}.

This self-coupled quantum evolution can give rise to several classes of
flavor instabilities, including fast flavor instability (FFI) driven by
neutrino self-interactions as a refractive
effect~\cite{Zaizen2021,Ehring2023,Fiorillo2024,Fiorillo2024b,Wang2025a,Johns2025,Xiong2026,Zaizen2026,Mezzacappa2026},
collisional flavor instability (CFI) arising from the interplay between
self-interactions and
collisions~\cite{Johns2021,Johns2022,Lin2023,Xiong2022,Kato2023,Liu2026,Takahashi2026},
many-body effects~\cite{Kost2024,Laraib2025}, and slower flavor modes
involving the vacuum
scale~\cite{Banerjee2011,Padilla-Gay2025}.  These phenomena
can operate on extremely small length and time scales, far shorter than
those of the global astrophysical system, making their self-consistent
coupling to neutrino radiation hydrodynamics computationally
prohibitive~\cite{Sawyer2005,Chakraborty2016,Abbar2017,Richers2019,Nagakura2023c}.

BNS mergers are favorable environments for the flavor conversion processes
because of the high neutrino number densities, complex geometries,
anisotropic radiation fields, and extended semitransparent disks. As
resolving the quantum-kinetic scales in a global BNS simulation remains
prohibitively expensive, different studies have employed local QKE
calculations~\cite{Richers2019,Grohs2022,Urquilla2026}, applied linear stability analysis
(LSA)~\cite{Wu2017b,George2020,Froustey2023,Kneller2024,Froustey2026}, or
incorporated approximate flavor-conversion
treatments~\cite{Li2021prl,Just2022,Fernandez2022,Qiu2025,Qiu2025b,Lund2025},
showing significant impacts on the dynamics, composition, nucleosynthesis, and
neutrino signatures. However, one important ingredient remains
unexplored: the muon degree of freedom. Most of these studies employ two-flavor
calculations, assuming only neutrons, protons, and electrons in matter, i.e.,
$npe$-matter, where
muon- and tau-flavor (anti)neutrinos participate in identical weak
interactions.
Although some works~\cite{Shalgar2025, Urquilla2026} have considered three-flavor QKE calculations, they have not consistently incorporated muons and muonic interactions in the matter background.
Recent studies have shown in either core-collapse supernovae
(CCSNe) or BNS simulations that matter with sufficiently high temperature and
density contains an appreciable fraction of muons owing to their charged-current
reactions from the large reservoir of $\nu_\mu$ and $\bar{\nu}_\mu$ generated by
pair processes
~\cite{Bollig2017,Fischer2020b,Gieg2024,Ng2024c,Pajkos2024,Gieg2026} (see also
related works~\cite{Loffredo2023, Hernandez2026}).
Some preliminary studies in CCSN calculations have demonstrated
that the inclusion of muons can trigger flavor mixing in the
$\mu\tau$ and $e\tau$ sectors~\cite{Capozzi2020,Liu2024}.
In contrast, the muon fraction in a BNS merger is around one order of magnitude
higher than in a CCSN~\cite{Ng2024c,Gieg2026}, meaning that
the missing $e\mu$ and $\mu\tau$ sectors in a three-flavor framework
could produce stronger and more complex flavor instabilities.

In this work, we investigate three-flavor neutrino flavor instabilities in long-lived BNS merger remnants using general-relativistic Boltzmann neutrino transport with three flavors, following a handoff from 3D merger simulations employing a grey moment-based transport scheme. Our study highlights the importance of including muons and muonic interactions, revealing flavor-instability properties and nonlinear behavior that cannot be captured by conventional two-flavor calculations.

The paper is organized as follows. In Sec.~\ref{sec:secII}, we introduce the framework of QKEs, the LSA for FFI and CFI, and the classical Boltzmann transport and weak-interaction physics. In Sec.~\ref{sec:secIII}, we describe the numerical methods and setup, including the handoff procedure, the treatment of muonic interactions, and the implicit coupling between neutrino radiation and matter composition. In Sec.~\ref{sec:secIV}, we present the global simulation results and compare the occurrence and properties of FFI and CFI between two- and three-flavor treatments, highlighting the widespread CFI in all three flavor sectors and the substantial changes in FFI morphology induced by muonic interactions. In Sec.~\ref{sec:QKE_results}, we perform local, isotropic, multi-energy three-flavor QKE calculations to investigate the nonlinear evolution of CFI, focusing on the coupling among flavor sectors and comparing the effects of including or omitting diagonal collision terms on the flavor evolution, asymptotic states, and relaxation timescales. 
We find that a full collision operator in QKE calculations 
is necessary to achieve the accurate asymptotic states.
Finally, in Sec.~\ref{sec:conclusion}, we summarize our findings and conclusions.

\section{Physical Framework}\label{sec:secII}
\subsection{Quantum kinetic framework and linearized stability analysis}\label{sec:qke-lsa}
We describe neutrinos and antineutrinos by Hermitian $3\times3$ density
matrices $\varrho$ and $\bar\varrho$ in flavor space. We work in a local
fluid-comoving inertial frame and neglect variations of the background
over the microscopic evolution scales. To focus on fast and collisional
flavor dynamics, we neglect the vacuum Hamiltonian. Within this local
approximation, the spatially uniform, direction-independent,
flavor-diagonal matter potential is removed by a common time-dependent
rotation in flavor space. The resulting mean-field quantum kinetic
equations are
\begin{align}
(\partial_t+\mathbf v\cdot\nabla)\varrho &= -i[H_{\nu\nu},\varrho]+C,
\\
(\partial_t+\mathbf v\cdot\nabla)\bar\varrho &= -i[H_{\nu\nu},\bar\varrho]+\bar C,
\label{eq:local-qke}
\end{align}
where $t$ is the local time, $\mathbf v=\mathbf v(\Omega_p)$ is the unit propagation-direction vector with $|\mathbf v|=1$, 
$\Omega_p=(\vartheta,\varphi)$ denotes the angular part of momentum-space of (anti)neutrino, and $C$ and $\bar C$ are the neutrino and antineutrino collision operators, respectively. We use the antineutrino convention in which the self-interaction Hamiltonian has the same form in both equations. The neutrino self-interaction Hamiltonian is
\begin{equation}
H_{\nu\nu}(\mathbf v)=\sqrt{2}G_F\int dV_p'\,(1-\mathbf v\cdot\mathbf
v')\left[\varrho(\varepsilon',\mathbf
  v')-\bar\varrho(\varepsilon',\mathbf v')\right]\,,
\label{eq:qke-self-interaction}
\end{equation}
where $dV_p'=(\varepsilon'^2\,d\varepsilon'\,d\Omega'_p)/(2\pi)^3$ is the
momentum-space volume element, $\varepsilon$ is the neutrino energy, 
and $G_F$ is the Fermi constant. 
For a pair of distinct flavors
$a,b\in\{e,\mu,\tau\}$, the corresponding $2\times2$ submatrices of the
full density matrices are
\begin{equation}\label{eq:density_matrix}
\varrho^{(ab)}
=
\begin{pmatrix}
f_{\nu_a} & S_{ab} \\
S_{ab}^{*} & f_{\nu_b}
\end{pmatrix}\,,
\qquad
\bar{\varrho}^{(ab)}
=
\begin{pmatrix}
f_{\bar\nu_a} & \bar{S}_{ab} \\
\bar{S}_{ab}^{*} & f_{\bar\nu_b}
\end{pmatrix}\,,
\end{equation}
where the diagonal components, $f_{\nu_a}(\varepsilon,\Omega_p)$ and $f_{\bar\nu_a}(\varepsilon,\Omega_p)$, describe the flavor-dependent neutrino and antineutrino distributions for flavor $a$, respectively.  The off-diagonal components $S_{ab}$
and $\bar S_{ab}$ represent the flavor coherence between flavors $a$ and
$b$ in the neutrino and antineutrino sectors, respectively, and the
asterisk denotes complex conjugation.

Neutrino flavor instabilities can be investigated through a linear stability analysis (LSA) of the QKEs about a locally frozen, flavor-diagonal background. Treating $S_{ab}$ and $\bar S_{ab}$ as small perturbations and adopting the normal-mode dependence $\exp[-i(\omega t-\mathbf K\cdot\mathbf x)]$ leads to a dispersion relation. For a real wave vector $\mathbf K$, solutions with $\operatorname{Im}\omega>0$ correspond to exponentially growing flavor coherence~\cite{Banerjee2011,Izaguirre2017}.

Within the vacuum-free framework introduced above, we focus on two classes of instability. FFI is driven by neutrino self-interactions in the collisionless limit~\cite{Sawyer2016,Izaguirre2017}, whereas CFI arises from their interplay with the energy- and species-dependent collisional damping of flavor coherence~\cite{Johns2021,Lin2023,Liu2023}. For fixed background distributions and collision rates, the linear stability properties in this approximation are independent of the neutrino mass ordering. The specific diagnostic approximations adopted for FFI and CFI are described below.

Many previous studies adopt an effective two-flavor treatment, in which the muon- and tau-type neutrinos are assumed to be indistinguishable and represented by a common heavy-lepton flavor, $\chi$. In this case, only the $(e\chi)$ sector is considered. Here, we retain all three physical flavors and analyze the $(e\mu)$, $(e\tau)$, and $(\mu\tau)$ off-diagonal sectors. These sectors decouple at linear order around a flavor-diagonal background under the adopted collision approximation, but need not remain independent during nonlinear three-flavor evolution. 
This allows us to quantify the effects of $npe\mu$ matter and the corresponding muonic interactions on flavor instabilities.

Throughout, $a,b\in\{e,\mu,\tau\}$ label neutrino flavors, and $ab$
denotes the corresponding flavor sector.
Unless stated otherwise, all neutrino momentum-space labels and LSA
quantities are defined in the local fluid-comoving orthonormal frame,
without assuming axial symmetry in momentum space. 
\subsubsection{Fast flavor instability}\label{sec:FFI}
In the fast limit, collision terms are neglected. The characteristic local
neutrino self-interaction scale is $\mu_{\nu\nu} = \sqrt{2}G_F n_\nu$, where
$n_\nu$ denotes the neutrino number density in the comoving frame.
In hot environments surrounding
compact objects, such as the accreting HMNS and PNS cores,
the FFI growth rate can exceed the vacuum oscillation frequency by many orders of
magnitude. Furthermore, we drop the explicit energy dependence of the FFI within
the vacuum-free, collisionless approximation, allowing the neutrino
distributions to be integrated over energy, giving the lepton number (LN)
angular distribution of flavor $a$ as

\begin{equation}
\mathcal G_a(\Omega_p)
=
\sqrt{2}G_F
\int
\frac{\varepsilon^2 d\varepsilon}{2\pi^2}
\left[
f_{\nu_a}(\varepsilon,\Omega_p)
-
f_{\bar\nu_a}(\varepsilon,\Omega_p)
\right]\,.
\label{eq:flavor-lepton-angular-distribution}
\end{equation}
The
corresponding $a$LN--$b$LN angular distribution for each $(ab)$ flavor
sector is defined as
\begin{equation}
\mathcal G_{ab}(\Omega_p)
:=
\mathcal G_a(\Omega_p)
-
\mathcal G_b(\Omega_p)\,,
\label{eq:pairwise-flavor-lepton-angular-distribution}
\end{equation}
and an $a$LN--$b$LN angular crossing occurs when $\mathcal
G_{ab}(\Omega_p)$ changes sign as a function of momentum direction,
enabling FFI. The growth rate of the fastest-growing mode is of order
$\mathcal{O}(\mu_{\nu\nu})$. To be specific, we estimate the FFI growth
rate using the geometric-mean prescription~\cite{Morinaga2019}
\begin{align}
\sigma_{ab}^{\rm FFI}
&=
\sqrt{
\left(-
\int_{\mathcal G_{ab}<0}
\frac{d\Omega_p}{4\pi}\,
\mathcal G_{ab}
\right)
\left(
\int_{\mathcal G_{ab}>0}
\frac{d\Omega_p}{4\pi}\,
\mathcal G_{ab}
\right)
} \nonumber
\\
&=
\sqrt{I^-_{\rm FFI} I^+_{\rm FFI}}\,, 
\label{eq:ffi-growth-rate}
\end{align}
where the two integrals measure the total positive $I^+_{\rm FFI}\ge0$
and negative $I^-_{\rm FFI}\ge0$ components of the $a$LN--$b$LN angular
crossing over the full momentum-space solid angle. Here,
$\sigma_{ab}^{\rm FFI}$ is an estimator rather than a maximum obtained by
solving the full dispersion relation.

\subsubsection{Collisional flavor instability}\label{sec:CFI}
Unlike FFI, collisions play an indispensable role in CFI. The instability arises from the interplay between neutrino self-interactions and flavor-dependent collisional effects on the off-diagonal components of the neutrino density matrices. CFI is expected in dense neutrino environments where both the neutrino self-interaction potential and neutrino--matter interaction rates remain appreciable.

The growth rate of ordinary CFI is typically controlled by the
collisional scale, $\sigma^{\rm CFI}\sim\mathcal O(\Gamma)$, where
$\Gamma$ denotes a characteristic collisional decoherence rate. In the
usual regime $\Gamma\ll\mu_{\nu\nu}$, this scale is smaller than the
characteristic fast scale $\mu_{\nu\nu}$. The actual FFI growth rate,
however, also depends on the angular crossing and can be substantially
smaller than $\mu_{\nu\nu}$.

For the present CFI survey, we consider spatially homogeneous
perturbations, $K=0$, of an isotropized background, $f_{\nu_a}^{\rm
  iso}(\varepsilon) = \int{d\Omega_p}\, f_{\nu_a}(\varepsilon,\Omega_p) /
{4\pi}$, with analogous definitions for antineutrinos. This approximation
targets the dense regions of interest and provides a direct connection to
the local isotropic QKE calculations. We do not establish that these
modes maximize the growth rate over all wave-vectors and angular
perturbations.

Several approximate prescriptions have been developed to diagnose CFI without
solving the full multi-energy dispersion relation
~\cite{Lin2023,Xiong2022,Liu2023}. Energy-averaging (grey)
prescriptions based on ratios of signed spectral moments can become numerically
ill-conditioned when cancellations make the corresponding denominator small,
producing artificially large effective collision rates (e.g., see
comparisons in~\cite{Froustey2026} and the discussions in~\cite{Liu2026}). We
thus adopt the new approximate energy-integration method of
Ref.~\cite{Liu2026}. The method acts on the energy-resolved neutrino spectra
and energy-dependent collision rates before reducing them to effective positive
and negative spectral sectors. The resulting dispersion relation is energy
integrated, but retains the effects of the spectral asymmetries and energy
dependence of the collision rates through these effective quantities. This
construction also avoids the cancellation-induced numerical singularities that
can arise in simpler grey prescriptions. The resulting roots provide an
approximate energy-integrated diagnostic.

For the
local QKE prescription, we adopt the reduced collision operator
\begin{equation}
 \tilde C[\varrho]
=
\mathfrak J-\frac12\{\mathfrak K,\varrho\}\,,
\label{eq:qke-reduced-collision}
\end{equation}
where $\mathfrak
J=\operatorname{diag}(j_{\nu_e},j_{\nu_\mu},j_{\nu_\tau})$ and $\mathfrak
K=\operatorname{diag}(\Gamma_{\nu_e},
\Gamma_{\nu_\mu},\Gamma_{\nu_\tau})$. Here, $j_{\nu}=j^{\rm CC}_\nu +
\sum_X j_\nu^X$ is the total emissivity and
$\Gamma_{\nu}=\kappa_{a,\nu}^{*,\rm CC}+\sum_X\kappa_{a,\nu}^{*,X}$ is
the total stimulated absorption opacity (see the definitions of
$\kappa^{*,X}_{a,\nu}$ and $j^{X}_{\nu}$ in
Sec.~\ref{sec:implicit-composition}, while $\kappa_{a,\nu}^{*,\rm CC}$
and $j^{\rm CC}_\nu$ are the corresponding quantities for charged-current
interactions), summed over all interactions in
Tab.~\ref{tab:weak-interactions}, excluding the elastic and inelastic
scattering contributions.

The components of $\tilde C$ are
\begin{align}
\tilde C_{aa}
&=j_{\nu_a}-\Gamma_{\nu_a}\varrho_{aa},
\\
\tilde C_{ab}
&=-\Gamma_{ab}\varrho_{ab},
\qquad a\ne b,
\end{align}
with $\Gamma_{ab}=(\Gamma_{\nu_a}+\Gamma_{\nu_b})/2$. Analogous expressions apply to antineutrinos.

Using $[\mathcal Q]_\pm:=\max(\pm \mathcal Q,0)$, the positive and negative effective spectral weights $\mathcal W_{ab,\pm}$ and their corresponding collision rates are
\begin{equation}
\begin{aligned}
\mathcal W_{ab,\pm}
={}&
\sqrt{2}G_F
\int\frac{\varepsilon^2d\varepsilon}{2\pi^2}
\left(
[\Delta f_{ab}]_\pm
+
[\Delta\bar f_{ab}]_\mp
\right)\,,
\\
\Gamma_{ab,\pm}
={}&
\frac{\sqrt{2}G_F}{\mathcal W_{ab,\pm}}
\int\frac{\varepsilon^2d\varepsilon}{2\pi^2}
\left(
[\Delta f_{ab}]_\pm\Gamma_{ab}
+
[\Delta\bar f_{ab}]_\mp\bar\Gamma_{ab}
\right)\,,
\end{aligned}
\label{eq:cfi-effective-sectors}
\end{equation}
where $\Delta f_{ab}=f_{\nu_a}^{\rm iso}-f_{\nu_b}^{\rm iso}$
($\Delta\bar f_{ab}$ for antineutrinos), and the second expression
applies for $\mathcal W_{ab,\pm}>0$; a vanishing $\mathcal W_{ab,\pm}$
corresponds to an absent effective sector.

The four quantities entering the reduced dispersion relation are then
\begin{equation}
\begin{aligned}
G_{ab}
&=\frac{\mathcal W_{ab,+}+\mathcal W_{ab,-}}{2}\,,
&
A_{ab}
&=\frac{\mathcal W_{ab,+}-\mathcal W_{ab,-}}{2}\,,
\\
\gamma_{ab}
&=\frac{\Gamma_{ab,+}+\Gamma_{ab,-}}{2}\,,
&
\alpha_{ab}
&=\frac{\Gamma_{ab,+}-\Gamma_{ab,-}}{2}\,.
\end{aligned}
\label{eq:cfi-effective-parameters}
\end{equation}
Here, $G_{ab}$ characterizes the overall self-interaction strength,
$A_{ab}$ the signed imbalance between the two effective spectral sectors,
$\gamma_{ab}$ their mean damping rate, and $\alpha_{ab}$ their
collision-rate asymmetry.

The solutions can be separated into an isotropy-preserving (IP) branch, which preserves the isotropy of the flavor-coherence perturbation, and an isotropy-breaking (IB) branch, which develops a nonvanishing flavor-coherence flux. Their roots can be written compactly as
\begin{equation}
\omega_{r,ab}^{\pm}
=
-\xi_r A_{ab}
-i\gamma_{ab}
\pm
\sqrt{
\xi_r^2 A_{ab}^2
-\alpha_{ab}^2
+2i\xi_rG_{ab}\alpha_{ab}
}\,,
\label{eq:cfi-roots}
\end{equation}
where $r\in\{{\rm IP},{\rm IB}\}$ labels the branch, with
$\xi_{\rm IP}=1$ and $\xi_{\rm IB}=-\frac{1}{3}$. Within this approximate, spatially homogeneous diagnostic, we define the CFI growth rate as the largest nonnegative growth rate among the IP and IB branches,
\begin{equation}
\sigma_{ab}^{\rm CFI}
:=
\max\left[
0,\,
\max_{r\in\{{\rm IP},{\rm IB}\}}
\left\{
\operatorname{Im}\omega_{r,ab}^{+},
\operatorname{Im}\omega_{r,ab}^{-}
\right\}
\right]\,.
\label{eq:cfi-growth-rate}
\end{equation}

One of the most significant features of CFI is its resonance-like
enhancement~\cite{Liu2023,Xiong2023b,Wang2025}. To make the conditions
and characteristic growth rates of ordinary and resonance-like CFI
(R-CFI) explicit, we expand the above solutions in the two limiting
regimes,
\begin{equation}
\sigma_{r,ab}^{\rm CFI}
\simeq
\begin{cases}
\displaystyle
\left[
-\gamma_{ab}
+\frac{
\left|
G_{ab}\alpha_{ab}
\right|
}{
\left|
A_{ab}
\right|
}
\right]_+
&
c_r^2 A_{ab}^2\gg c_r|G_{ab}\alpha_{ab}|\,,
\\[10pt]
\displaystyle
\left[
-\gamma_{ab}
+
\sqrt{
c_r
\left|
G_{ab}\alpha_{ab}
\right|
}
\right]_+
&
c_r^2 A_{ab}^2\ll c_r|G_{ab}\alpha_{ab}|\,,
\end{cases}
\label{eq:cfi-limiting-growth}
\end{equation}
where $c_r=|\xi_r|$, so that $c_{\rm IP}=1$ and $c_{\rm IB}=1/3$. These
conditions specify the asymptotic hierarchies used in the two
expansions. In the non-resonant regime, the approximate growth rate is
positive when $|G_{ab}\alpha_{ab}|/|A_{ab}|>\gamma_{ab}$. In the
resonance regime, the growth requires
$\sqrt{c_r|G_{ab}\alpha_{ab}|}>\gamma_{ab}$. So, resonance-like
enhancement requires a sufficiently small effective-sector imbalance
{$A_{ab}$} and a nonzero collision-rate asymmetry {$\alpha_{ab}$} that
overcomes the mean damping.

The precise position of a growth-rate maximum depends on the simultaneous
variation of $G_{ab}$, $A_{ab}$, $\gamma_{ab}$, and $\alpha_{ab}$. The
relations $|A_{ab}|\simeq|\alpha_{ab}|$ for IP modes and
$|A_{ab}|\simeq3|\alpha_{ab}|$ for IB modes characterize particular
fixed-rate parameter scans~\cite{Liu2023}.  For $G_{ab}\sim\mu_{\nu\nu}$
and $|\alpha_{ab}|\sim\gamma_{ab}\sim\Gamma$, the resonance-like scale is
schematically $\sigma^{\rm R-CFI}\sim\sqrt{\mu_{\nu\nu}\Gamma}$. For
$\Gamma\ll\mu_{\nu\nu}$, this scale lies parametrically above $\Gamma$
and below $\mu_{\nu\nu}$. It does not necessarily lie below the FFI
growth rate.
So, their relative importance must be assessed at various locations.

\subsection{Classical Neutrino Transport and Collisional Physics}
\label{sec:weak_interactions}
Accurately diagnosing FFI and CFI requires the
flavor-dependent neutrino phase-space distributions and, for CFI,
the corresponding collisional rates.
In our calculations, the diagonal
components of the density matrices in Eq.~\eqref{eq:density_matrix}
are obtained from classical general-relativistic Boltzmann transport,
whereas the off-diagonal flavor coherences are introduced only in the
QKE and LSA. Here, ``classical'' refers to flavor-diagonal kinetic
transport without flavor coherence. We discuss the classical transport framework and weak interactions adopted in this work.

Classical moment schemes such as M1 are computationally efficient and are
therefore widely used in multidimensional neutrino-radiation hydrodynamics
simulations~\cite{Skinner2019,Musolino2023,Cheong2024c,Endeve2026}. However,
they evolve the first two moments of the neutrino distributions.
This limitation is crucial for FFI,
whose occurrence depends sensitively on the detailed angular distributions,
which depend on the chosen analytical closure in moment
schemes~\cite{Cornelius2025}. These schemes could therefore underestimate the occurrence of FFI~\cite{Akaho2026}.
Various formulations have been developed to tackle this problem,
including quantum moment closures~\cite{Froustey2023,Kneller2024,Froustey2026}, which capture
the characteristic FFI and CFI growth rates with improved accuracy.

Nevertheless, the resulting instability analysis remains dependent on the
chosen closure and retains limited angular information. On the other
hand, full Boltzmann transport avoids both moment closure and angular
reconstruction of the radiation field used in
Refs.~\cite{Akaho2023,Kawaguchi2025,Nagakura2025}. Monte Carlo transport
also resolves the full momentum space, but the finite number of packets
introduces statistical noise in the
distributions~\cite{Foucart2023,Lund2025}. This can obscure the small
angular and spectral differences relevant for FFI and CFI, particularly
near weak or thermal equilibrium. Unless sufficiently many packets are
used, this noise is more severe in optically thick regions, where
interactions are stiff ~\cite{Kawaguchi2023b,Foucart2026}.

These considerations motivate the use of full momentum-space Boltzmann
transport to construct the smooth flavor-diagonal neutrino background
entering the LSA. For each neutrino flavor, the distribution function
obeys the general-relativistic Boltzmann equation describing advection in
configuration and momentum spaces:
\begin{equation}
p^\mu\frac{\partial f_{\nu_a}}{\partial x^\mu}
-
\Gamma^{i}_{\mu\nu}p^\mu p^\nu
\frac{\partial f_{\nu_a}}{\partial p^i}
=
p^{\hat 0} \mathcal C_{\nu_a}[f]\,,
\label{eq:boltzmann}
\end{equation}
with a corresponding equation for $f_{\bar\nu_a}$. Here, $p^{\hat 0}$ is
the comoving neutrino energy, $\mathcal C_{\nu_a}[f]$ denotes the
collision operator in the comoving frame, $p^\mu$ is the neutrino
four-momentum, and $\Gamma^{i}_{\mu\nu}$ denotes the Christoffel symbols.
The numerics of the Boltzmann solver are described in
Sec.~\ref{sec:snapshot-handoff}.

\begin{table}[t]
\centering
\caption{Weak interactions implemented in \texttt{Weakhub} and considered
  in this work. The labels denote charged-current processes (CC), elastic
  scattering (ES), neutrino-pair production and annihilation (PP),
  inelastic neutrino--lepton scattering (IS), lepton-flavor
  exchange/conversion (LFE/C), and inverse lepton decay (ILD). For the
  plasma process, T, A, M, and L denote the transverse, axial, mixed, and
  longitudinal modes, respectively. Here $p$ and $n$ denote a free proton
  and neutron, $N\in\{n,p\}$, $A$ denotes a heavy nucleus, and $A_{\rm
    light}$ denotes a light nuclear cluster, including $^2\mathrm{H}$,
  $^3\mathrm{H}$, $^3\mathrm{He}$, and $^4\mathrm{He}$. Unless a flavor
  is specified, $\nu$/$\bar{\nu}$ denotes that the interaction applies to
  any neutrino/antineutrino flavor. The six-species set contains all
  four-species interactions plus the additional muonic interactions in
  the lower block. The ILD treatment is extended with the vacuum decay
  rate described in App.~\ref{app:ILD_extended}. }
\label{tab:weak-interactions}
\scriptsize
\renewcommand{\arraystretch}{1.15}
\setlength{\tabcolsep}{12pt}
\begin{tabular}{@{}ll@{}}
\hline\hline
Label & Weak interaction \\
\hline
CC & $\nu_e+n\rightleftharpoons p+e^-$, \quad
$\bar\nu_e+p\rightleftharpoons n+e^+$ \\
CC & $\nu_e+A'(N+1,Z-1)\rightleftharpoons A(N,Z)+e^-$; \;
$\bar\nu_e+e^-+p\rightleftharpoons n$ \\
PP & $e^-+e^+\rightleftharpoons\nu+\bar\nu$, \quad
$A^*\rightleftharpoons A+\nu+\bar\nu$ \\
PP & $N+N\rightleftharpoons N+N+\nu+\bar\nu$; \;
$\gamma_{\rm T/A/M/L}^*\rightleftharpoons\nu+\bar\nu$ \\
ES & $\nu+N\rightleftharpoons\nu+N$, \quad
$\nu+A\rightleftharpoons\nu+A$ \\
ES & $\nu+A_{\rm light}\rightleftharpoons\nu+A_{\rm light}$ \\
IS & $e^\mp+\nu\rightleftharpoons e^\mp+\nu$ \\
\hline
\multicolumn{2}{c}{Additional six-species muonic interactions} \\
\hline
CC & $\nu_\mu+n\rightleftharpoons p+\mu^-$, \quad
$\bar\nu_\mu+p\rightleftharpoons n+\mu^+$ \\
IS & $\mu^\mp+\nu\rightleftharpoons\mu^\mp+\nu$ \\
LFE & $\nu_\mu+e^-\rightleftharpoons\nu_e+\mu^-$; \;
$\bar\nu_\mu+e^+\rightleftharpoons\bar\nu_e+\mu^+$ \\
LFC & $\bar\nu_e+e^-\rightleftharpoons\bar\nu_\mu+\mu^-$; \;
$\nu_e+e^+\rightleftharpoons\nu_\mu+\mu^+$ \\
ILD & $\nu_\mu+\bar\nu_e+e^-\rightleftharpoons\mu^-$; \;
$\bar\nu_\mu+\nu_e+e^+\rightleftharpoons\mu^+$ \\
\hline\hline
\end{tabular}
\end{table}

Flavor instabilities depend sensitively on the properties and populations
of the neutrino radiation fields, which are in turn strongly shaped by
weak interactions in compact-object environments. Various classes of
interactions are included in our calculations and summarized in
Tab.~\ref{tab:weak-interactions}. The corresponding opacities and
interaction kernels depend on the local thermodynamic state and neutrino
phase-space variables defined in the fluid-comoving frame~\cite{Ng2024a}.

The upper interaction set in Tab.~\ref{tab:weak-interactions}, including
charged-current (CC) reactions, elastic scattering (ES),
neutrino--antineutrino pair processes (PP), and inelastic scattering (IS)
between neutrinos and electrons/positrons
~\citep{Bruenn85,Mezzacappa1993b,Guo2020,Ng2024a}, is included in both
the two-flavor and three-flavor frameworks. For the three-flavor
calculations in $npe\mu$ matter, we additionally include the lower
interaction set consisting of muonic CC reactions, IS between neutrinos
and muons/antimuons, lepton-flavor exchange/conversion (LFE/C), and
inverse lepton decay (ILD). LFE/C and ILD can directly redistribute
neutrinos between the electron- and muon-flavor (anti)neutrinos and
change the composition of the corresponding charged leptons. At low
neutrino energies and under representative PNS-neutrinosphere conditions
with density $\rho\sim10^{13}~{\rm g\,cm^{-3}}$ and temperature $T$ of a
few $\rm{MeV}$, LFE/C can dominate the opacity for $\nu_e$ and
$\bar\nu_\mu$, whereas ILD can dominate for $\bar\nu_e$ and
$\nu_\mu$~\citep{Guo2020,Sugiura2022}. Besides, because the tauon rest
mass is much larger than the thermal energy scale of the merger remnant,
tauons, antitauons, and tauonic charged-current reactions are neglected.

We apply various corrections and treatments to the interactions listed in
Tab.~\ref{tab:weak-interactions}. Except for the $\beta$ processes
involving heavy nuclei, which are approximated following
Ref.~\cite{Bruenn85}, the CC reactions are evaluated with the 2D integral
approach within a full-kinematics
framework~\cite{Guo2020,Alford2021b,Ng2024a}, retaining the energy and
momentum transfer of all particles, weak-magnetism corrections, and
mean-field medium modifications from the EOS. We do not include nucleon
form factors so that the 2D integral approach agrees with the direct 4D
integration to within $0.1\%$ for neutrino energies up to at least
$\varepsilon=150\,{\rm MeV}$~\cite{Guo2020}. For neutrino--nucleon ES,
we include nucleon recoil, weak magnetism, virial-response
corrections~\cite{Horowitz2016}, and a strange-quark contribution with
$g_s=-0.1$~\citep{Horowitz2002}. For nucleon-nucleon bremsstrahlung, we
adopt the normalized dynamical structure function tabulated in
Ref.~\cite{Guo2019}. Compared with the commonly used one-pion-exchange
approach~\cite{Hannestad1997}, this $\chi$EFT $T$-matrix treatment
includes non-diagonal contributions and final-state Pauli blocking and
remains applicable for arbitrary nucleon degeneracy. The resulting rate
is enhanced at low density and low neutrino energy by the resonant
low-energy nucleon-nucleon interaction, but is suppressed at higher
density ($\rho\sim{\rm few}\times10^{12}\,{\rm g\,cm^{-3}}$) or energy,
where the $T$-matrix decreases with nucleon relative momentum and Pauli
blocking becomes important~\cite{Guo2019}. Finally, the kernel
calculations of LFE/C and ILD follow Refs.~\cite{Bolligthesis,Guo2020}.
\section{Numerical Methods and Setup}\label{sec:secIII}
\subsection{Merger-Remnant Handoff to the Boltzmann and Quantum Kinetic Solvers}\label{sec:snapshot-handoff}
We numerically solve the 2.5D axisymmetric general-relativistic Boltzmann
neutrino-radiation equations in a conformally flat spacetime through a
handoff procedure~\cite{Ng2024b}. The initial remnants were modeled
using fully general-relativistic grey moment-based neutrino-radiation
simulations. We first hand off the simulation outcomes of equal-mass,
irrotational BNS merger systems from Ref.~\cite{Ng2024c} (see the setup
therein), with a total binary gravitational mass of $2.5~M_{\odot}$.
These BNS merger systems are expected to form long-lived remnants. The
initial simulations were performed with
\texttt{FIL-M1}~\citep{Most2019b,Musolino2023}, a fully
general-relativistic hydrodynamics (GRHD), grey, moment-based (M1)
neutrino-radiation code, in 3D Cartesian geometry with zero magnetic
field. The simulations evolved three and five neutrino species in $npe$
and $npe\mu$ matter, respectively~\cite{Ng2024c}. For these systems at
$\bar t\simeq30\,{\rm ms}$ after merger, where $\bar t$ denotes the time
measured after the moment of merger, we perform a coordinate
transformation and azimuthal averaging of the hydrodynamic and metric
fields. The resulting snapshots are then handed off to the
multidimensional GRMHD Boltzmann-neutrino-radiation code
\texttt{Gmunu}~\citep{Cheong2021}. The evolution is continued
axisymmetrically in cylindrical coordinates $(R,z,\phi)$ with a
conformally flat spacetime, where the metric is evolved by solving the
conformal-flatness-condition equations~\cite{Cheong2021,Ng2024b}.

The radiation fields in \texttt{Gmunu} are initialized and evolved with
the general-relativistic Boltzmann solver~\cite{Offermans2026}, which directly evolves the distribution
$f_{\nu_a}(x^\mu;\varepsilon,\Omega_p)$ with configuration space
described in the Eulerian frame and the full momentum space measured in the orthonormal fluid-comoving frame.
A velocity-dependent tetrad relates the Eulerian-frame spacetime basis to the orthonormal fluid-comoving momentum basis.
Using the fluid-comoving frame for momentum space provides several advantages. It is convenient for microphysical calculations, as the local neutrino energy and angular distributions are directly available in the comoving frame. Besides, the interaction rates described in Sec.~\ref{sec:weak_interactions} are computationally expensive to evaluate on the fly and are typically tabulated as functions of the local thermodynamic state and comoving neutrino energy.
This choice naturally matches the tabulated interaction rates and momentum-space discretization, avoiding repeated Lorentz transformations and numerical remapping of the discretized neutrino distributions between Eulerian and comoving momentum grids, and thereby reducing interpolation errors in the collision source terms.

The resulting momentum-space advection accounts for gravitational redshift, Doppler shifts, relativistic aberration, and
frame-dragging contributions.
The Boltzmann solver evolves four species in the $npe$ models,
$\{\nu_e,\bar\nu_e,\nu_{\chi},\bar\nu_{\chi}\}$, where $\nu_{\chi}$ and
$\bar\nu_{\chi}$ each represent two degenerate heavy-lepton flavors,
and all six neutrino species in the $npe\mu$ models,
$\{\nu_e,\bar\nu_e,\nu_\mu,\bar\nu_\mu,\nu_\tau,\bar\nu_\tau\}$.
The interactions in Table~\ref{tab:weak-interactions} are implemented and supplied through \texttt{Weakhub}~\cite{Ng2024a}.

Following Ref.~\cite{Ng2024c}, we extend \texttt{Gmunu} by including the
thermodynamic contributions of ideal Fermi gases of $e^\mp$ and $\mu^\mp$
as functions of $\rho$, $T$, and $Y_l$, where $l\in\{e,\mu\}$ and
$Y_l=Y_{l^-}-Y_{l^+}$, subject to the charge-neutrality condition for the
proton fraction, $Y_p=Y_e+Y_\mu$. To avoid double counting, we first
subtract the original $e^\mp$ contribution from the EOS and retain the
nuclear EOS table as a function of $(\rho,T,Y_p)$. We then add the
separately tabulated ideal-Fermi-gas contributions of $e^\mp$ and
$\mu^\mp$. As muons demuonize at low densities and temperatures, we can
omit the muonic degree of freedom in the approximate Helmholtz
EOS~\cite{Ng2026b}, which is applied in regions outside the density and
temperature ranges of the nuclear EOS or outside the nuclear statistical
equilibrium regime.

After the handoff from the 3D simulations to the 2D setup, we discard the
inherited grey M1 radiation variables, initialize the Boltzmann radiation
fields with $f_\nu=0$, and relax them for $2\,{\rm ms}$ while holding the
matter and metric fields fixed. This prevents sudden changes in the
initialized neutrino radiation fields from dramatically affecting the
hydrodynamics and composition. Unlike Ref.~\cite{Nagakura2025}, which
assesses flavor instabilities after relaxing the radiation field on a
frozen matter background, we further evolve the matter, metric, and
radiation fields self-consistently for $8\,{\rm ms}$. This allows the
coupled matter--neutrino system to readjust and reduces perturbations and
numerical artifacts from the change in radiation scheme, dimensional
reduction, and handoff, yielding a hydrodynamically relaxed
configuration. We then perform LSA on the resulting $npe$ and $npe\mu$
configurations to characterize their global flavor-instability
properties.

Finally, we complement the global survey with local, energy-dependent QKE calculations at three representative locations,
selected to probe distinct combinations of unstable CFI sectors
and their growth-rate hierarchies (see Sec.~\ref{sec:QKE_results} for details).
These calculations resolve the microscopic flavor-evolution timescales. Resolving them over the macroscopic evolution timescale of a global simulation would be computationally prohibitive.
The calculations examine the growth and nonlinear coupling of the flavor sectors, asymptotic behavior, and collisional repopulation to guide future subgrid treatments
for global simulations.

As we focus on CFI properties, we initialize the density
matrices from the angle-averaged distributions and introduce
small flavor-coherence perturbations.
We retain all three flavors and impose spatial homogeneity
and isotropy throughout the local evolution, thereby isolating
the IP modes of CFI.
The energy-dependent density matrices are evolved directly,
rather than replacing the spectra with the effective sectors
used in the approximate CFI diagnostic.

To isolate the contributions of the diagonal and off-diagonal
components of the collision operator in
Eq.~\eqref{eq:qke-reduced-collision}, we consider three types of collision terms:
off-diagonal terms only (\texttt{QKE-noDC}), the full collision operator
(\texttt{QKE-fullC}), and diagonal terms only (\texttt{Classical}).
The local matter state and energy-dependent emissivities and opacities
are held fixed at their extracted values, while the neutrino
self-interaction Hamiltonian is updated from the evolving density
matrices.
\subsection{Numerical Treatment of the Dilute-Muon Limit}
\label{sec:dilute-muon}

A particular numerical difficulty arises when matter demuonizes toward a
very small net muon fraction. Existing simulations necessarily represent
$Y_\mu$ over a finite numerical range, with lower boundaries typically of
order $Y_\mu\sim10^{-5}$--$10^{-4}$, while muon-rich matter can reach
$Y_\mu\sim10^{-3}$--$10^{-2}$ in dense and hot regions
~\cite{Bollig2017,Fischer2020b,Gieg2026}. Although such lower bounds
appear numerically small, they still remain many orders of magnitude
above the physical values of $Y_\mu$ actually reached in transparent
matter after demuonization in order to attain $\beta$-equilibrium, i.e.,
\begin{equation}
\mu_n-\mu_p-\mu_\mu
\to
\mu_n-\mu_p-\mu_e\,,
\label{eq:npmu-eqm}
\end{equation}
where $\mu_n$, $\mu_p$ and $\mu_l$ ($l\in\{e,\mu\}$) denote the relativistic
chemical potential of neutron, proton and 
charged lepton, respectively.
Under these conditions, and together with the weak-equilibrium condition
$\mu_\mu\rightarrow\mu_e$ through processes such as muon decay, $\mu^-
\to e^- + \nu_\mu + \bar{\nu}_e$, the muon fraction keeps vanishing, i.e.,
$Y_\mu\rightarrow0$.

Indeed, in the nondegenerate Boltzmann limit, the net muon fraction of an
ideal muon gas is
\begin{equation}
Y_\mu\simeq
\frac{2(m_\mu c^2)^2T}
{\pi^2n_{\rm B}(\hbar c)^3}
K_2\!\left(\frac{m_\mu c^2}{T}\right)
\sinh\!\left(\frac{\mu_\mu}{T}\right)\,,
\label{eq:boltzmann-muon-fraction}
\end{equation}
where $n_{\rm B}$ is the baryon number density and $K_2$ is the modified
Bessel function of the second kind. For example, near the typical
low-end values in density and temperature of the simulation, requiring
$|\mu_\mu|\lesssim1\,{\rm MeV}$ corresponds to $Y_\mu\sim10^{-22}$ at
$(\rho,T)=(10^{10}\,{\rm g\,cm^{-3}},2\,{\rm MeV})$, and to
$Y_\mu\sim10^{-85}$ at $(\rho,T)=(10^{4}\,{\rm g\,cm^{-3}},0.5\,{\rm
  MeV})$. Conversely, even a seemingly negligible value as small as
$Y_\mu\sim10^{-6}$ in a typical
simulation can still have $\mu_\mu\simeq80$--$100\,{\rm MeV}$
(this is clearly visible in Fig.~3.5 of Ref.~\cite{Bolligthesis}). Thus,
simply lowering the numerical floor from $Y_\mu\sim10^{-4}$ to $10^{-6}$
does not allow one to recover the physical $\mu_\mu \rightarrow \mu_e$
dilute limit.

On the other hand, setting $Y_\mu=0$ as the strict lower bound is also
numerically unfeasible, as small contributions to $Y_\mu$ can be
repeatedly discarded when enforcing composition bounds during
reconstruction, primitive recovery, and weak-interaction source updates,
leading to violations of muon-lepton-number conservation. As a result,
resolving this dilute low-$\rho$ and low-$T$ limit directly is
numerically impractical. 
That said, a leptonic EOS is needed to cover $Y_\mu$ from $\sim10^{-1}$ down to $10^{-20}$–$10^{-90}$ with sufficient interpolation accuracy to retain the smooth transition $\mu_\mu\rightarrow\mu_e$ as $Y_\mu\rightarrow0$. Such vanishing values are far below the typical numerical accuracy of conservative-variable updates and stiff weak-interaction source terms in GRMHD simulations.

The unresolved dilute limit becomes particularly problematic through the equilibrium neutrino degeneracy $\eta^{\rm eq}_{\nu}$ entering detailed balance. For muonic CC reactions, the collision source drives the neutrino distribution toward the equilibrium target $f_\nu^{\rm
  eq}(\eta_\nu^{\rm eq})$ and can be written as
\begin{equation}
\mathcal C_{\nu_\mu/\bar{\nu}_\mu}^{\rm CC}
=
\kappa_a^*
\left[
f_{\nu_\mu/\bar{\nu}_\mu}^{\rm eq}
\!\left(\eta_{\nu_\mu/\bar{\nu}_\mu}^{\rm eq}\right)
-
f_{\nu_\mu/\bar{\nu}_\mu}
\right]\,,
\label{eq:CC_DB}
\end{equation}
where $ \eta_{\nu_\mu}^{\rm eq} = (\mu_\mu+\mu_p-\mu_n)/T$ is the equilibrium 
degeneracy parameter for $\nu_\mu$.
Thus, a large $\mu_\mu$ can lead to an extremely large $|\eta_{\nu_\mu/\bar{\nu}_\mu}^{\rm eq}|$, yielding an unphysical and false demuonization target in the unresolved low-$Y\mu$ regime down to the EOS lower bound.
A lower EOS-table bound or an atmospheric floor
in $Y_\mu$, together with the unresolved $\mu_\mu$ end-point, can
act as a persistent reservoir for $\nu_\mu$ emission through the
effectively inexhaustible equilibrium target
$f_{\nu_\mu/\bar{\nu}_\mu}^{\rm eq} (\eta_{\nu_\mu/\bar{\nu}_\mu}^{\rm
  eq})$. The emitted $\nu_\mu$ can subsequently be reabsorbed and
increase $Y_\mu$ again to allow artificial muonization and
demuonization, thus violating lepton number conservation.
In addition, if the unresolved low-$\rho$ and low-$T$ matter
retains large values of $\mu_\mu$ the matter cannot relax toward the correct neutrinoless equilibrium [see Eq.~\eqref{eq:npmu-eqm}]. This behavior is evident in the $\mu_\mu$ profiles computed for CCSNe~\cite{Bolligthesis} and 
the incorrect equilibrium found outside the disks of BNS remnants~\cite{Ng2024c,
  Gieg2026}.
We observe that this problem leads to similar qualitative behavior in both grey-M1~\cite{Ng2024c} and 
energy-dependent Boltzmann calculations. 

Different prescriptions have been adopted to address this subtle problem. 
Ref.~\cite{Bolligthesis} allows negative $Y_\mu$ to ensure continuous evolution through $Y_\mu=0$ to recover $\mu_\mu=0$. However, source-driven oscillations across $Y_\mu=0$ can, through charge neutrality, push $Y_p$ outside the admissible range of the baryonic EOS, inducing oscillatory corrections in $Y_e$, $Y_\mu$, and $Y_p$.
Ref.~\cite{Ng2024c} instead suppresses $\eta_{\nu_\mu/\bar{\nu}_\mu}^{\rm eq}$ using optical-depth thresholds, recovering the muonless equilibrium for CC and PP interactions outside the semitransparent region.
Ref.~\cite{Gieg2026} retains the local $\eta_{\nu_\mu/\bar{\nu}\mu}^{\rm eq}$ everywhere, computing the PP opacity from the kernel with a thermal, weak-equilibrium partner distribution and the emissivity through the CC detailed balance (Eq.~\eqref{eq:CC_DB}).
This may drive $\nu_\mu$ and $\bar{\nu}_\mu$ toward unphysical asymmetric equilibrium targets set by unresolved high-$\mu_\mu$ values, artificially altering $Y_\mu$ despite the original PP interactions preserving the matter composition. 
More generally, suppressing muonic weak interactions using too high $\rho$ or $T$ thresholds can artificially reduce the separation between the $\nu_\mu$ and $\bar{\nu}_\mu$ neutrinospheres, artificially suppressing flavor instabilities.

While the different prescriptions described above tackle several
numerical manifestations of the muonic problem with different degrees of
success, none of them resolves directly the dilute limit
$Y_\mu\rightarrow0$ with $\mu_\mu\rightarrow\mu_e$ and the problem of the
incorrect neutrinoless $np\mu$ equilibrium. 
Therefore, we invent a regularization for the unresolved
dilute regime by modifying the tabulated $\mu_\mu^{\rm tab}$ with
\begin{equation}
\begin{aligned}
\mu_\mu &=
\begin{cases}
\mu_\mu^{\rm tab}, & \hspace{-0.5em}\mu_\mu^{\rm
  tab}<\mu_e\\ \mu_e+(\mu_\mu^{\rm tab}-\mu_e)\tanh\!\left[
  \frac{\max(Y_\mu-Y_{\mu,{\rm thr}},0)} {\Delta_{Y_\mu}} \right]\,,
& \hspace{-0.5em}\mu_\mu^{\rm tab}\geq\mu_e
\end{cases}
\end{aligned}
\label{eq:mumu-regularization}
\end{equation}
Here, we choose $\Delta_{Y_\mu}=5\times10^{-5}$ and $Y_{\mu,{\rm
    thr}}=6\times10^{-4}$, slightly above the lower table bound of
$Y_\mu^{\rm min} = 5\times10^{-4}$. For $\mu_\mu^{\rm tab}\geq\mu_e$,
this prescription smoothly drives $\mu_\mu\to\mu_e$ as $Y_\mu\to
Y_{\mu,{\rm thr}}$, with $\mu_\mu=\mu_e$ for $Y_\mu\leq Y_{\mu,{\rm
    thr}}$. States with $\mu_\mu^{\rm tab}<\mu_e$ remain unmodified,
including at and below the abundance threshold. For $Y_\mu-Y_{\mu,{\rm
    thr}}\gg\Delta_{Y_\mu}$, the tabulated value
$\mu_\mu\simeq\mu_\mu^{\rm tab}$ is recovered in either branch.

This method allows demuonizing matter to approach the 
accessible dilute limit $Y_\mu\simeq Y_{\mu,{\rm thr}}$ with
$\mu_\mu\simeq\mu_e$, facilitating the recovery of the neutrinoless
$np\mu$- and $npe$-$\beta$-equilibria. Therefore, in the cold,
neutrino-transparent limit with $\mu_{\nu_\mu}^{\rm eq}\to0$ and
$\varepsilon/T\gg1$, the detailed-balance emissivity
$j_{\nu_\mu}=\kappa_a^*f_{\nu_\mu}^{\rm eq}
\simeq\kappa_a^*\exp(-\varepsilon/T)$ becomes exponentially suppressed,
removing spurious persistent $\nu_\mu$ emission and preventing residual
numerical muons from becoming an inexhaustible reservoir during
demuonization. Conversely, the prescription does not switch off
absorption of incident neutrinos, that is, $C_{\nu_\mu}^{\rm
  CC}\simeq-\kappa_a^*f_{\nu_\mu}$ when $f_{\nu_\mu}\gg f_{\nu_\mu}^{\rm
  eq}$, thus allowing muonization from initially muonless matter wherever
the reaction is kinematically allowed. At the same time, it preserves
the tabulated value $\mu_\mu^{\rm tab}$ when $\mu_e>\mu_\mu^{\rm tab}$,
as it can occur during core bounce of a collapsing progenitor, thus
allowing muonization at sufficiently high $\rho$, $T$, and incident
neutrino energies. Overall, because this treatment can accommodate
muonization in core collapse and the merger of BNS, as well as
demuonization of accreting/ejecting matter, it may represent the optimal
solution for dilute low-$\rho$ and low-$T$ matter.

\subsection{Linearized Collisional Sources and Implicit Radiation--Composition Coupling}
\label{sec:implicit-composition}

The neutrino collision terms become stiff in dense matter and necessarily
require an implicit treatment, which is much slower than an explicit
update. The simplest treatment is the single-species, single-group
(SSSG) scheme~\cite{Cheong2023,Offermans2026}, in which each neutrino
species and energy group is evolved independently during the implicit
update. The SSSG scheme scales linearly with the total number of
radiation degrees of freedom, $\mathcal O\!\left( N =N_{\rm
  spec}N_\varepsilon N_\vartheta N_\varphi \right)$, where $N_{\rm spec}$
is the number of neutrino species, $N_\varepsilon$ the number of energy
groups, and $N_\vartheta$ and $N_\varphi$ the numbers of angular bins,
respectively. 
While effective, this approach can only treat source terms
expressed in terms of emissivities and opacities, so that kernel-based
interactions, e.g., PP and IS, that couple different energies or species
cannot be included.

Another approach, the multispecies, multigroup (MSMG) treatment consistently
retains all of these kernel-based couplings, including the full cross-energy, momentum-space, and species couplings~\cite{Offermans2026}. The implicit source update becomes a
nonlinear system of $N$ radiation unknowns at every spatial cell. This
approach is particularly expensive for kernel-based interactions, which
couple a target state $(\nu_a,\varepsilon,\Omega_p)$ to an partner state
$(\nu_b,\varepsilon',\Omega'_p)$ through the corresponding microphysical
kernels~\cite{Fischer2020b}. A global Newton--Raphson-type treatment
requires repeated evaluations of the collision integrals, the
construction of a generally nondiagonal $N\times N$ Jacobian, and the
solution of the associated linear system at every iteration. A direct
factorization alone scales as $ \mathcal O(N^3) = \mathcal O\!\left(
N_{\rm spec}^3 N_\varepsilon^3 N_\vartheta^3 N_\varphi^3 \right)$. These
iterative nonlinear treatments remain costly because the large coupled
collision operator must be repeatedly evaluated. Although
Ref.~\cite{Offermans2026} developed more efficient methods for the full
MSMG treatment, its cost with a small number of angular bins is still
dominated by an $N_{\rm spec}^3 N_\varepsilon^3$ term, in contrast with
the linear contributions in $N_{\rm spec}$ and $N_\varepsilon$ in the
SSSG treatment. As a result, the MSMG treatment remains prohibitively
expensive for multi-D simulations with Boltzmann transport at practical
momentum-space resolutions.

Various approximations have been introduced to reduce kernel-based
interactions to effective opacities and emissivities by assuming thermal
and weak-equilibrium partner distributions, e.g.,
Refs.~\cite{Burrows2006b,Oconnor2015}. Such approximations can reproduce
important features of PP in certain regimes compared with the full
kernel-based treatment~\cite{Oconnor2015,Cheong2024c}. However, for
strongly inelastic interactions such as IS, these approximations can
substantially reduce the accuracy and fail to retain the inelastic nature
of the interaction. To counter this, we introduce an accurate and
efficient approximation -- linearized single-species, single-group
(SSSG-lin) treatment. First, the general form of a kernel-based
collision term can be written as
\begin{equation}
\mathcal C_{\nu_a}^{X}
=
\int dV'_p
\left[
R_X^{+}\,
\mathcal B_X^{+}(f'_{\nu_b})
(1-f_{\nu_a})
-
R_X^{-}\,
\mathcal B_X^{-}(f'_{\nu_b})
f_{\nu_a}
\right]\,,
\label{eq:generic-kernel-source}
\end{equation}
where $X\in\{{\rm PP,IS,LFE/C,ILD}\}$, $f'_{\nu_b}:=
f_{\nu_b}(\varepsilon',\Omega'_p)$ denotes the partner distribution, and
$R_X^{+}(\varepsilon,\varepsilon',\omega)$ and
$R_X^{-}(\varepsilon,\varepsilon',\omega)$ are 
the corresponding gain and loss kernels for the target neutrino $\nu_a$.
Here, $\omega$ is the cosine of the angle between
the directions of the target and partner neutrinos. The kernels are
represented by their zeroth- and first-order Legendre
moments~\cite{Bruenn85, Ng2024a} and the partner occupation/blocking
factors are given by
\begin{equation}
\begin{aligned}
\mathcal B_X^{+}(f'_{\nu_b})
&=
\begin{cases}
f'_{\nu_b}, & X\in\{{\rm IS,LFE/C}\},\\
1-f'_{\nu_b}, & X\in\{{\rm PP,ILD}\},
\end{cases}
\\
\mathcal B_X^{-}(f'_{\nu_b})
&=
\begin{cases}
1-f'_{\nu_b}, & X\in\{{\rm IS,LFE/C}\},\\
f'_{\nu_b}, & X\in\{{\rm PP,ILD}\}\,.
\end{cases}
\end{aligned}
\label{eq:generic-opponent-factors}
\end{equation}

Monte-Carlo radiation schemes can become noisy in optically thick regions
when the number of packets is limited. Their partner distribution
requires an interpolation between a local-thermal-equilibrium (LTE)
distribution and a lagged distribution (see Eq.~(22) in
Ref.~\cite{Foucart2026}). Instead, we use as the partner distribution the
full transported distribution from the preceding
substep~\cite{Sumiyoshi2012,Kawaguchi2025,Foucart2026},
\begin{equation}
f'_{\nu_b}
\longrightarrow
f_{\nu_b}^{\rm lag}(\varepsilon',\Omega'_p)\,,
\label{eq:sssg-lagged-f}
\end{equation}
without replacing it by an LTE or weak-equilibrium distribution. The
corresponding effective emissivity and stimulated absorption opacity are
calculated independently from the $R_X^{\pm}$ kernels as
\begin{equation}
\begin{aligned}
j_\nu^{X}
& :=
\int dV'_p\,
R_X^{+}\,
\mathcal B_X^{+}
\!\left(f_{\nu_b}^{\rm lag}\right),
\\
\kappa_{a,\nu}^{*,X}
& :=
\int dV'_p
\left[
R_X^{+}\,
\mathcal B_X^{+}
\!\left(f_{\nu_b}^{\rm lag}\right)
+
R_X^{-}\,
\mathcal B_X^{-}
\!\left(f_{\nu_b}^{\rm lag}\right)
\right]\,,
\end{aligned}
\label{eq:sssg-lin-coefficients}
\end{equation}
rather than deriving the emissivity from the opacity through a CC-like
detailed-balance relation involving a prescribed target distribution
$f^{\rm eq}_\nu(\eta^{\rm eq}_{\nu})$~\cite{Oconnor2015,Gieg2026}. This
avoids introducing an artificial neutrino--antineutrino asymmetry that
could feed back on the matter composition. At the same time, the
detailed-balance relations between $R_X^{+}$ and $R_X^{-}$ are still
imposed~\citep{Guo2020,Ng2024a}, e.g., $\Phi_{\mathrm{PP}}^{+} =
\Phi_{\mathrm{PP}}^{-} {\rm e}^{-(\varepsilon+\varepsilon')/T}$ for
PP. Under these conditions, the collision source then reduces to $
\mathcal C_{\nu_a}^{X} = j_{\nu_a}^{X} - \kappa_{a,\nu_a}^{*,X}f_{\nu_a}$.

We note that SSSG-lin not only avoids the global $\mathcal O(N^3)$ MSMG
Jacobian inversion, but also reduces the nonlinear coupled system to
independent linear radiation systems for each species and energy
group. Furthermore, in SSSG-lin, lepton-number conservation can be
violated at a small level because of the lagged partner
distribution~\cite{Foucart2026}. We keep the contributions of IS and PP
in $\nu_e$ and $\bar\nu_e$, but do not couple their source terms to the
matter composition. However, the intrinsically small timestep of the
Boltzmann solver~\cite{Offermans2026} largely suppresses these
conservation errors. Also, muonic interactions become negligible in
low-$\rho$, low-$T$ regions as $Y_\mu$ vanishes, except for demuonization
driven primarily by muon decay (see App.~\ref{app:ILD_extended} for
details). In practice, we apply a smooth density--temperature suppression
factor 
to the LFE/C and CC rates, with
thresholds $\rho=10^{11}\,{\rm g\,cm^{-3}}$ and
$T=2.5\,{\rm MeV}$ 
(see Eq.~(9.31) in \cite{Bolligthesis}).
A comparison of
the SSSG-lin and MSMG treatments is presented in
App.~\ref{app:hot-star-demuonization}, and shows a close agreement in the
hydrodynamical quantities with differences within $\lesssim5\%$,

Another difficulty arises when the radiation--matter source becomes
sufficiently stiff that $Y_e$ or $Y_\mu$ changes substantially during the
implicit steps. In many radiation-hydrodynamics schemes, the hydrodynamic
state is held fixed while the radiation field is solved implicitly, and
the composition is updated only afterwards~\cite{Foucart2015a,
  Radice2022, Schianchi2023, Cheong2023}. However, when the radiation
timescale becomes much shorter than the evolution timestep, this
approximation can produce large composition overshoots or oscillations,
particularly in regions where $Y_\mu$ varies more rapidly than $Y_e$.
Some radiation treatments reduce this stiffness by first diagnosing rapid
equilibration and then modifying the rates at an estimated
beta-equilibrium state~\cite{Radice2022, Gieg2026} or employing an
expensive fully implicit update of the hydrodynamic and radiation
variables~\cite{Kuroda2016,Laiu2021}.

To avoid modifying the input rates and increase accuracy, we here
introduce an additional but computationally affordable nested implicit
treatment of $Y_e$ and $Y_\mu$ together with the radiation sources,
forming a two-dimensional nonlinear problem in which the interaction
rates are evaluated consistently with the trial composition. The
self-consistent composition is determined from the solution of the
nonlinear equation
\begin{equation}
\bm F(\bm y)
=
\mathcal M(\bm y)-\bm y
=
\bm 0\,,
\end{equation}
where $\mathcal M$ is the radiation--composition update map and
\hbox{$\bm y=(Y_e,Y_\mu)^{\rm T}$}. We solve this two-variable nonlinear
system with a Broyden method~\cite{Press1992}. More specifically, for
each trial composition, $\mathcal M(\bm y)$ performs an inexpensive
primitive recovery of $Y_e$ and $Y_\mu$, recomputes the EOS and
interaction rates while holding the remaining hydrodynamic quantities
$\rho$, $T$ and $v^i$ fixed and performs the complete inner implicit
SSSG-lin radiation solve-step. We normally employ a normalized residual
tolerance of $10^{-6}$, an iteration tolerance of $10^{-8}$, and a
maximum of five outer iterations. The original fixed-composition implicit
predictor is retained if the nested solver fails.

To reduce the computational costs, the composition-nested implicit solver
is by-passed when the source is not sufficiently stiff, i.e., when the
local proper-time interval of the IMEX stage is shorter than both the
electron- and muon-flavor equilibration timescales of
\begin{equation}
\tau_\ell
= \min_{\substack{
q\in\{\nu_\ell,\bar\nu_\ell\}\\
\varepsilon,\Omega_p}}
\Gamma_{{\rm eq},q}^{-1}\,,
\qquad
\Gamma_{{\rm eq},q}
:=
\left[
\kappa_{a,q}^{*}
\left(
\kappa_{a,q}^{*}+\kappa_{{\rm s},q}
\right)
\right]^{1/2}\,,
\label{eq:implicit-composition-trigger-main}
\end{equation}
where $\ell\in\{e,\mu\}$ and $\kappa_{\rm s}$ is the ES opacity. Overall, this approach is able to
stably capture the rapid feedback of the interactions on $Y_e$ and
$Y_\mu$ at a substantially lower cost than a fully implicit update of the
hydrodynamic and radiation variables. 

Combining the nested implicit treatment with SSSG-lin reduces the computational cost by factors of 15 and 24 compared with MSMG for the $npe$ and $npe\mu$ tests, respectively, in App.~\ref{app:hot-star-demuonization}.



%
\begin{figure*}
    \includegraphics[width=0.9\textwidth]{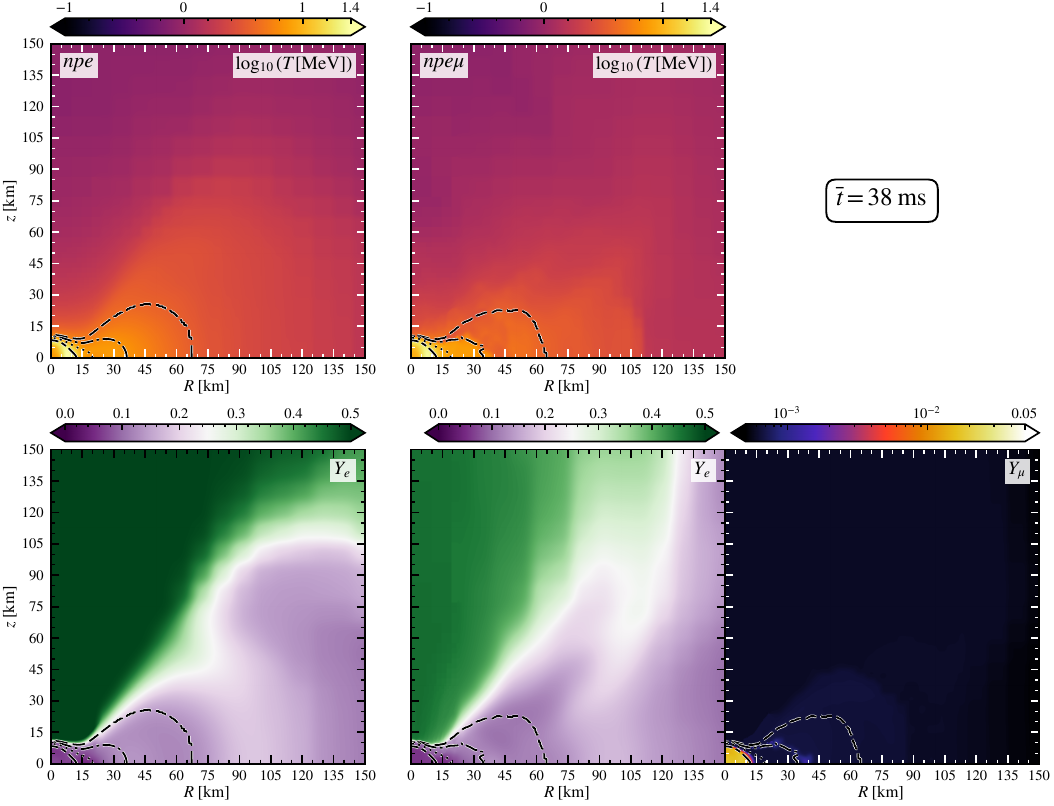}
    \caption{ 
      Meridional slices of the temperature $T$ and composition ($Y_e$ and
      $Y_\mu$) in the $npe$ (left) and $npe\mu$ (right) models,
      respectively. The snapshots are shown at a postmerger time of $\bar
      t=38~{\rm ms}$.  The four black contour styles denote rest-mass
      densities of $\rho=10^{14},\,10^{13},\,10^{12}$, and $10^{11}\,{\rm
        g\,cm^{-3}}$. }
    \label{fig:fig1}
\end{figure*}
%
\subsection{Numerical setup}
\label{sec:numerical-setup}
For each post-handoff model, we retain the same nuclear EOS as in the corresponding parent 3D simulation: DD2~\cite{Typel:2009sy}.
As reported in Ref.~\cite{Ng2024c}, muonic effects are milder for the DD2 EOS than for softer EOSs, motivating our choice to start with a case exhibiting smaller differences between $npe$ and $npe\mu$ matter.
The leptonic thermodynamic tables follow the
same format as in Ref.~\cite{Ng2024c}.

The configuration-space domain imposes axisymmetry, with reflection symmetry across $z=0$. We employ seven refinement levels. The finest level has a resolution of
$\Delta R=\Delta z= 150\,{\rm m}$ and three finest levels cover
$(R,z)\leq(22,12)$, $(38,20)$, and $(60,30)\,{\rm km}$,
respectively. The remaining coarser levels extend the computational
domain to $310\,{\rm km}$ in both directions.
The momentum-space grid uses \((N_\varepsilon,N_\vartheta,N_\varphi)=(16,10,6)\), with spherical coordinates \(\Omega_p=(\vartheta,\varphi)\).
The first energy group is centered at
$1\,{\rm MeV}$ with a width of $2\,{\rm MeV}$, while the remaining
group centers are distributed logarithmically up to $\sim300\,{\rm MeV}$.
The angular grid covers the full solid angle,
$0\leq\vartheta\leq\pi$ and $0\leq\varphi<2\pi$.

\texttt{Weakhub} uses a hybrid strategy that combines
tabulated rates for computationally expensive processes, such as full-kinematics CC, LFE/C and ILD, with on-the-fly calculations for inexpensive interactions, such as ES and
nucleon--nucleon bremsstrahlung. This approach trades the
computational cost against the memory requirements of the interaction tables.

The \texttt{Weakhub} table resolutions are chosen separately for the different
interaction channels. The electron- and muon-flavor CC tables use
$75\times55\times50\times15$ and
$48\times45\times40\times40$ points, respectively, in
$(\rho,T,Y_e,Y_\mu)$ for each neutrino energy.
Both grids are sampled in $(\log\rho,\log T,Y_e,\log Y_\mu)$,
with uniform spacing for electron-flavor CC and finer sampling
for muon-flavor CC at
$\rho>10^{11}\,\mathrm{g\,cm^{-3}}$, $T>1\,\mathrm{MeV}$,
and $Y_e\leq0.3$, reflecting the reduced importance of
muon-flavor CC interactions in low-density and low-temperature regions.
For each neutrino-energy pair $(\varepsilon,\varepsilon^{\prime})$
and each of the zeroth- and first-order Legendre moments,
the LFE/C and ILD kernels use
$35\times40\times40$ points in
$(T/\mathrm{MeV},\eta_e,\eta_\mu)
\in[0.5,150]\times[0.1,100]\times[0.1,100]$,
where $\eta_l=\mu_l/T$, with logarithmic spacing along all three axes.

The hydrodynamic and Boltzmann equations are evolved using a
second-order finite-volume discretization~\cite{Cheong2021,Offermans2026}. Advection in both
configuration and momentum space employs third-order piecewise parabolic reconstruction~\cite{Colella84} and the Harten--Lax--van Leer approximate Riemann solver~\cite{Harten83}. Time integration uses
the third-order IMEX-CB3a integrator~\cite{Cavaglieri2015}.

\begin{figure*}
    \includegraphics[width=0.8\textwidth]{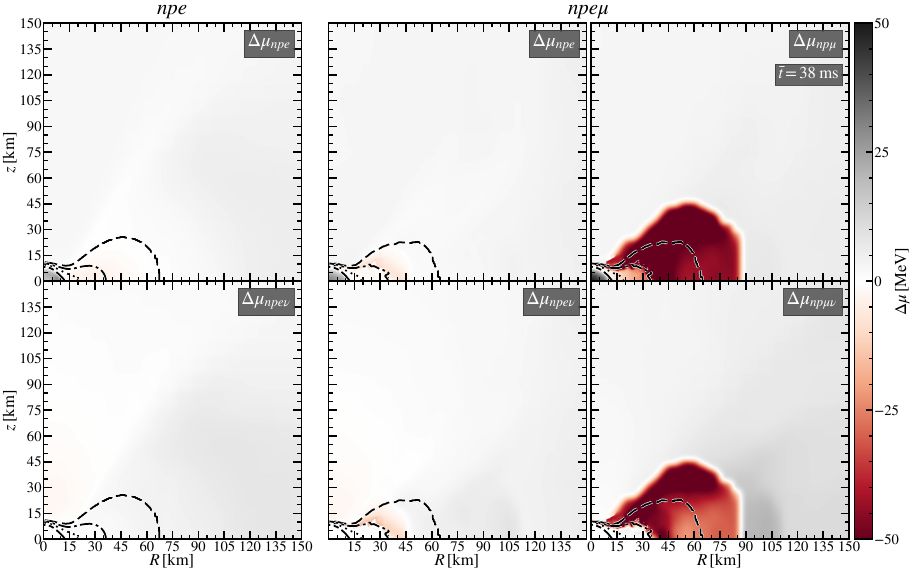}
    \caption{
    Meridional distributions of the
      out-of-weak-equilibrium chemical potentials, defined in
      Eq.~\eqref{eq:out-of-eqm_Delta}, for the $npe$ (left) and $npe\mu$
      (middle and right) models at $\bar t=38\,{\rm ms}$. The panels show
      the corresponding $\Delta\mu$ quantities indicated by the labels,
      with the color scale in MeV. The contour lines are the same as
      those defined in Fig.~\ref{fig:fig1}. }
    \label{fig:out-of-eqm}
\end{figure*}

Based on the $npe\mu$ global simulation results at
$\bar t = 38\,{\rm ms}$, we perform local QKE simulations with both collision prescriptions described in
Sec.~\ref{sec:snapshot-handoff}, retaining all
$N_{\varepsilon}=16$ energy groups and the original
quadrature weights of the input Boltzmann distributions.
For the \texttt{QKE-noDC} and \texttt{QKE-fullC} cases, small flavor coherences are introduced through random unitary rotations with a rotation parameter $\delta_{\rm seed}=10^{-7}$ at the beginning of the calculations.
For the evolution, we employ an adaptive, eighth-order DOP853 explicit Runge–Kutta time integrator, with the $H_{\nu\nu}$ recomputed at each substep.
Lastly, we adopt a relative tolerance of $10^{-10}$ and absolute tolerances of $10^{-14}$ for the diagonal occupations and $10^{-20}$ for the real and imaginary components of the flavor coherences.

\section{Results of the Global simulations}\label{sec:secIV}
\subsection{Hydrodynamical and out-of-equilibrium conditions}\label{sec:hydro_and_delta}
After evolving the handoff configuration for $8~\rm{ms}$ from $\bar t=30\,{\rm ms}$ to substantially damp perturbations,
we discuss the hydrodynamical and out-of-weak-equilibrium
conditions in $npe$ and $npe\mu$ models.

Figure~\ref{fig:fig1} shows the resulting thermodynamic and composition
profiles at $\bar t=38\,{\rm ms}$. The density structures of the two
models remain very similar, as the inclusion of muons does not modify the
bulk structure of the remnant. At $\bar t=30~\rm{ms}$, the maximum
temperatures are $T_{\rm max}=25.3$ and $23.6\,{\rm MeV}$ for the $npe$
and $npe\mu$ models, respectively, and only oscillate within
$0.5~\rm{MeV}$ during the evolution. Even after evolution with Boltzmann
transport to nearly $40~\rm{ms}$ after merger, the $npe\mu$ model remains
less electronized than the $npe$ model in the disk and polar
regions~\cite{Ng2024c}.

The remnant retains similar hydrodynamical properties to those inherited from the parent grey-M1 simulation of Ref.~\cite{Ng2024c}, especially in the neutrino-trapping regions.
First, the central $Y_\mu$ changes only weakly during the evolution, and the core has the same trapped-neutrino hierarchy as reported~\cite{Ng2024c} in Fig.~\ref{fig:nnu_hierarchy}, indicating that the previous grey-M1 calculations quantitatively captured the thermodynamic and trapped-neutrino state of the
dense core, despite the substantial differences in dimensionality, spatial and momentum-space resolution, and radiation treatment.

However, the largest differences arise in the disk, where
$Y_\mu$ decreases from $\sim10^{-3}$ to values of order $\gtrsim 6\times 10^{-4}$ for $\rho<10^{12}~\rm{g~cm^{-3}}$.
Outside $\rho=10^{10}~\rm{g~cm^{-3}}$, matter demuonizes
to $Y_{\mu,\rm thr}$ with $\mu_\mu \to \mu_e$.
The stronger demuonization reflects the more accurate transport, including full momentum dependence, the additional muonic interactions LFE/C and ILD, and the recovery of the dilute-muon limit.
The consistency of demuonization is validated in App.~\ref{app:hot-star-demuonization}.

To quantify how much the system deviates from weak equilibrium, for $l\in\{e,\mu\}$ we first define
\begin{align}
\Delta\mu_{npl} & := \mu_n-\mu_p-\mu_l,\\ \Delta\mu_{npl\nu} &=
\mu_n-\mu_p-\mu_l-\mu_{\bar\nu_l},
\label{eq:out-of-eqm_Delta}
\end{align}
where $\Delta\mu_{npl}=0$ corresponds to neutrinoless beta equilibrium and $\Delta\mu_{npl\nu}=0$ to
neutrino-trapping beta equilibrium with the
$\mu_{\bar\nu_l}$ inverted from the neutrino number density ($n_\nu$)~\cite{Espino2024b}.

\begin{figure*}
    \hspace*{-0.02\textwidth}
    \includegraphics[width=\textwidth]{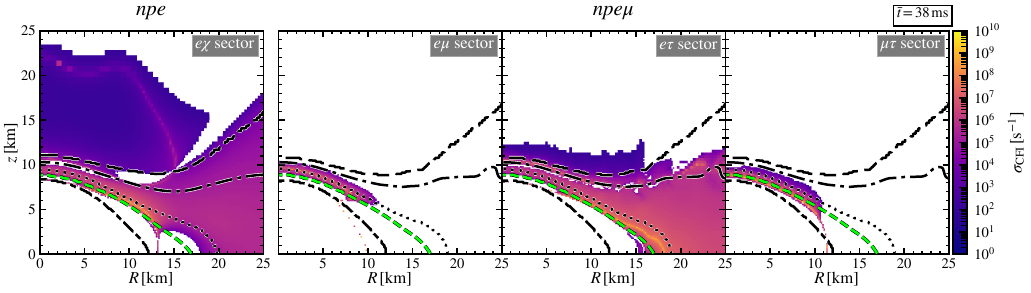}
    \caption{
      Meridional distributions of the CFI growth rate
      $\sigma_{\rm CFI}$ for the $e\chi$ sector in the $npe$ model (left)
      and the $e\mu$, $e\tau$, and $\mu\tau$ sectors in the $npe\mu$
      model (right) at $\bar t=38\,{\rm ms}$. The lime dashed lines
      correspond to the location of the energy sphere of
      $\nu_{\chi/\tau}$ associated with $\tau_\nu = 1$ in
      Eq.~\eqref{eq:effective-energy-sphere}; for the remnant matter
      inside this line, the CFI is highly suppressed. }
    \label{fig:CFI_sigma}
\end{figure*}
%

As discussed below, the dense core and inner disk also recover both electronic and muonic trapped weak equilibrium.
Figure~\ref{fig:out-of-eqm} shows that the dense remnant extending to the inner disk at $\rho>10^{12}\,{\rm g\,cm^{-3}}$ has attained both electronic and muonic neutrino-trapping equilibria, in contrast to the
trapped equilibrium for $\rho \gtrsim 10^{13}\,{\rm g\,cm^{-3}}$ in Ref.~\cite{Ng2024c} at $\bar t = 7~\rm{ms}$.

However, the semitransparent transition region in the outer disk has
largely negative values of $\Delta\mu_{np\mu}$ and $\Delta\mu_{np\mu\nu}$, which are attributed to the
high value of $\mu_\mu$ with nonvanishing $Y_\mu \sim 8 \times 10^{-4}$ in Fig.~\ref{fig:fig1} and the
significantly reduced $\mu_n - \mu_p$ at low $\rho$.
The muonic matter is advecting outward and
muonic rates in the outer disk are significantly weakened and remove muon lepton number more slowly.
For example, even when muon decay dominates the inverse process, the rate is still reduced
by the electron and neutrino final-state blocking.
These outward-moving muons persist temporarily
because of the sufficiently long demuonization timescale
$\tau_{\rm demu}
:=
-Y_\mu/\dot{Y}_\mu \sim 10^{-3}-10^{-2}~\rm{s}
$.
A similar behavior was identified by Ref.~\cite{Espino2024b} as the ``translucent'' regime, where neutrinos remain trapped by scattering while insufficient interaction rates allow deviations from weak equilibrium; the deviation is larger in both magnitude and spatial extent in the $np\mu\nu$ sector.

With decreasing $\rho$ and $T$, $\tau_{\rm demu}$ decreases rapidly farther outward, and $Y_\mu$
asymptotically approaches $Y_{\mu,\rm thr}$ with $\mu_\mu\simeq\mu_e$ extending to the edge of the domain.
This forms a muonless and neutrinoless $\beta$-equilibrium with all $\Delta\mu \approx 0$ in Fig.~\ref{fig:out-of-eqm}.
This behavior directly demonstrates that the unblocked muonic interactions, especially muon decay,
drive instant demuonization toward the
regularized $Y_{\mu, \rm thr}$ without being bounded by the
large value of $\mu_\mu$ that would enforce a false $\beta$ equilibrium.

\subsection{Properties of collisional flavor instability}\label{sec:CFI_results}
%
\begin{figure}
    \includegraphics[width=0.48\textwidth]{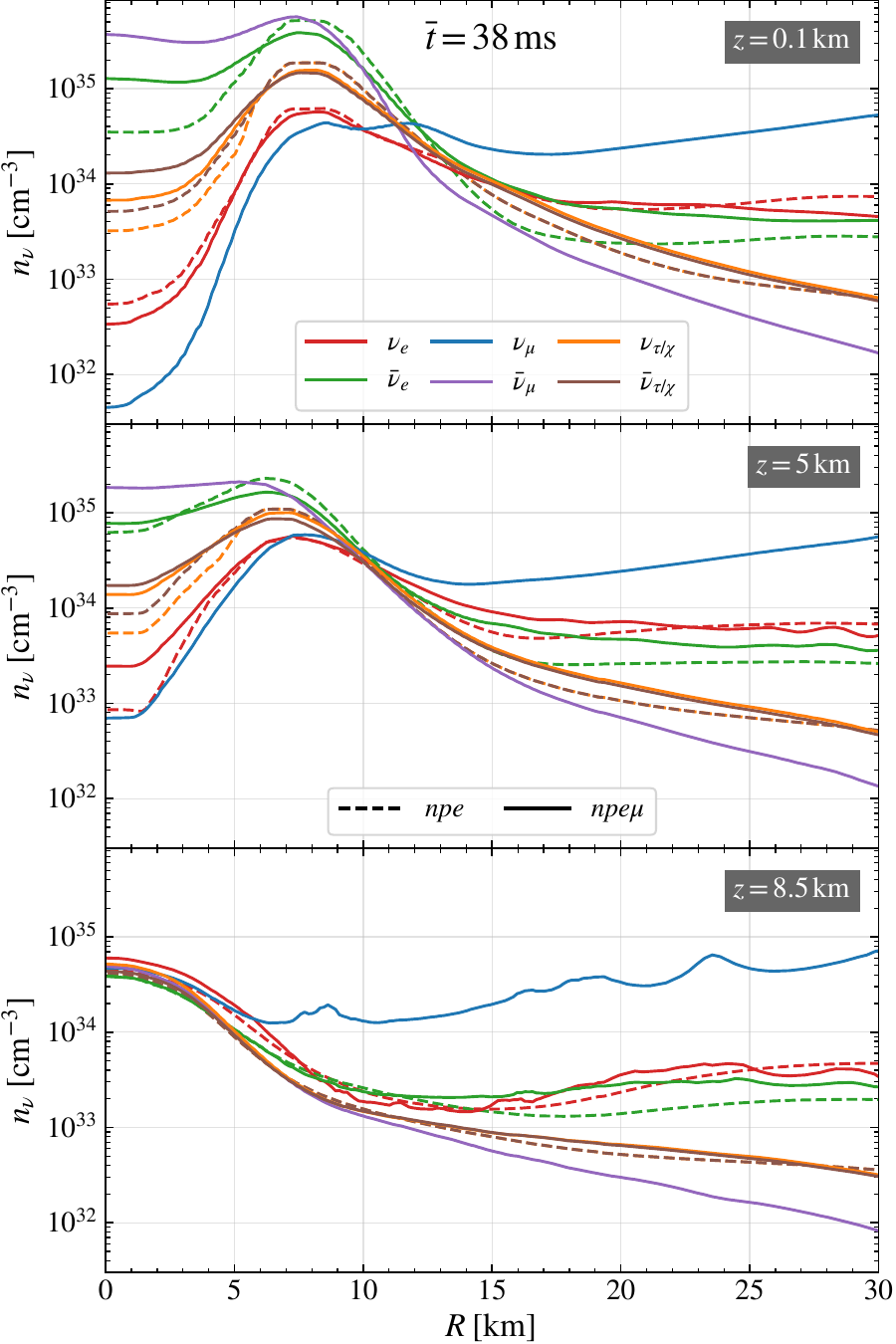}
    \caption{Radial profiles of the comoving-frame neutrino number
      density $n_\nu$ at $\bar t=38\,{\rm ms}$, extracted at $z=0.1$,
      $5$, and $8.5,{\rm km}$ from top to bottom. Solid and dashed curves
      correspond to the $npe\mu$ and $npe$ models, respectively. }
    \label{fig:nnu_hierarchy}
\end{figure}
We now investigate how the muon degree of freedom modifies the occurrence and properties of CFI. Figure~\ref{fig:CFI_sigma} shows the morphology and growth rates, focusing on the remnant and inner disk where the strongest CFI occurs, while the much weaker CFI farther out is not emphasized here. Strong CFI develops over broad regions outside the densest part of the remnant and extends to the inner disk. This differs completely from CCSN calculations, where CFI occurs at $\rho\sim10^{10}$--$10^{12}\,{\rm g,cm^{-3}}$ with much smaller growth rates of $\lesssim{\cal O}(10^3)\,{\rm s^{-1}}$~\cite{Xiong2022,Liu:2023vtz,Akaho2023,Wang2025},
while the higher-density regions may contain mostly dotted CFI structures~\cite{Akaho2023,Wang2025}. In the merger remnants, the broad bands commonly reach $\sigma_{\rm CFI}\gtrsim10^5\,{\rm s^{-1}}$. These stronger CFIs are consistent with the hotter, denser, and more neutron-rich environment, which supplies larger neutrino number densities and collisional rates. High, widespread, and persistent CFI in postmerger disks has also been reported in Refs.~\cite{Nagakura2025,Froustey2026}.

The $e\tau$ sector in the $npe\mu$ model has a structure similar to that
of the $e\chi$ sector in the $npe$ model, indicating that the asymmetric
features of the electron-flavor neutrinos and rates are robust. Also, CFI
occurs simultaneously in all three sectors with similar growth rates, and
two additional broad bands in the $e\mu$ and $\mu\tau$ sectors appear
mainly at polar angles of $\theta\lesssim60^\circ$. All sectors persist
throughout the evolution with mild changes. Figure~\ref{fig:CFI_sigma}
also contains much narrower and stronger R-CFI structures in all sectors,
which we discuss further below.

An interpretation of the CFI occurrences is provided by the
$\nu_{\chi/\tau}$ thermalization surface. We calculate the optical depth
with an effective thermalization opacity of $\kappa_{\rm eff} = \sqrt{
  \bar{\kappa}_a^* \left( \bar{\kappa}_a^*+\bar{\kappa}_s \right) } ,$
where $\bar{\kappa}_a^*$ is the $f$-weighted absorption opacity, which
includes all kernel-based interactions, and $\bar{\kappa}_s$ is the
$f$-weighted scattering opacity. The optical depth is given as
\begin{equation}
\tau_\nu(R,z)
=
\min_{\Gamma}
\int_{\Gamma}
\kappa_{\mathrm{eff},\nu}(R',z')\,
d\ell_{\mathrm{proper}}\,,
\label{eq:effective-energy-sphere}
\end{equation}
where $dl_{\rm proper}=\psi^2\sqrt{dR^2+dz^2}$ and
$\min_\Gamma\int_\Gamma$ selects the minimum optical-depth integral among
all paths $\Gamma$ from $(R,z)$ to an outer boundary. We define the
$\tau_{\nu_{\chi/\tau}} = 1$ surface as the energy sphere of
$\nu_{\chi/\tau}$, represented by the lime dashed contour in
Fig.~\ref{fig:CFI_sigma}. The $\nu_{\chi/\tau}$ energy sphere lies in the
deepest region among all species, and neutrinos inside it are
sufficiently thermalized by PP. As a result, all CFI sectors are highly
suppressed by thermal and pair equilibria~\cite{Liu2024,Nagakura2025},
which approximately explains why most CFI occurs outside the
$\nu_{\chi/\tau}$ energy sphere in both models, except for a very narrow
$\mu\tau$-sector region beneath it. This could be a useful macroscopic
indicator of CFI.

To understand the reasons for the CFI occurrences and sector-dependence, we look at the species-dependent neutrino populations in Fig.~\ref{fig:nnu_hierarchy}. At $z=0.1$ and $5\,{\rm km}$ profiles, the trapped hierarchy in the $npe\mu$ model retains the order of
$
n_{\bar\nu_\mu}
>
n_{\bar\nu_e}
>
n_{\nu_\tau}
\simeq
n_{\bar\nu_\tau}
>
n_{\nu_e}
>
n_{\nu_\mu} ,
$
consistent with Ref.~\cite{Ng2024c} in the early postmerger phase. The small difference between $n_{\nu_{\chi/\tau}}$ and $n_{\bar\nu_{\chi/\tau}}$ is caused by recoil and weak-magnetism corrections to the ES~\cite{Horowitz2002}. Farther outward, however, the electron- and muon-flavor neutrino and antineutrino populations exchange order, and the population hierarchy becomes
$
n_{\nu_\mu}
>
n_{\nu_e}
>
n_{\bar\nu_e}
>
n_{\bar\nu_\mu}.
$
The large $\nu_\mu$--$\bar\nu_\mu$ asymmetry is directly connected to the negative $\Delta\mu_{np\mu}$ (positive $\mu^{\rm eq}_{\nu_\mu}$) in Fig.~\ref{fig:out-of-eqm}, providing a tendency to emit $\nu_\mu$ during demuonization, as discussed in Sec.~\ref{sec:hydro_and_delta}. Similarly, in the $npe$ model, finite $\Delta\mu_{npe}$ in the inner disk explains the larger asymmetry between $n_{\nu_e}$ and $n_{\bar\nu_{e}}$ in Fig.~\ref{fig:nnu_hierarchy}.

The population hierarchy explains why the large muon-neutrino asymmetry does not generate broad $e\mu$ or $\mu\tau$ CFI throughout the disk, particularly in the low-latitude region of Fig.~\ref{fig:CFI_sigma}. At low latitudes ($z=0.1$ and $5\,{\rm km}$), the (anti)neutrino-population ordering in Fig.~\ref{fig:nnu_hierarchy} exhibits a reversal along the $R$-direction. Taking the $\mu\tau$ sector as an example, the neutrino populations in the inner region correspond predominantly to $\Delta f_{\mu\tau}<0$ and $\Delta\bar f_{\mu\tau}>0$, such that $W_{\mu\tau,-}$ dominates, whereas farther outward both $\Delta f_{\mu\tau}$ and $\Delta{\bar f}_{\mu\tau}$ exchange signs and $W_{\mu\tau,+}$ becomes dominant. Thus, only one effective sector dominates on either side of the transition, yielding $G_{\mu\tau}/|A_{\mu\tau}|\simeq1$. As discussed in Sec.~\ref{sec:CFI}, the asymmetric collisional driving is then insufficient to overcome the mean damping, i.e.,
$
\frac{G_{\mu\tau}}{|A_{\mu\tau}|}\,
|\alpha_{\mu\tau}|
<
\gamma_{\mu\tau},
$
and the $\mu\tau$-sector CFI is significantly suppressed despite the large $\nu_\mu$--$\bar\nu_\mu$ asymmetry. The $e\mu$ sector exhibits the same behavior. 
In contrast, at high latitudes (the $z=8.5,{\rm km}$ profile), the population ordering is different, and the reversal of $\Delta f_{ab}$ and $\Delta \bar{f}_{ab}$ along the $R$ direction does not occur, allowing the other two CFI sectors to develop.
Moreover, the higher $e\tau$ growth rate in the $npe\mu$ case than in the $npe$ case in the inner disk in Fig.~\ref{fig:CFI_sigma} can be explained by the smaller electron-flavor asymmetry in the $npe\mu$ model, which produces a smaller effective-sector imbalance $|A_{e\tau}|$ and thus a larger $G/|A|$ that more readily overcomes the mean damping.

We now focus on the very narrow, high-growth-rate structures caused by resonance-like CFI (R-CFI) in Fig.~\ref{fig:CFI_sigma}. R-CFI is present in all three physical flavor sectors, with apparent growth rates reaching $\sigma_{\rm CFI}\sim10^8$--$10^{10}\,{\rm s^{-1}}$. In both models, the R-CFI layers lie within extremely narrow regions associated with sign changes of $A_{ab}$~\cite{Xiong2022,Liu2023}. From the inner to the outer region, $A_{e\mu}$ changes from positive to negative, whereas $A_{e\tau}$ and $A_{\mu\tau}$ change from negative to positive. These narrow sign-changing regions can also be traced by regions where $n_\nu\simeq n_{\bar\nu}$~\cite{Nagakura2025}. So, the very thin R-CFI layers correspond closely to the $n_\nu\simeq n_{\bar\nu}$ regions shown in Fig.~\ref{fig:nnu_hierarchy}. For example, at $z=0.1\,{\rm km}$, $n_{\nu_e}\simeq n_{\bar\nu_e}$ at $R\simeq13\,{\rm km}$ in the $npe$ model and $n_{\nu_\mu}\simeq n_{\bar\nu_\mu}$ at $R\simeq11\,{\rm km}$ in the $npe\mu$ model correspond to the $e\chi$ and $\mu\tau$ R-CFI layers, respectively.
Macroscopically, the different opacities and decoupling locations of neutrinos and antineutrinos can modify their relative populations across the trapped-to-semitransparent transition and help bring $A_{ab}$ close to zero, producing the condition $n_\nu\simeq n_{\bar\nu}$.

R-CFI is expected to occur in optically thick regions {when the neutrino spectra depart from exact equilibrium}~\cite{Liu2024, Nagakura2025}. We indeed observe R-CFI in the $e\mu$ and $\mu\tau$ sectors of the $npe\mu$ case and in the $e\chi$ sector of the $npe$ case deep inside regions where the neutrinos are nearly in LTE (within the $\nu_\tau$ energy sphere), weak equilibrium (see Sec.~\ref{sec:hydro_and_delta}), and pair equilibrium ($\mu_{\nu_a}+\mu_{\bar\nu_a}\simeq0$).
This does not contradict the argument in
Ref.~\cite{Liu2024}; for Fermi--Dirac
spectra satisfying $\mu_{\nu_a}+\mu_{\bar\nu_a}=0$
for each flavor, at least one effective spectral sector
vanishes, and the present CFI diagnostic is stable.\footnote{
Although the CFI growth-rate prescription of
Ref.~\cite{Liu2026} differs from that of
Ref.~\cite{Liu2024} in the definitions of the effective
spectral sectors, both predict no CFI in exact equilibrium.}
The apparent R-CFI in these nearly equilibrated regions
requires deviations from exact equilibrium,
potentially in the spectral tails; whether these deviations
are physical or numerical remains to be established.

However, these R-CFI growth rates are not fully resolved at the present spatial resolution. Previous studies have shown that R-CFI can occupy extremely thin spatial layers and can be missed or strongly underestimated at insufficient resolution~\cite{Akaho2023,Wang2025}. Therefore, some R-CFI structures appear fragmented or dotted in Fig.~\ref{fig:CFI_sigma}, and their growth rates could be underestimated. The identification of these structures as R-CFI is nevertheless supported by the corresponding $A_{ab}$ sign changes within the CFI-unstable bands of all three sectors in the present diagnostic.

\begin{figure*}
    \includegraphics[width=\textwidth]{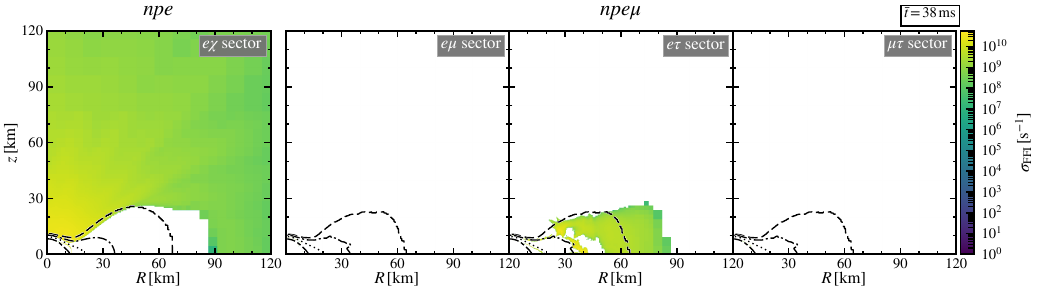}
    \caption{
      Meridional distributions of the FFI growth rate
      $\sigma_{\rm FFI}$ for the $e\chi$ sector in the $npe$ model (left)
      and the $e\mu$, $e\tau$, and $\mu\tau$ sectors in the $npe\mu$
      model (right) at $\bar t=38\,{\rm ms}$. }
    \label{fig:ffi_result}
\end{figure*}
The six-species $npe\mu$ analysis identifies CFI and R-CFI in all three physical flavor sectors, with sector-dependent spatial distributions and growth rates. These results highlight the importance of retaining the muonic neutrino--antineutrino asymmetry and the differences between muon- and tau-flavor populations and collision rates, which are not captured by the effective heavy-lepton treatment used in the $npe$ model.

\subsection{Properties of fast flavor instability}\label{sec:FFI_results}
%
\begin{figure*}
    \hspace*{+0.02\textwidth}
    \includegraphics[width=\textwidth]{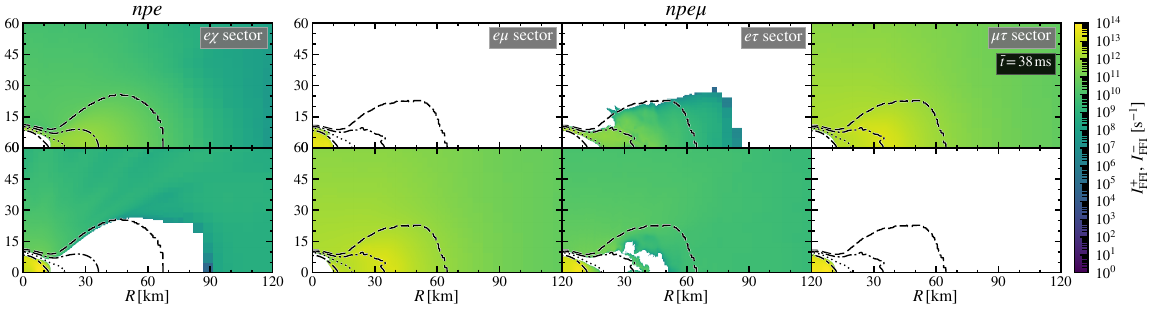}
    \caption{
      Meridional distributions of the positive (top) and
      negative (bottom) angular components, $I_{\rm FFI}^{+}$ and $I_{\rm
        FFI}^{-}$, defined in Eq.~\eqref{eq:ffi-growth-rate}, for the
      $e\chi$ sector in the $npe$ model (left) and the $e\mu$, $e\tau$,
      and $\mu\tau$ sectors in the $npe\mu$ model (right) at $\bar
      t=38,{\rm ms}$. A location with both finite $I_{\rm FFI}^{+}$ and
      $I_{\rm FFI}^{-}$ indicates an $a{\rm LN}-b{\rm LN}$ crossing and
      hence the possibility of FFI. }
    \label{fig:Iposneg}
\end{figure*}
We next investigate how the muon degree of freedom modifies the properties of FFI in the merger remnant. 
Unlike CFI, the occurrence and spatial distribution of FFI are strongly affected by muonic interactions, producing qualitatively different instability patterns between the two- and three-flavor models.
Figure~\ref{fig:ffi_result} shows spatially extensive FFI in the $e\chi$ sector, occurring mainly in the low-density regions extending from the outer disk to the nearly atmospheric region of the $npe$ model, with growth rates reaching $\sigma_{\rm FFI}\sim10^{9}$--$10^{10}\,{\rm s^{-1}}$. These rates are comparable to those of the R-CFI structures and are several orders of magnitude larger than the ordinary-CFI growth rates discussed in Sec.~\ref{sec:CFI_results}. This widespread $e\chi$ FFI outside the dense region is consistent with previous BNS-merger and postmerger-disk studies that found $e$LN angular crossings throughout the disk and surrounding low-density regions~\cite{George2020,Just2022,Froustey2023}. The $npe$ case here is also broadly similar to Ref.~\cite{Nagakura2025}, owing to the similar setup and the same EOS. The $npe\mu$ model, however, exhibits a strong sector dependence: FFI in the $e\tau$ sector is concentrated in the outer disk at $\rho<10^{11}\,{\rm g\,cm^{-3}}$ but is absent from most of the atmospheric region, whereas the $e\mu$ and $\mu\tau$ sectors are strongly suppressed. For all sectors, thin layers or dotted FFI structures also appear close to the remnant core at $\rho>10^{13}\,{\rm g\,cm^{-3}}$. Such narrow layers have also been found in CCSN simulations~\cite{Glas2019}; resolving their detailed structure and growth rates may require very high spatial resolution.

To explain the different FFI morphologies between the two models, we examine how muonic interactions modify the positive and negative components of the flavor-lepton-number angular distributions.
Figure~\ref{fig:Iposneg} separately shows the positive and negative angular components, $I_{\rm FFI}^{+}$ and $I_{\rm FFI}^{-}$, defined in Eq.~\eqref{eq:ffi-growth-rate}. FFI requires an overlap of both components at the same spatial location, corresponding to a crossing in the $a{\rm LN}-b{\rm LN}$ angular distribution. Let us first focus on the $e\tau(\chi)$ sector in the two models. Since the $\tau(\chi)$-flavor LN remains small, $\mathcal{G}_{e\tau}\approx\mathcal{G}_e$, and similarly for the $e\chi$ sector.

For the $npe$ model, the negative $\Delta\mu_{npe}$ in the disk corresponds to a positive equilibrium chemical potential for $\nu_e$, i.e., $\mu_{\nu_e}^{\rm eq}>0$, providing a tendency toward $\nu_e$-rich emission. The resulting excess of $\nu_e$ produces a very spatially extended $I_{\rm FFI}^{+}$ in Fig.~\ref{fig:Iposneg}. At the same time, $\bar\nu_e$ emitted from the remnant core and redistributed into different directions mainly through scattering contributes to $I_{\rm FFI}^{-}$ extending throughout the outer regions, thus producing $e\chi$-sector FFI mainly outside the disk.

A similar behavior was identified in Ref.~\cite{Nagakura2025} through the
early-time $\nu_e$-diffusion mechanism, in which $\nu_e$-rich radiation
from the disk competes with $\bar\nu_e$ radiation from the HMNS to
generate ELN angular crossings.  However, this diagnostic alone cannot
explain the $e\tau$-sector FFI in the $npe\mu$ model, despite the larger
$\mu_{\nu_e}^{\rm eq}$ in the disk, because the situation becomes more
nonlinear. In addition to the $\bar\nu_e$ emission from the dense core,
muonic interactions directly reprocess the electron-flavor
neutrinos. Demuonizing matter drives the production of $\bar\nu_e$
through, e.g., muon decay, $\mu^-\rightarrow e^-+\bar\nu_e+\nu_\mu$, and
LFC. Subsequent transport and scattering redirect and reprocess the
$\bar\nu_e$ momentum distribution, especially within the disk around
$\rho>10^{11}\,{\rm g\,cm^{-3}}$, thereby producing $I_{\rm FFI}^{-}$ in
the inner regions. At the same time, LFE can absorb $\nu_e$ in reactions
involving muons and produce $\nu_\mu$, while the modified population
hierarchy and hydrodynamical profiles further reduce the difference
between $n_{\nu_e}$ and $n_{\bar\nu_e}$, as shown in
Fig.~\ref{fig:nnu_hierarchy}. The depletion of $\nu_e$, together with the
additional $\bar\nu_e$, reduces $\mathcal{G}_{e\tau}$ in angular
directions that would otherwise be positive and can convert some of them
into negative directions, thereby spatially confining $I_{\rm
  FFI}^{+}$. The resulting $e{\rm LN}-\tau{\rm LN}$ angular distribution
is highly nonlinear in muonic matter, producing $e\tau$-sector FFI in the
denser disk regions without extending far into the outer low-density
atmosphere. This different FFI morphology could potentially lead to a
different composition of the disk ejecta in muonic matter.

While demuonization reshapes the $e\tau$-sector FFI, it also suppresses the $e\mu$- and $\mu\tau$-sector instabilities through a different mechanism: the development of a large muon-lepton-number asymmetry.
In the outer disk, muonic interactions generate an excess of $\nu_\mu$ over $\bar\nu_\mu$, reversing the number-density hierarchy from $n_{\bar\nu_\mu}>n_{\nu_\mu}$ in the trapped core to $n_{\nu_\mu}>n_{\bar\nu_\mu}$ farther outward, as shown in Fig.~\ref{fig:nnu_hierarchy}. This excess is also broadly distributed across propagation directions, producing a predominantly positive $\mu$LN angular distribution and contributing to the one-sided angular distributions in the $e\mu$ and $\mu\tau$ sectors shown in Fig.~\ref{fig:Iposneg}.
As discussed in Sec.~\ref{sec:hydro_and_delta}, the highly negative $\Delta\mu_{np\mu}$, corresponding to $\mu_{\nu_\mu}^{\rm eq}>0$, identifies the $\nu_\mu$-rich and demuonizing regions, particularly toward lower densities and temperatures. The depletion of $\bar\nu_\mu$ during demuonization, including through $\bar\nu_\mu+p\rightarrow n+\mu^+$, further increases the difference between $n_{\nu_\mu}$ and $n_{\bar\nu_\mu}$, as also shown in Fig.~\ref{fig:nnu_hierarchy}. The resulting
magnitude of $\mu$LN dominates over both those of $e$LN and $\tau$LN, leading to a one-sided $I_{\rm FFI}^{-}$ in the $e\mu$ sector, i.e.,
$\mathcal G_{e\mu} \approx -\mathcal G_{\mu}$,
and a one-sided $I_{\rm FFI}^{+}$ in the $\mu\tau$ sector,
i.e., $\mathcal G_{\mu\tau} \approx \mathcal G_{\mu}$,
thereby strongly suppressing FFI in both sectors.

\begin{figure}[!t]
   \centering
   \includegraphics[width=0.52\textwidth]{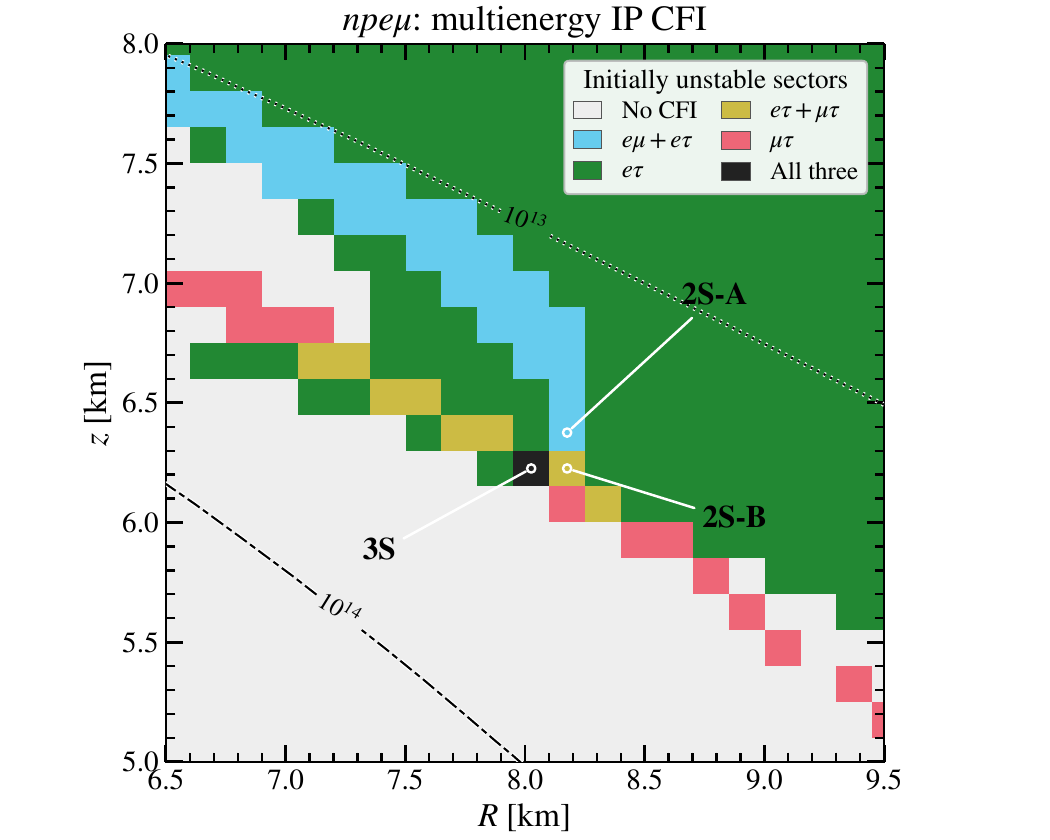}
   \caption{
    The three representative locations selected for local QKE calculations
    in a magnified CFI sector-combination map of the $npe\mu$ remnant of Fig.~\ref{fig:CFI_sigma}:
    two locations with two unstable flavor sectors (2S-A and 2S-B)
    and one with all three sectors unstable (3S).
    The three locations are sufficiently close to have similar hydrodynamic conditions, allowing a fairer comparison.
    Their properties are summarized in Table~\ref{tab:qke_points}.
    Colors distinguish the combinations of unstable CFI flavor sectors.
    The classification is based on positive growth rates obtained from
    spatially homogeneous, isotropy-preserving, multienergy linearized
    equations evaluated on the angular-averaged neutrino distributions.
    The contour lines represent $\rho = 10^{13}$ and $10^{14}\,\rm{g~cm^{-3}}$.
   }
   \label{fig:CFI_sector_map}
\end{figure}
%
The muonic remnant scenario differs significantly from the muonic CCSN scenario reported in Ref.~\cite{Capozzi2020}.
In CCSNe, $\mu\tau$-sector FFI can arise from continuous muonization in the hot PNS combined with demuonization at different radii and times as the stellar envelope continues to collapse, producing a more angle-dependent but smaller-magnitude $\mu$LN.
In addition, the extended stellar envelope provides substantial backward-scattered neutrino populations that can reverse the sign of $a{\rm LN}-b{\rm LN}$ between forward and backward directions and enable $\mu\tau$-sector FFI in the outer regions~\cite{Capozzi2020}. As a result, the CCSN situation is not directly analogous to the long-lived merger-remnant case, where the disk and wind are predominantly demuonizing and no comparable extended backward-scattering reservoir is present in the merger remnant.

Finally, the FFI distributions vary during the $8\,{\rm ms}$ evolution, consistent with FFI being local, transient, and highly variable across different stages of BNS-merger evolution~\cite{Nagakura2025}. Longer and more self-consistent simulations are therefore required to determine the long-term behavior of these instabilities. In particular, $e\mu$- and $\mu\tau$-sector FFI may become more favorable during the earlier merger phase, when strong muonization is still ongoing~\cite{Ng2024c,Gieg2024}.

\section{Results of the Local QKE simulations}
\label{sec:QKE_results}

%
\begin{table}[t]
\centering
\caption{
Initial CFI growth rates $\sigma_{ab}$ of the fastest-growing isotropy-preserving modes at the three locations selected in Fig.~\ref{fig:CFI_sector_map} for the local QKE calculations.
The three locations are sufficiently close to have similar hydrodynamic conditions, allowing a fairer comparison.
Zero indicates that sector is CFI-stable. 
Daggers mark the particular unstable sector 
at the location near the resonance-like CFI limit, which is 
not fully resolved by the limited resolution; the criterion for near-resonance is that a resonance condition indicator satisfies $(A_{ab})^2/|G_{ab}\,\alpha_{ab}|<200$.
}

\label{tab:qke_points}
\small
\setlength{\tabcolsep}{5pt}
\begin{ruledtabular}
\begin{tabular}{lccc}
Point & $\sigma_{e\mu}~[{\rm s^{-1}}]$ & $\sigma_{e\tau}~[{\rm s^{-1}}]$ & $\sigma_{\mu\tau}~[{\rm s^{-1}}]$ \\
\hline
2S-A & $3.94\times10^{6}$ & $3.68\times10^{6}$ & $0$ \\
2S-B & $0$ & $6.27\times10^{8}\,{}^{\dagger}$ & $1.42\times10^{7}\,{}^{\dagger}$ \\
3S & $4.88\times10^{6}\,{}^{\dagger}$ & $2.59\times10^{7}$ & $4.19\times10^{7}$ \\
\end{tabular}
\end{ruledtabular}
\end{table}

{In this section, we investigate the nonlinear evolution of overlapping CFIs in $npe\mu$ matter. While FFI occurs predominantly in the $e\tau$ sector (Fig.~\ref{fig:ffi_result}), CFI develops simultaneously in multiple flavor sectors over broad regions of the remnant (Fig.~\ref{fig:CFI_sigma}). The prediction of the asymptotic states of CFIs is on ongoing endeavor. Previous analytical studies of CFI asymptotic states have focused on simplified homogeneous, isotropic, monochromatic two-flavor systems~\cite{Froustey2025}, while a recent separate work of one of us has formulated analytical solutions of asymptotic states of single-sector CFI with either off-diagonal damping alone or with additional diagonal collisional relaxation~\cite{LiuCFIInPreparation}. 
Here, we perform local three-flavor QKE calculations to examine the nonlinear coupling among overlapping CFI sectors by neglecting the FFI contributions and the resulting flavor redistribution, providing insights for future CFI subgrid models.

We select three locations with overlapping CFI sectors, 
The three points, denoted 2S-A, 2S-B, and 3S, lie near $(R,z)\simeq(8.1,6.3)\,{\rm km}$ 
and their locations are close enough to have similar hydrodynamic conditions, allowing a fairer comparison.
To compare different CFI configurations, Fig.~\ref{fig:CFI_sector_map} shows the selected locations and their combinations of unstable flavor sectors, while Table~\ref{tab:qke_points} lists the corresponding initial growth rates $\sigma_{ab}$ of the fastest-growing IP modes by a multienergy LSA. 
The three locations exhibit distinct instability patterns: $e\mu$ and $e\tau$ sectors with comparable growth rates (2S-A); $e\tau$ and $\mu\tau$ sectors with relatively greater proximity to the R-CFI regime (2S-B); and all three sectors simultaneously unstable (3S).
For each point, with a fixed hydrodynamical background, 
we compare three approaches: \texttt{QKE-noDC}, \texttt{QKE-fullC} and \texttt{Classical} (see details in Sec.~\ref{sec:snapshot-handoff}). 
The \texttt{QKE-noDC} case omits diagonal collision terms to isolate the effects of CFI from the classical collisional sources already included in global simulations, thereby avoiding double counting when treating them separately in constructing a subgrid model.
In contrast, \texttt{QKE-fullC} retains all components of the collision operator, allowing collisional repopulation of the neutrino distributions alongside the development of CFI-driven flavor conversion.
\texttt{Classical} case retains only the diagonal collision terms, recovering classical neutrino transport without the development of any flavor coherence.
Note that all calculations are evolved until the distributions of all flavors have reached their asymptotic states.

\begin{figure*}[t]
    \centering
    \includegraphics[width=\textwidth]{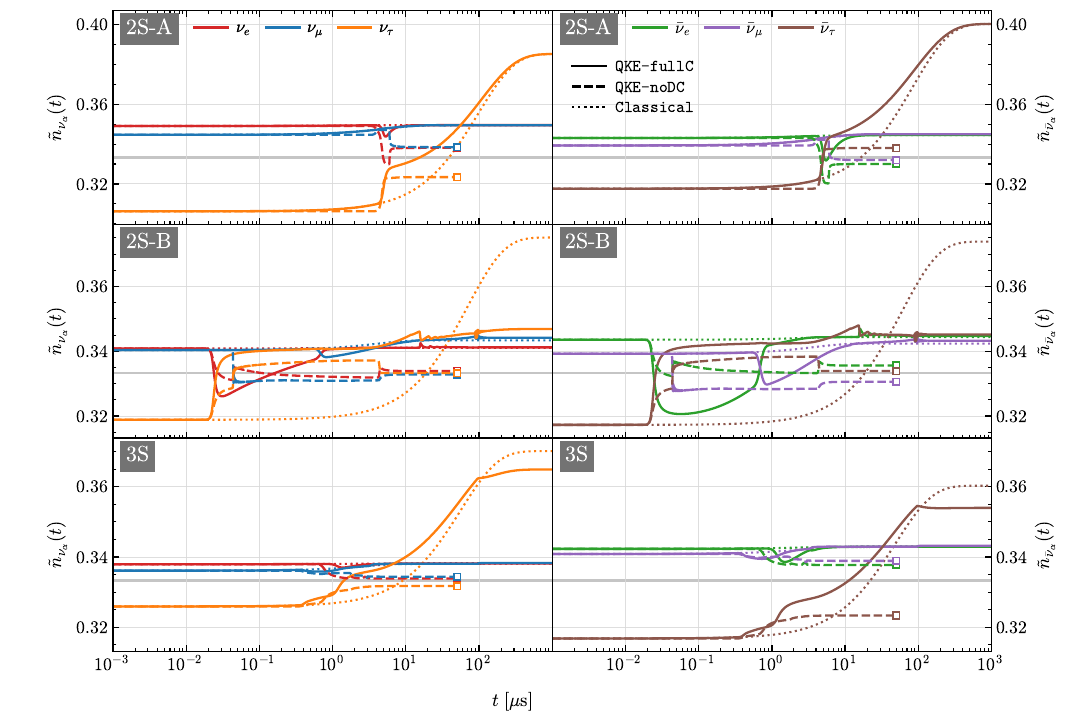}
    \caption{
The evolution of normalized number-density fraction of neutrinos $\tilde n_{\nu_\alpha}(t)$ (left panels) and antineutrinos $\tilde n_{\bar\nu_\alpha}(t)$ (right panels), defined in Eq.~\eqref{eq:qke-normalized-densities}, at the selected locations 2S-A, 2S-B, and 3S (top to bottom). Colors distinguish neutrino flavors; solid, dashed, and dotted curves denote \texttt{QKE-fullC}, \texttt{QKE-noDC}, and \texttt{Classical}, respectively. Open squares mark the \texttt{QKE-noDC} endpoints, as their results attain 
asymptotic states very quickly. Gray horizontal lines indicate $1/3$, corresponding to the flavor-equipartition point for \texttt{QKE-noDC} calculations.
}
\label{fig:QKE_populations}
\end{figure*}
We first present the time evolution of number density of (anti)neutrinos
for three flavor in Fig.~\ref{fig:QKE_populations}. We define the
normalized number-density fraction as
\begin{equation}
\tilde n_{\nu_\alpha}(t) := \frac{n_{\nu_\alpha}(t)}{\sum_\beta
  n_{\nu_\beta}(0)}\,,
\label{eq:qke-normalized-densities}
\end{equation}
where
$n_{\nu_\alpha}(t)=\int\varepsilon^2d\varepsilon\,f_{\nu_\alpha}(\varepsilon,t)/(2\pi^2)$,
with an analogous antineutrino definition. The denominators are fixed
initial totals shared by all three prescriptions at each point.
\texttt{QKE-noDC} separately conserves the total neutrino and
antineutrino numbers, whereas the diagonal collision sources in
\texttt{QKE-fullC} and \texttt{Classical} can change these totals due to
emission and absorption.

In Fig.~\ref{fig:QKE_populations}, a sector with higher $\sigma_{ab}$ generally undergoes earlier flavor conversion, e.g., in 2S-B, exponential changes occur at $t\sim 10^{-2}\,\mu{\rm s}$.
Focusing particularly on the \texttt{QKE-noDC}, the flavor populations approach their asymptotic states more rapidly than in \texttt{QKE-fullC} and \texttt{Classical}. Its neutrino populations tend toward approximate flavor-equipartition, with nearly equal number densities among the three flavors, whereas the antineutrino populations retain larger flavor-dependent differences in 2S-B and 3S, but not in 2S-A.
At the point closest to resonance, 2S-B, all three neutrino flavors approach $\simeq 1/3$. Mode-dependent approximate flavor equipartition and swap behaviors have also been found in homogeneous, isotropic, multienergy two-flavor CFI calculations with the \texttt{QKE-noDC} type collision treatment~\cite{Zaizen2025}.
At all three locations, the asymptotic states obtained with \texttt{QKE-noDC} differ from those obtained with the other two cases. These differences highlight the importance of diagonal collisional repopulation in determining the late-time flavor distributions and suggest that its interplay with off-diagonal flavor coherence should be carefully considered when modeling flavor conversions driven by CFIs. The influence of diagonal collision terms on the early nonlinear evolution of two-flavor CFI has also been demonstrated~\cite{Kato2023}.

Compared to \texttt{Classical}, flavor coherence in \texttt{QKE-fullC} leads to different neutrino and antineutrino populations during the intermediate nonlinear evolution, with the largest differences occurring in the $\tau$-flavor sector. These differences persist into the asymptotic states, where \texttt{QKE-fullC} generally predicts lower $\nu_\tau$ and $\bar\nu_\tau$ populations, particularly in 2S-B and 3S. In contrast, the electron- and muon-flavor populations remain relatively close to their classical counterparts and reach asymptotic states much more quickly than those in $\tau$-flavor, consistent with relaxation driven by charged-current reactions.
At the two locations with relatively weaker CFI, 2S-A and 3S, both \texttt{QKE-fullC} and \texttt{Classical} require $\gtrsim 10^2\,\mu{\rm s}$ to approach the asymptotic state of $\tau$-flavor (anti)neutrino. This timescale is comparable to the characteristic local dynamical timescale of BNS merger remnants, suggesting that neutrino transport and background evolution may intervene before the asymptotic states are reached, posing additional challenges for CFI subgrid modeling.
}


Next, we discuss the CFI-driven growth of flavor coherence in all three sectors.
We first define the normalized root-mean-square coherence for each sector as
\begin{equation}
\begin{aligned}
\tilde S_{ab}(t) &:= \frac{S_{ab}^{\mathrm{rms}}(t)}{S_{\mathrm{ref}}}
\\ &=\left[\frac{\sum_i w_i\left(|S_{ab}(\varepsilon_i,t)|^2+|\bar
    S_{ab}(\varepsilon_i,t)|^2\right)}{\sum_i
    w_i\operatorname{Tr}\!\left[\varrho_i^2(0)+\bar\varrho_i^2(0)\right]}\right]^{1/2}\,,
\end{aligned}
\label{eq:qke-rms-coherence}
\end{equation}
where $w_i=\varepsilon_i^2\Delta\varepsilon_i$ are the original energy
weights and $S_{ab}$ and $\bar S_{ab}$ are the neutrino and antineutrino
non-diagonal coherence defined in Eq.~\eqref{eq:density_matrix}. The
dimensionless reference coherence $S_{\rm ref}\approx10^{-2}$ is
evaluated from the full initial density matrices and remains fixed
throughout the evolution, with the same value used for all flavor pairs
and both QKE prescriptions at each location. This measure combines their
squared magnitudes without cancellation between energy-dependent
phases. The normalization is fixed by the initial density matrices and
shared by \texttt{QKE-noDC} and \texttt{QKE-fullC} at each point.

\begin{figure}[t]
    \hspace*{-0.05\textwidth}
    \includegraphics[width=0.5\textwidth]{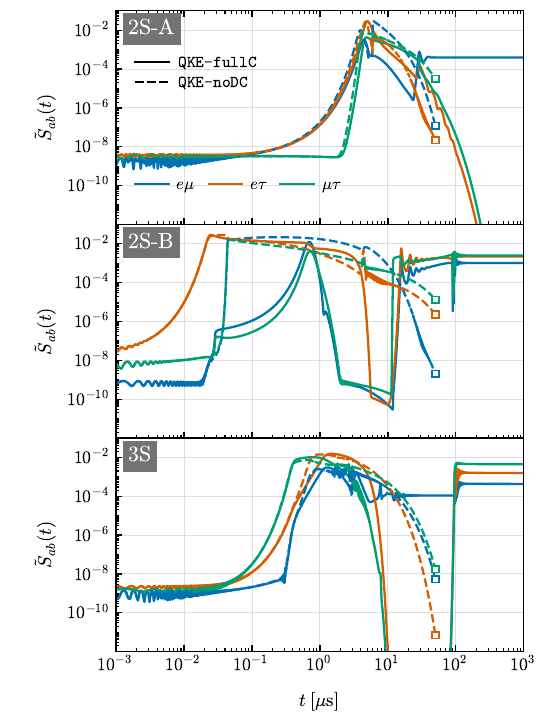}
    \caption{Time evolution of normalized root-mean-square coherence
      $\tilde S_{ab}(t)$ defined in Eq.~\eqref{eq:qke-rms-coherence} for
      2S-A, 2S-B, and 3S from top to bottom. Colors distinguish the three
      flavor pairs; solid and dashed curves denote \texttt{QKE-fullC} and
      \texttt{QKE-noDC}, respectively.
    }
    \label{fig:QKE_coherence}
\end{figure}
%

Figure~\ref{fig:QKE_coherence} shows the time evolution of the normalized coherence $\tilde{S}_{ab}$ at all three selected locations for both \texttt{QKE-fullC} and \texttt{QKE-noDC}.
During the early stage, the coherence growth depends on the initial growth rate $\sigma_{ab}$ of each sector listed in Tab.~\ref{tab:qke_points}; for example, in 3S, sectors with larger $\sigma_{ab}$ have larger $\tilde{S}_{ab}$.
For both approaches, once the initially unstable sectors develop sufficiently large flavor coherence, nonlinear couplings become important, and the initially stable sector also undergoes exponential growth (e.g., the $\mu\tau$ sector in 2S-A).
This can be explained by the coupling of flavor coherences via the Hamiltonian commutator, which contains products involving the third flavor. It implies that the evolution cannot be represented by independent two-flavor calculations or inferred solely from the initial growth rates.
This nonlinear activation of all sectors could be a distinct feature of CFI in three-flavor calculation, while for FFI that tends to erase angular crossings 
rather than generating new ones, new instabilities are unlikely to be regenerated.
We emphasize the properties of three-flavor FFI should also be further studied in realistic three-flavor environments.

After reaching their peaks, the coherences in \texttt{QKE-noDC} decrease monotonically, and the calculations are stopped once the three-flavor asymptotic states are reached. Whereas, \texttt{QKE-fullC} can exhibit transient decoherence in one or more sectors, followed by sudden regrowth and approximately constant coherence magnitudes at later times as the asymptotic states form (see Fig.~\ref{fig:QKE_coherence}). For examples, \texttt{QKE-fullC} retains appreciable coherence only in the $e\mu$ sector in 2S-A, but in all three sectors in 2S-B and 3S. Thus, the number of initially unstable sectors determines neither the transient sector participation nor the surviving coherence structure.

These quasisteady coherent configurations explain the departures from
\texttt{Classical} in Fig.~\ref{fig:QKE_populations}. For
\texttt{QKE-fullC}, the diagonal component of the QKE gives
\begin{equation}
\partial_t
f_{\nu_\alpha}=-i[H_{\nu\nu},\varrho]_{\alpha\alpha}+j_{\nu_\alpha}-\Gamma_{\nu_\alpha}f_{\nu_\alpha}\simeq0\,,
\label{eq:qke-late-balance}
\end{equation}
with an analogous antineutrino relation. Coherent flavor transfer can
balance a nonzero collisional source, rather than requiring
$f_{\nu_\alpha}=j_{\nu_\alpha}/\Gamma_{\nu_\alpha}$ as in the
\texttt{Classical} equilibrium of the fixed-coefficient model. The
$\tau$-flavor populations can therefore remain displaced from their
\texttt{Classical} values, while their proximity to \texttt{Classical} in
2S-A mainly reflects the absence of appreciable $e\tau$ and $\mu\tau$
coherences at late times. A similar balance was also identified in
two-flavor CFI models~\cite{Froustey2025,LiuCFIInPreparation}.

Figures~\ref{fig:QKE_populations} and \ref{fig:QKE_coherence} reveal several important challenges for future CFI subgrid models. The \texttt{QKE-noDC} and \texttt{QKE-fullC} prescriptions differ not only in their asymptotic flavor populations and surviving coherences, but also during the nonlinear evolution and in the timescales required to reach these states. In particular, \texttt{QKE-noDC} tends toward flavor equipartition and may overestimate the degree of flavor equilibration compared with \texttt{QKE-fullC}. This demonstrates that diagonal collisional repopulation affects the nonlinear flavor evolution rather than merely the final relaxation. Therefore, a CFI subgrid model should not simply add an independently approximated flavor-conversion source on top of the classical collision operator in global simulations. Instead, the nonlinear coupling between coherent flavor conversion and diagonal collisional repopulation should be treated consistently, replacing the corresponding classical collision sources to avoid double counting. Moreover, the relaxation timescales can be comparable to those of neutrino transport and background evolution, so local asymptotic states may not be reached before the background changes. Appreciable flavor coherence can also survive in quasisteady states. Both the finite-time evolution and surviving coherence must therefore be considered when developing reliable CFI subgrid models.

\section{Conclusion}\label{sec:conclusion}
We have investigated fast and collisional neutrino flavor instabilities in long-lived binary neutron star (BNS) merger remnants using six-species general-relativistic Boltzmann neutrino transport in axisymmetry, following a handoff from 3D grey-M1 merger simulations. The transport includes kernel-based muonic interactions and employs a novel and efficient linearization for those collisional sources. The Boltzmann transport evolves the full neutrino momentum space to avoid analytical closures and Monte Carlo sampling noise.

The evolution of the $npe\mu$ case retains hydrodynamical properties, muon fractions, and trapped-neutrino populations similar to those of the parent grey-M1 simulations~\cite{Ng2024c}, except a stronger demuonization in the disk. The evolution naturally recovers neutrino-trapping equilibrium, a semitransparent disk undergoing demuonization
and cold weak equilibrium in the transparent regions. 
Including muonic interactions breaks the conventional muon–tau (anti)neutrino degeneracy in two-flavor simulations, enabling independent investigation of all three flavor sectors.

Our linear stability analysis reveals strong and widespread 
collisional flavor instability (CFI) in all three sectors, with growth rates much higher than those reported in core collapse supernova (CCSN). 
A large asymmetry between $\nu_\mu$ and $\bar{\nu}_\mu$ alone does not guarantee CFI, because collisional driving must overcome damping.
Narrow, strong resonance-like CFI (R-CFI) structures appear in all three sectors, although their peak growth rates remain unresolved. We found that an effective $\nu_\tau$ energy sphere provides a useful indicator separating thermalized from most CFI-active regions. These features are qualitatively insensitive to the EOS.

For the Fast flavor instability (FFI) occurrences in $npe$-matter, it reproduces the extended $e\chi$-sector FFI outside the disk found in previous studies. However, in $npe\mu$ matter, demuonization produces a strong $\nu_\mu$ excess and predominantly one-sided angular distributions in the $e\mu$ and $\mu\tau$ sectors, strongly suppressing their FFI. Muonic interactions also modify the electron-flavor angular crossings, shifting the $e\tau$-sector FFI toward the denser disk regions. This different morphology may affect the neutrino irradiation and composition of outer-disk outflows. 

Additionally, 
to investigate the subsequent nonlinear evolution of flavor instabilities and their asymptotic states, providing insights for future CFI subgrid models, 
we have solved the first local three-flavor quantum kinetic equations (QKEs) using a self-consistently evolved muonic BNS remnant background and the corresponding muonic rates at a few 
selected locations.
For QKEs with full collision operator, sectors with higher linear growth rates develop coherence earlier, but nonlinear coupling excites all three-sector coherences even when only two sectors are initially unstable, demonstrating that independent two-flavor treatments cannot capture the subsequent evolution.
It also produces very different tau-flavor populations compared to 
classical evolution.
Whereas, QKEs omitting diagonal collision terms, which avoids double-counting source terms in global simulations, tend toward nearly equal flavor populations, overestimating flavor equilibration.

These findings have important implications for future CFI subgrid modeling. Unlike FFI-type subgrid models motivated by rapid relaxation toward asymptotic states, e.g.,~\cite{Nagakura2023c,Xiong2024}, CFI exhibits a broader range of nonlinear relaxation timescales and a stronger dependence on collisional repopulation. Although the initial coherence growth can be rapid, the subsequent evolution may take $10^{2}$-$10^{3}~\mu\rm{s}$ to reach 
the asymptotic state, comparable to characteristic local dynamical timescales. 
A reliable CFI subgrid model should account for nonlinear coupling among all flavor sectors and the combined effects of coherent evolution and diagonal collisional repopulation. 
Rather than introducing an independent flavor-conversion source term to the neutrino transport with classical collision terms, a consistent effective treatment should replace the corresponding classical collision update to avoid double counting and recover the appropriate asymptotic states of the underlying neutrino quantum kinetics. 

Current BNS merger models remain inadequate, even from the perspectives of classical and quantum three-flavor neutrino transport.
We should pursue several directions for future work: (i) incorporating kernel-based rates into the QKE collision operator; (ii) performing self-consistent 3D merger simulations to investigate flavor instabilities during early muonization and the later evolution of long-lived remnants; and (iii) the evolutions of multi-sector activated FFI remain 
to be studied, especially the rapid muonization during the merger stage.

Our findings in BNS merger remnants may also have implications for CCSN simulations. The nonlinear coupling among flavor sectors demonstrated here could also occur in CCSNe, potentially affecting neutrino signatures, heating, and explosion dynamics, thus motivating the development of three-flavor neutrino transport in CCSN simulations.
 \section*{Acknowledgments}
The authors are grateful to Ryuichiro Akaho, Adam Burrows, Eirik Endeve, Francois Foucart, Henrique Gieg, Debraj Kundu, Zoha Laraib, Bronson Messer, Vassilios Mewes, Anthony Mezzacappa, Hiroki Nagakura, David Radice, Samantha Rath, Sherwood Richers, Erick Urquilla, Tianshu Wang, and Masamichi Zaizen for insightful discussions.
H.H.Y.N. is supported by a Croucher Fellowship from the Croucher Foundation.
J.L. is supported by Grant-in-Aid for JSPS Fellows (26KJ2052).  A.O. is
supported by the Research Foundation Flanders (FWO) research project G086722N
and FWO International Research Infrastructure (IRI) projects I002123N and
I000725N.  S.Y. is supported by Grant-in-Aid for Scientific Research
(25K01006).  E.R.M. acknowledges support by the U.S. National Science
Foundation under Grant No. PHY-2541792, and is also supported by a Research
Fellowship from the Sloan Foundation and a William H. Hurt Scholarship at the
California Institute of Technology.  L.R. acknowledges support by the ERC
Advanced Grant ``JETSET: Launching, propagation and emission of relativistic
jets from binary mergers and across mass scales'' (Grant No. 884631) and by the
Walter Greiner Gesellschaft zur F\"orderung der physikalischen
Grundlagenforschung e.V.  through the Carl W. Fueck Laureatus Chair.
The global simulations were performed on the NSF Frontera supercomputer under
grant AST21006.  Additional simulations were performed on the local ITP
Supercomputing Clusters Iboga and Calea.  The local QKE simulations were
performed using local HPC clusters at Yamada Laboratory, Waseda University.

\appendix
\onecolumngrid
\noindent
\begin{figure}[!h]
    \centering
    \includegraphics[width=0.48\textwidth]{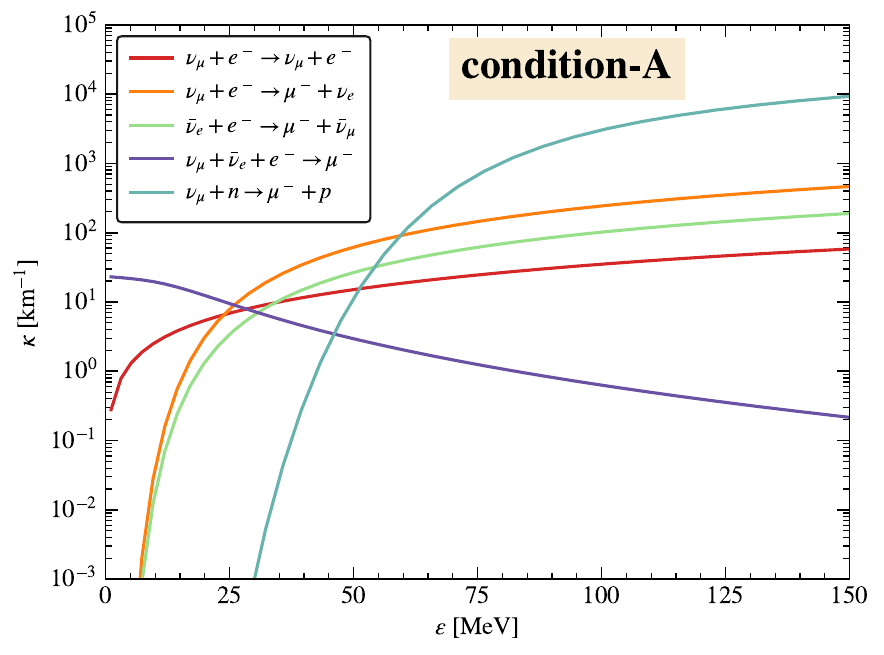}
    \hfill
    \includegraphics[width=0.48\textwidth]{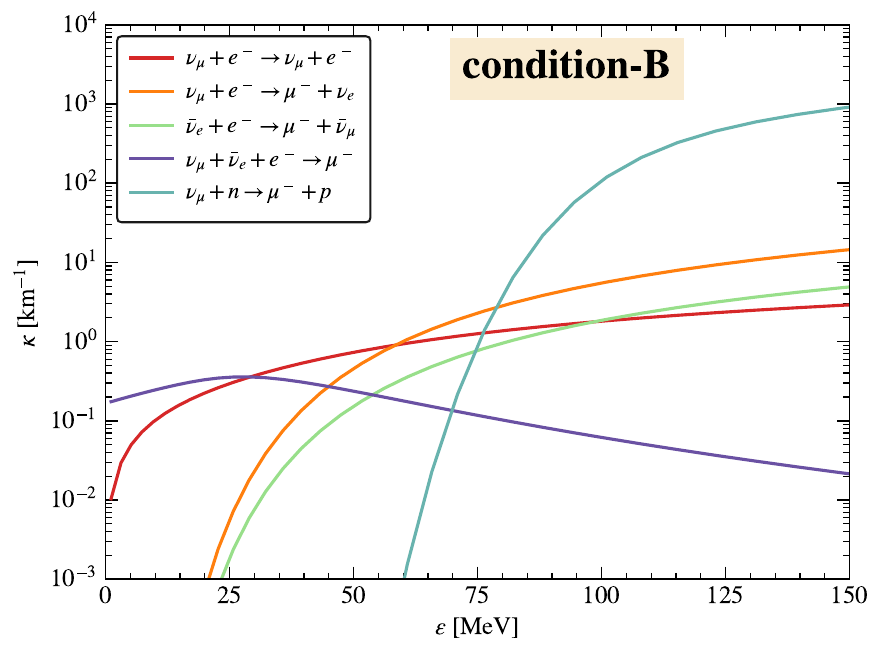}
    \caption{Inverse mean free paths (Eq.~\eqref{eq:guo-opacity-convention}) computed from \texttt{Weakhub} for various interactions under conditions A (left) and B (right) of Ref.~\cite{Guo2020}, compared with the reference results in Fig.~5 of Ref.~\cite{Guo2020}. Shown are inelastic scattering (IS) between neutrinos and electrons, lepton-flavor exchange/conversion producing $\mu^-$, inverse lepton decay (ILD), and charged-current (CC) absorption on neutrons.
    The \texttt{Weakhub} implementation closely reproduces the reference results over the energy range.}
    \label{fig:guo-opacity-benchmark}
\end{figure}
\twocolumngrid
\section{Validations of the muonic Boltzmann transport}
\label{app:reduced-kernels}

\subsection{Opacity Spectra of muonic Weak Interactions}
\label{app:opacity-benchmark}

We validate the energy-dependent spectra of the muonic rates against benchmark conditions A and B of
Ref.~\cite{Guo2020}.
Figure~\ref{fig:guo-opacity-benchmark} compares the resulting
inverse mean free paths directly with Fig.~5 of
Ref.~\cite{Guo2020}.
Following Ref.~\cite{Guo2020}, the CC opacity excludes
the stimulated absorption correction, i.e., $\kappa_a$, while
for the kernel-based processes, we compute the effective inverse mean free path as~\cite{Guo2020} with the same convention,
\begin{equation}
\kappa^{X}(\varepsilon)
={}\frac{2\pi}{c(2\pi\hbar c)^3}
\int_0^\infty d\varepsilon'\,\varepsilon'^2 \Phi_{X,0}^{-}(\varepsilon,\varepsilon')
f^{\prime}(\varepsilon')\,,
\label{eq:guo-opacity-convention}
\end{equation}
where $X\in\{\mathrm{IS},\mathrm{LFE/C},\mathrm{ILD}\}$.  Following
Ref.~\cite{Guo2020}, we take $f^{\prime}(\varepsilon')=1$ for IS and
LFE/C, corresponding to an unoccupied final neutrino state in the
opacity, and $f^{\prime}(\varepsilon')=f_{\nu_b}^{\rm
  eq}(\varepsilon',\mu_{\nu_b}^{\rm eq})$ for ILD, which means the
partner neutrino is an absorbed initial-state particle. Here,
$\Phi_{X,0}^{-}$ denotes the zeroth-order Legendre coefficient of the
corresponding $R^-_X{}(\varepsilon,\varepsilon',\omega)$.

The close agreement with the results of Ref.~\cite{Guo2020} across both
benchmark conditions and all reaction channels validates the
implementation of muonic rates in \texttt{Weakhub}, providing confidence
in the microphysics employed in our transport calculations.  In
particular, we recover the characteristic hierarchy identified in
Ref.~\cite{Guo2020}, with ILD becoming important for low-energy $\nu_\mu$
and CC absorption dominating toward higher energies.  Similar behavior
and the importance of these muonic channels have also been found in
subsequent CCSN and PNS studies~\cite{Fischer2020b,Sugiura2022}.

\subsection{Hot neutron star tests in $npe$- and $npe\mu$-matter}
\label{app:hot-star-demuonization}
\begin{figure*}
    \includegraphics[width=\textwidth]{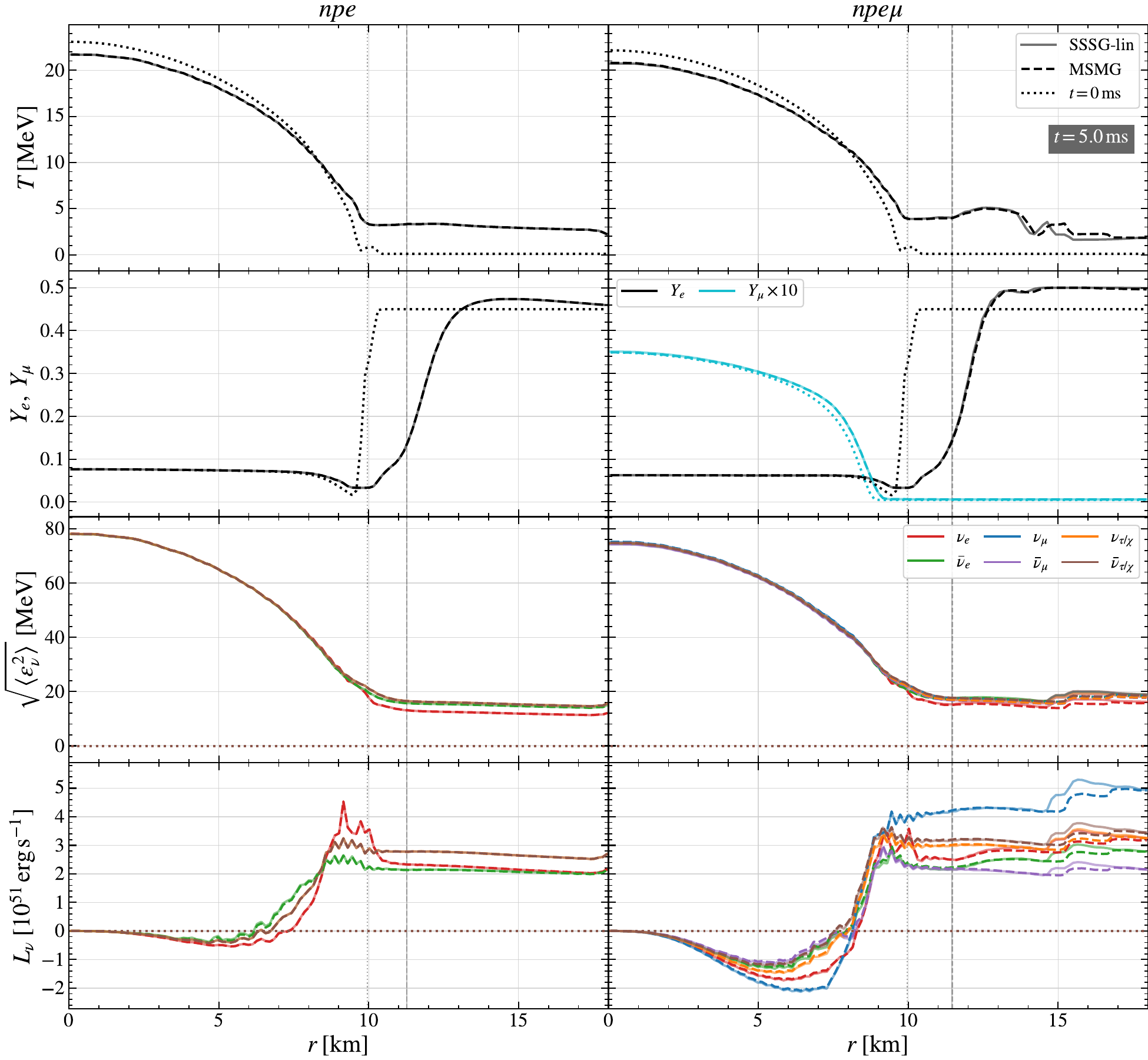}
    \caption{
    Radial profiles of the 1D hot-neutron-star tests in $npe$ matter (left) and $npe\mu$ matter (right) at $t=5\,{\rm ms}$. From top to bottom, we show temperature $T$, the charged-lepton fractions $Y_e$ and $Y_\mu$, the neutrino root-mean-squared energies $\sqrt{\langle\varepsilon_\nu^2\rangle}$, and neutrino luminosities $L_\nu$. For the $npe\mu$ model, $Y_\mu$ is rescaled by a factor of 10. Solid and dashed curves correspond to SSSG-lin and MSMG, respectively, while dotted curves show the initial profiles at $t=0$. The colors in the lower two rows distinguish the evolved neutrino species.
    The vertical grey contour lines correspond to $\rho=10^{11}~\rm{g~cm^{-3}}$.
    The SSSG-lin treatment closely reproduces the results of the fully coupled MSMG treatment.
    }
    \label{fig:hotns1d}
\end{figure*}
We perform a spherically symmetric hot neutron star test to validate the approximate dilute-muon treatment in Sec.~\ref{sec:dilute-muon} and the SSSG-lin solver in Sec.~\ref{sec:implicit-composition} against the fully coupled MSMG treatment, which retains the full cross-energy, momentum-space, and species couplings~\cite{Cheong2023,Offermans2026}.

We consider both $npe$ and $npe\mu$ matter, evolving four and six
neutrino species, respectively. The initial data are computed with
XNS~\cite{Bucciantini2014} using the SFHo EOS, with a central density of
$6.7\times10^{14}\,{\rm g\,cm^{-3}}$ and a constant specific entropy of
$1\,k_{\rm B}$, and then initialized in neutrinoless $\beta$-equilibrium,
giving a maximum temperature of $T\simeq23~{\rm MeV}$ at the center. We
use $N_\varepsilon=16$ and $N_\vartheta=10$, and the finest spatial
resolution of $144\,{\rm m}$ covering the inner $15\,{\rm km}$ with four
refinement levels. The MSMG treatment used here is 15 and 24 times more
expensive than SSSG-lin for the $npe$ and $npe\mu$ tests, respectively.

Figure~\ref{fig:hotns1d} shows that the star remains dynamically stable
during the evolution, with the matter profiles remaining smooth despite
an outward relaxation of the stellar surface. More importantly, SSSG-lin
closely matches the MSMG results throughout the simulation domain, with
maximum relative differences of $\lesssim5\%$ in the hydrodynamical
quantities and $\lesssim12\%$ in the neutrino signals, both occurring at
radii $\gtrsim14\,\mathrm{km}$; elsewhere, the differences remain below
$\sim2\%$. The small timestep required by the Boltzmann solver further
increases the accuracy of SSSG-lin. Despite the slight neutrino-number
non-conservation of the SSSG-lin scheme, the test demonstrates that it
retains the physics of kernel interactions with high accuracy while
substantially reducing the cost. Furthermore, when evaluating the
collisional integrals in SSSG-lin with the partner distribution, we find
no significant difference between employing the full $f^{\rm lag}_\nu$
and using angle-averaged $\langle f^{\rm lag}_\nu\rangle$ with only the
$\Phi_0$ kernel used in Ref.~\cite{Foucart2026}, with differences below a
few percent in all neutrino signatures. The validity of this
approximation in multidimensional simulations and at higher angular
resolution will be investigated in future work.

As the initially muon-rich surface oscillates and ejects
muonic matter outward to low-$\rho$ regions,
the ejected matter consistently demuonizes quickly and recovers
the asymptotic value $Y_{\mu,\rm thr}$ with $\mu_\mu\simeq\mu_e$, and
$\Delta\mu_{npe}\simeq\Delta\mu_{np\mu}$ approaches $\mu_n-\mu_p$.
This test reproduces the physically expected behavior of muons in transparent regions, supporting the validity of the dilute-muon treatment and the muonic physics implemented in \texttt{Gmunu}.

Consequently, the evolution approaches the intended dilute-muon branch rather than terminating at a finite $Y_\mu$ floor associated with $\mu_\mu\sim m_\mu c^2$.
In tests without the dilute-muon regularization,
spurious $\nu_\mu$ and $\bar\nu_\mu$ emission and reabsorption
drive up $Y_\mu$, and the resulting changes in $Y_\mu$ indirectly alter
$Y_e$ in the ejecta and distort the neutrino signatures.




\section{Muon decay fallback}\label{app:ILD_extended}
Muon decay provides an important channel for demuonization in
transparent matter~\cite{Bolligthesis, Guo2020}. As muon-rich matter expands
or is advected into the dilute, neutrino-transparent regime, the
final-state blocking becomes weak and negative muons decay into
electrons and neutrinos on the intrinsic vacuum timescale,
$\tau_{\rm vac}\simeq2.2\times10^{-6}\,{\rm s}$~\cite{MuLan2011}, which is much
shorter than the characteristic millisecond hydrodynamical timescale
of the merger remnant.

The inverse-lepton-decay (ILD) channel
is normally evaluated using the tabulated kernels
$R^{\pm}_{\rm ILD}(\mu_e/T, \mu_\mu/T, T)$ over the corresponding
neutrino-energy range ($0.5$--$300\,{\rm MeV}$ for this work).
In the transparent limit,
however, these quantities can leave the tabulated domain at low $\rho$ and $T$, for example, for
$\mu_e/T<0.1$, $\mu_\mu/T<0.1$, or
$T<0.5\,{\rm MeV}$, as well as because of the limited spectral coverage over the assigned neutrino-energy range.
Also, the finite tabulation and transport-energy resolution may poorly resolve the low-energy end of muon decay near those regions.
Simply clamping $\mu_e/T$, $\mu_\mu/T$, or $T$ to their table
boundaries does not in general guarantee the correct sign or magnitude of the demuonization rate.

These numerical limitations should neither suppress the physical vacuum
decay of muons nor prevent transparent matter from demuonizing on a
physically reasonable timescale toward the asymptotic value
$Y_\mu \simeq Y_{\mu,\rm thr}$ (see Sec.~\ref{sec:dilute-muon}).
So, we evaluate an effective forward-decay fallback on the fly
whenever the required state lies outside the ILD table.
The fallback decay source is switched off once $Y_\mu$ reaches $Y_{\mu,\mathrm{thr}}$.

The fallback is normalized by the measured free-muon lifetime
$\tau_{\rm vac}=2.2\times10^{-6}\,{\rm s}$~\cite{MuLan2011}, with
$\Gamma_0=\rho Y_\mu/(m_u\tau_{\rm vac})$.
We denote the energies of $\nu_\mu$ and $\bar\nu_e$ by
$\varepsilon$ and $\varepsilon'$, respectively, such that the
final-electron energy is
$E_e=m_\mu-\varepsilon-\varepsilon'$.
The maximum neutrino energy is
$\varepsilon_{\max}=(m_\mu^2-m_e^2)/(2m_\mu)$.

\begin{figure*}
    \includegraphics[width=\textwidth]{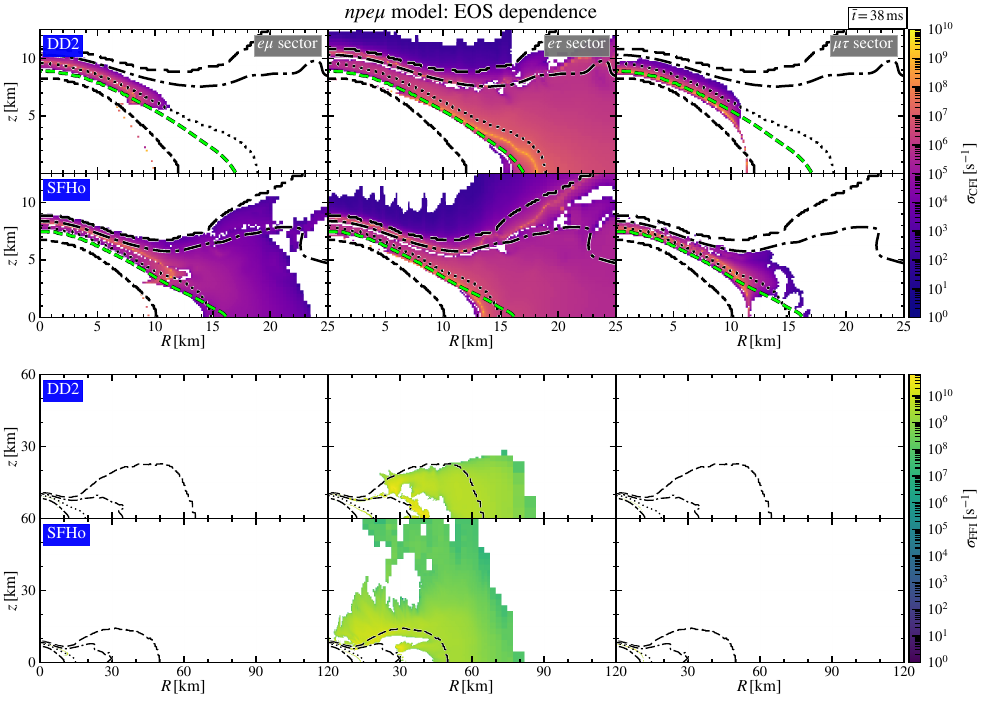}
    \caption{
    Meridional distributions of the CFI and FFI growth rates
    for all three sectors at $\bar t=38\,{\rm ms}$ for the fiducial case and another case with
    a softer SFHo EOS~\cite{Steiner2013} in $npe\mu$ matter.
    The black
      contour lines are density contours at $\rho=10^{14}$, $10^{13}$,
      $10^{12}$ and $10^{11}~\rm{g~cm^{-3}}$, while the lime dashed
      contour corresponds to the energy-sphere location of $\nu_\tau$ at
      $\tau_{\nu_\tau} = 1$ (see Eq.~\eqref{eq:effective-energy-sphere}).
    }
    \label{fig:eos_dependence}
\end{figure*}
Only energy pairs in the three-body kinematic domain
$\mathcal K$ are retained, satisfying
$0<\varepsilon,\varepsilon'\leq\varepsilon_{\max}$,
$E_e\geq m_e$, and
\[
 \left|
 1+\frac{m_\mu^2-m_e^2
 -2m_\mu(\varepsilon+\varepsilon')}
 {2\varepsilon\varepsilon'}
 \right|\leq1 .
\]
For these pairs, we adopt the normalized rest-frame spectral
weight
$P(\varepsilon,\varepsilon')
=\widetilde P(\varepsilon,\varepsilon')/
 \sum_{\mathcal K}\widetilde P$, with
$\widetilde P
=\varepsilon'(\varepsilon_{\max}-\varepsilon')
 \Delta\varepsilon\,\Delta\varepsilon'$,
where $\Delta\varepsilon$ and $\Delta\varepsilon'$ are the corresponding
$\nu_\mu$ and $\bar\nu_e$ energy-bin widths, respectively.

The corresponding blocked muon-decay source for $\nu_\mu$ is
\begin{equation}
\begin{aligned}
\mathcal C_{\nu_\mu}^{\rm vac}
(\varepsilon,\Omega_p)
={}&
\frac{\Gamma_0}{W(\varepsilon)}
\sum_{\varepsilon'\,:\,(\varepsilon,\varepsilon')\in\mathcal K}
P(\varepsilon,\varepsilon')
\left[1-f_e(E_e,\mu_e,T)\right]
\\
&\times
\left[
1-\left\langle
f_{\bar\nu_e}^{\rm lag}
(\varepsilon',\Omega'_p)
\right\rangle_{\Omega'_p}
\right]
\left[
1-f_{\nu_\mu}(\varepsilon,\Omega_p)
\right] \,,
\end{aligned}
\label{eq:ild_vac}
\end{equation}
where $ W(\varepsilon) = \frac{\varepsilon^2\,\Delta\varepsilon}{2\pi^2}
$ denotes the discrete momentum-space quadrature weight of the energy
group at $\varepsilon$, integrated over all angular bins.

We find that this fix is relevant only in very dilute, cool regions far
outside the merger remnant, at $\rho \lesssim 10^9 ~\rm{g~cm^{-3}}$.
This fix has no appreciable impact on the flavor-conversion results.

\section{EOS-dependence of the flavor instabilities}
\label{sec:eos_dependence}
%
\begin{figure*}
    \hspace*{-0.02\textwidth}
    \includegraphics[width=\textwidth]{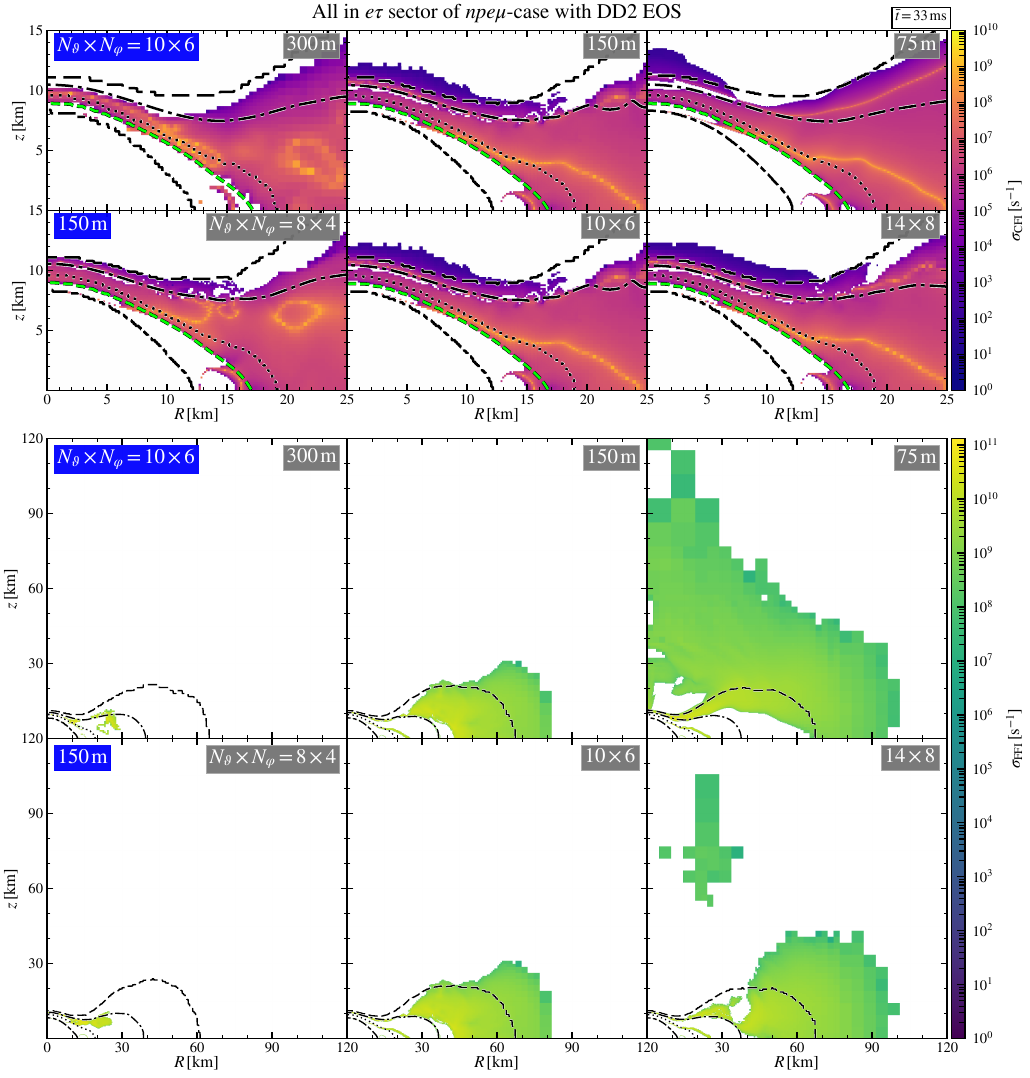}
    \caption{ Meridional distributions of the CFI and FFI growth rates in
      the $e\tau$ sector of the $npe\mu$ model with the DD2 EOS at $\bar
      t=33\,\mathrm{ms}$. For each type of instability, the first row
      compares the finest spatial resolutions of $300$, $150$, and
      $75\,\mathrm{m}$ at fixed $N_\vartheta\times N_\varphi=10\times6$,
      while the second row compares the momentum-space angular
      resolutions of $8\times4$, $10\times6$, and $14\times8$ at fixed
      $150\,\mathrm{m}$. All simulations use $N_\varepsilon=16$. 
    }
    \label{fig:convergence}
\end{figure*}

Here, we examine the EOS dependence of the flavor instabilities in $npe\mu$ matter using the softer SFHo EOS~\cite{Steiner2013}.
We perform another set of simulations with the SFHo EOS
using the same handoff procedure for the SFHo remnant in Sec.~\ref{sec:numerical-setup} from Ref.~\cite{Ng2024c}.
Figure~\ref{fig:eos_dependence} shows its resulting FFI and CFI growth rates for all three flavor sectors.

The $e\tau$-sector FFI exhibits EOS dependence.
The $e\tau$ FFI in SFHo extends over a larger volume, particularly toward higher latitudes and the polar region, while the disk-associated component around $\rho\sim10^{11}\,{\rm g\,cm^{-3}}$ remains present. The hotter and denser SFHo remnant sustains larger neutrino populations and stronger emission and scattering, leading to more efficient reprocessing and redirection of $\nu_e$
and $\bar{\nu}_e$ and larger angular variations.
This causes more $e{\rm LN}$ angular crossings.
On the other hand, the disk in the SFHo remnant also enters the
demuonization stage, in which the dominant magnitude of $\mu$LN
suppresses the other two FFI sectors, as in the DD2 case.

In comparison, CFI is less sensitive to the EOS variation.
Slightly broader CFI regions remain in all three sectors, as
the hotter and denser remnant provides larger reaction rates modifying both the relative collisional driving and damping. Importantly, the $\nu_\tau$ energy sphere remains a robust macroscopic indicator
for separating CFI-inactive thermalized regions from
CFI-active ones.
The narrow R-CFI structures also persist at locations similar to those in the DD2 case, where the neutrino--antineutrino number-density asymmetry becomes very small and is determined by the offset between the $\nu$ and $\bar\nu$ neutrinospheres.
The morphology of CFI is very robust against
the EOS variation, even though their precise locations shift due to the EOS stiffness.

We emphasize that this comparison inherits different merger histories from the parent simulations: the hotter and denser SFHo merger undergoes stronger early muonization than DD2~\cite{Ng2024c}, so part of the differences here may reflect dependence on the prior
hydrodynamic evolution.

\section{Resolution and convergence tests}
\label{sec:convergence}
To assess the resolution dependence of the flavor instabilities, we
perform additional $npe\mu$ simulations with the DD2 EOS. Starting from
the fiducial production setup with a finest spatial resolution of
$150\,\mathrm{m}$ and a momentum-space angular resolution of
$N_\vartheta\times N_\varphi=10\times6$, we vary the finest spatial
resolution among $300$, $150$, and $75\,\mathrm{m}$ at fixed
$N_\vartheta\times N_\varphi=10\times6$, and the momentum-space angular
resolution among $8\times4$, $10\times6$, and $14\times8$ at a fixed
spatial resolution of $150\,\mathrm{m}$. All simulations retain the same
$N_\varepsilon=16$ energy groups. Due to computational limitations, these
models are evolved for $3\,\mathrm{ms}$ after the $2\,\mathrm{ms}$ initial
radiation relaxation, instead of the $8\,\mathrm{ms}$ used in the runs
presented in the main text.

Figure~\ref{fig:convergence} compares the CFI and FFI growth rates in the
$e\tau$ sector at $\bar t=33\,\mathrm{ms}$. The distributions of broad CFI
remain quantitatively and geometrically similar across different spatial
and momentum-space angular resolutions. However, the trends and
structures of the narrow R-CFI are poorly resolved in the
$300\,\mathrm{m}$ or $8\times4$ runs. At the fiducial resolution, the
R-CFI structures are very similar to the $75\,\mathrm{m}$ or
$N_\vartheta\times N_\varphi=14\times8$ runs, except for the upper narrow
R-CFI layer around $z\simeq10\,\mathrm{km}$, which can only be clearly
resolved in the $150\,\mathrm{m}$ case.

FFI occurrences exhibit strong resolution dependence. Coarse spatial and
momentum-space angular resolutions significantly underestimate their
extent. The high-spatial-resolution case ($75~\rm{m}$) reveals more
extensive FFI outside the disk, especially toward the polar regions, but
fewer occurrences in the outer part of the disk. The case with
$N_\vartheta\times N_\varphi=14\times8$ also shows FFI regions extending
farther outward, but has similar occurrences in the outer disk compared
with the fiducial case. Higher spatial resolution better resolves
distinct directional neutrino beams and reduces spatial numerical
diffusion, while a higher number of momentum-space angular bins better
captures angular anisotropies, forward-peaked distributions, and narrow
sign changes in $\mathcal G_{e\tau}(\Omega_p)$. Similarly,
Ref.~\cite{Akaho2026b} found that coarse momentum-space angular
resolution underestimates FFI occurrences. The FFI diagnostics 
in Ref.~\cite{Akaho2026b} does not achieve quantitative consistency. 
All in all,
further simulations with higher resolutions are necessary to establish
the true properties of FFI across different astrophysical systems.

\bibliography{aeireferences,additional_bib}

@misc{LiuCFIInPreparation,
  author = {Liu, Jiabao and others},
  note   = {In preparation}
}

@article{Liu:2023vtz,
    author = "Liu, Jiabao and Nagakura, Hiroki and Akaho, Ryuichiro and Ito, Akira and Zaizen, Masamichi and Yamada, Shoichi",
    title = "{Universality of the neutrino collisional flavor instability in core-collapse supernovae}",
    eprint = "2310.05050",
    archivePrefix = "arXiv",
    primaryClass = "astro-ph.HE",
    doi = "10.1103/PhysRevD.108.123024",
    journal = "Phys. Rev. D",
    volume = "108",
    number = "12",
    pages = "123024",
    year = "2023"
}

@article{Abbar2017,
    author = "Abbar, Sajad and Duan, Huaiyu",
    title = "{Fast neutrino flavor conversion: roles of dense matter and spectrum crossing}",
    eprint = "1712.07013",
    archivePrefix = "arXiv",
    primaryClass = "hep-ph",
    doi = "10.1103/PhysRevD.98.043014",
    journal = "Phys. Rev. D",
    volume = "98",
    number = "4",
    pages = "043014",
    year = "2018"
}

@article{Abbott2017_etal,
  title = {GW170817: Observation of Gravitational Waves from a Binary Neutron Star Inspiral},
  author = {{Abbott}, B.~P. and {Abbott}, R. and {Abbott}, T.~D. and {Acernese}, F. and
         {Ackley}, K. and {Adams}, C. and {Adams}, T. and {Addesso}, P. and
         {Adhikari}, R.~X. and {Adya}, V.~B. and et al.},
  collaboration = {LIGO Scientific Collaboration and Virgo Collaboration},
  journal = {Phys. Rev. Lett.},
         year = "2017",
        month = oct,
       volume = {119},
       number = {16},
          eid = {161101},
        pages = {161101},
          doi = {10.1103/PhysRevLett.119.161101},
archivePrefix = {arXiv},
       eprint = {1710.05832},
 primaryClass = {gr-qc},
       adsurl = {https://ui.adsabs.harvard.edu/abs/2017PhRvL.119p1101A}
}

@article{Abbott2017b,
   author = {{The LIGO Scientific Collaboration} and {the Virgo Collaboration} and {Abbott}, B.~P. and {Abbott}, R. and {Abbott}, T.~D. and {Acernese}, F. and {Ackley}, K. and a{Adams}, C. and {Adams}, T. and {Addesso}, P. and
         {Adhikari}, R.~X. and {Adya}, V.~B. and et al.},
  collaboration = {LIGO Scientific Collaboration and Virgo Collaboration},
  title={Multi-messenger Observations of a Binary Neutron Star Merger},
  journal={Astrophys. J. Lett.},
         year = "2017",
        month = oct,
       volume = {848},
       number = {2},
          eid = {L12},
        pages = {L12},
          doi = {10.3847/2041-8213/aa91c9},
archivePrefix = {arXiv},
       eprint = {1710.05833},
 primaryClass = {astro-ph.HE},
       adsurl = {https://ui.adsabs.harvard.edu/abs/2017ApJ...848L..12A}
}

@article{Akaho2023,
    author = "Akaho, Ryuichiro and Liu, Jiabao and Nagakura, Hiroki and Zaizen, Masamichi and Yamada, Shoichi",
    title = "{Collisional and fast neutrino flavor instabilities in two-dimensional core-collapse supernova simulation with Boltzmann neutrino transport}",
    eprint = "2311.11272",
    archivePrefix = "arXiv",
    primaryClass = "astro-ph.HE",
    doi = "10.1103/PhysRevD.109.023012",
    journal = "Phys. Rev. D",
    volume = "109",
    number = "2",
    pages = "023012",
    year = "2024"
}

@article{Akaho2026,
    author = "Akaho, Ryuichiro and Nagakura, Hiroki and Iwakami, Wakana and Furusawa, Shun and Harada, Akira and Okawa, Hirotada and Matsufuru, Hideo and Sumiyoshi, Kohsuke and Yamada, Shoichi",
    title = "{Bifurcated Impact of Neutrino Fast Flavor Conversion on Core-Collapse Supernovae Informed by Multiangle Neutrino Radiation Hydrodynamics}",
    eprint = "2601.08269",
    archivePrefix = "arXiv",
    primaryClass = "astro-ph.HE",
    doi = "10.1103/fksy-1jtw",
    journal = "Phys. Rev. Lett.",
    volume = "136",
    number = "19",
    pages = "191002",
    year = "2026"
}

@article{Akaho2026b,
    journal = {},

    author = "Akaho, Ryuichiro",
    title = "{Momentum Space Resolution Dependence in Boltzmann Neutrino Radiation Hydrodynamics Simulations of Core-collapse Supernovae}",
    eprint = "2607.12891",
    archivePrefix = "arXiv",
    primaryClass = "astro-ph.HE",
    month = "7",
    year = "2026"
}

@article{Alford2021b,
    author = "Alford, Mark and Harutyunyan, Arus and Sedrakian, Armen",
    title = "{Bulk viscosity from Urca processes: npe{\ensuremath{\mu}} matter in the neutrino-trapped regime}",
    eprint = "2108.07523",
    archivePrefix = "arXiv",
    primaryClass = "astro-ph.HE",
    doi = "10.1103/PhysRevD.104.103027",
    journal = "Phys. Rev. D",
    volume = "104",
    number = "10",
    pages = "103027",
    year = "2021"
}

@article{Alford2026,
    author = "Alford, Mark G. and Brodie, Liam and Foucart, Francois and Haber, Alexander",
    title = "{Thermalization of neutrinos in a neutron star merger simulation}",
    eprint = "2603.06788",
    archivePrefix = "arXiv",
    primaryClass = "astro-ph.HE",
    doi = "10.1103/37ys-qs4h",
    journal = "Phys. Rev. D",
    volume = "114",
    number = "4",
    pages = "043033",
    year = "2026"
}

@article{Baiotti2016,
      author         = "Baiotti, Luca and Rezzolla, Luciano",
      title          = "{Binary neutron-star mergers: a review of Einstein's
                        richest laboratory}",
      journal        = "Rept. Prog. Phys.",
      volume         = "80",
      year           = "2017",
      number         = "9",
      pages          = "096901",
      doi            = "10.1088/1361-6633/aa67bb",
      eprint         = "1607.03540",
      archivePrefix  = "arXiv",
      primaryClass   = "gr-qc",
      SLACcitation   = "%%CITATION = ARXIV:1607.03540;%%"
}

@article{Banerjee2011,
    author = "Banerjee, Arka and Dighe, Amol and Raffelt, Georg",
    title = "{Linearized flavor-stability analysis of dense neutrino streams}",
    eprint = "1107.2308",
    archivePrefix = "arXiv",
    primaryClass = "hep-ph",
    reportNumber = "MPP-2011-81, TIFR-TH-11-30",
    doi = "10.1103/PhysRevD.84.053013",
    journal = "Phys. Rev. D",
    volume = "84",
    pages = "053013",
    year = "2011"
}

@article{Bollig2017,
       author = {{Bollig}, R. and {Janka}, H. -T. and {Lohs}, A. and {Mart{\'\i}nez-Pinedo}, G. and {Horowitz}, C.~J. and {Melson}, T.},
        title = "{Muon Creation in Supernova Matter Facilitates Neutrino-Driven Explosions}",
      journal = {Phys. Rev. Lett.},
         year = 2017,
        month = dec,
       volume = {119},
       number = {24},
          eid = {242702},
        pages = {242702},
          doi = {10.1103/PhysRevLett.119.242702},
archivePrefix = {arXiv},
       eprint = {1706.04630},
 primaryClass = {astro-ph.HE},
       adsurl = {https://ui.adsabs.harvard.edu/abs/2017PhRvL.119x2702B}
}

@phdthesis{Bolligthesis,
	author = {Bollig, Robert G\"unther},
	title = {Muon Creation and Effects in Supernovae},
	year = {2018},
	school = {Technische Universit\"ut M\"unchen},
	pages = {184},
	note = {},
	url = {https://mediatum.ub.tum.de/1435391},
}

@article{Bruenn85,
	Adsurl = {http://adsabs.harvard.edu/abs/1985ApJS...58..771B},
	Author = {{Bruenn}, S.~W.},
	Doi = {10.1086/191056},
	Journal = {Astrophys. J., Supp.},
	Month = aug,
	Pages = {771-841},
	Title = {{Stellar core collapse - Numerical model and infall epoch}},
	Volume = 58,
	Year = 1985}

@article{Bucciantini2014,
    author = "Bucciantini, N. and Pili, A. G. and Del Zanna, L.",
    editor = "Pogorelov, N. V. and Audit, E. and Zank, G. P.",
    title = "{Solving the 3+1 GRMHD equations in the eXtended Conformally Flat Condition: the XNS code for magnetized neutron stars}",
    eprint = "1401.3101",
    archivePrefix = "arXiv",
    primaryClass = "astro-ph.HE",
    journal = "ASP Conf. Ser.",
    volume = "488",
    pages = "211--216",
    year = "2014"
}

@ARTICLE{Burns2020,
       author = {{Burns}, Eric},
        title = "{Neutron star mergers and how to study them}",
      journal = {Living Reviews in Relativity},
         year = 2020,
        month = dec,
       volume = {23},
       number = {1},
          eid = {4},
        pages = {4},
          doi = {10.1007/s41114-020-00028-7},
archivePrefix = {arXiv},
       eprint = {1909.06085},
 primaryClass = {astro-ph.HE},
       adsurl = {https://ui.adsabs.harvard.edu/abs/2020LRR....23....4B}
}

@ARTICLE{Burrows2006b,
   author = {{Burrows}, A. and {Reddy}, S. and {Thompson}, T.~A.},
    title = "{Neutrino opacities in nuclear matter}",
  journal = {Nuclear Physics A},
   eprint = {astro-ph/0404432},
     year = 2006,
    month = oct,
   volume = 777,
    pages = {356-394},
      doi = {10.1016/j.nuclphysa.2004.06.012},
   adsurl = {http://adsabs.harvard.edu/abs/2006NuPhA.777..356B}
}

@article{Capozzi2020,
    author = "Capozzi, Francesco and Abbar, Sajad and Bollig, Robert and Janka, H. -Thomas",
    title = "{Fast neutrino flavor conversions in one-dimensional core-collapse supernova models with and without muon creation}",
    eprint = "2012.08525",
    archivePrefix = "arXiv",
    primaryClass = "astro-ph.HE",
    reportNumber = "MPP-2020-226",
    doi = "10.1103/PhysRevD.103.063013",
    journal = "Phys. Rev. D",
    volume = "103",
    number = "6",
    pages = "063013",
    year = "2021"
}

@article{Cavaglieri2015,
  title={Low-storage implicit/explicit Runge--Kutta schemes for the simulation of stiff high-dimensional ODE systems},
  author={Cavaglieri, Daniele and Bewley, Thomas},
  journal={Journal of Computational Physics},
  volume={286},
  pages={172--193},
  year={2015},
  publisher={Elsevier}
}

@article{Chakraborty2016,
    author = "Chakraborty, Sovan and Hansen, Rasmus Sloth and Izaguirre, Ignacio and Raffelt, Georg",
    title = "{Self-induced neutrino flavor conversion without flavor mixing}",
    eprint = "1602.00698",
    archivePrefix = "arXiv",
    primaryClass = "hep-ph",
    doi = "10.1088/1475-7516/2016/03/042",
    journal = "JCAP",
    volume = "03",
    pages = "042",
    year = "2016"
}

@article{Cheong2021,
    author = {Cheong, Patrick Chi-Kit and Lam, Alan Tsz-Lok and Ng, Harry Ho-Yin and Li, Tjonnie Guang Feng},
    title = "{Gmunu: paralleled, grid-adaptive, general-relativistic magnetohydrodynamics in curvilinear geometries in dynamical space-times}",
    journal = {Monthly Notices of the Royal Astronomical Society},
    volume = {508},
    number = {2},
    pages = {2279-2301},
    year = {2021},
    month = {09},
    issn = {0035-8711},
    doi = {10.1093/mnras/stab2606},
    url = {https://doi.org/10.1093/mnras/stab2606},
    eprint = {https://academic.oup.com/mnras/article-pdf/508/2/2279/40566816/stab2606.pdf},
}

@article{Cheong2023,
doi = {10.3847/1538-4365/acd931},
url = {https://dx.doi.org/10.3847/1538-4365/acd931},
year = {2023},
month = {aug},
publisher = {The American Astronomical Society},
volume = {267},
number = {2},
pages = {38},
author = {Patrick Chi-Kit Cheong and Harry Ho-Yin Ng and Alan Tsz-Lok Lam and Tjonnie Guang Feng Li},
title = {General-relativistic Radiation Transport Scheme in Gmunu. I. Implementation of Two-moment-based Multifrequency Radiative Transfer and Code Tests},
journal = {The Astrophysical Journal Supplement Series},
}

@article{Cheong2024c,
  title = {Influence of neutrino-electron scattering and neutrino-pair annihilation on hypermassive neutron star},
  author = {Cheong, Patrick Chi-Kit and Foucart, Francois and Ng, Harry Ho-Yin and Offermans, Arthur and Duez, Matthew D. and Muhammed, Nishad and Chawhan, Pavan},
  journal = {Phys. Rev. D},
  volume = {111},
  issue = {4},
  pages = {043036},
  numpages = {17},
  year = {2025},
  month = {Feb},
  publisher = {American Physical Society},
  doi = {10.1103/PhysRevD.111.043036},
  url = {https://link.aps.org/doi/10.1103/PhysRevD.111.043036}
}

@article{Colella84,
	Author = {Colella, Phillip and Woodward, Paul R.},
	Doi = {DOI: 10.1016/0021-9991(84)90143-8},
	Issn = {0021-9991},
	Journal = {Journal of Computational Physics},
	Number = {1},
	Pages = {174-201},
	Title = {The Piecewise Parabolic Method (PPM) for gas-dynamical simulations},
	Url = {http://www.sciencedirect.com/science/article/B6WHY-4DD1PHM-SJ/2/13d69a59afba3d6a5d6bbf1144d860aa},
	Volume = {54},
	Year = {1984}}

@article{Cornelius2025,
    author = "Cornelius, Marie and Tamborra, Irene and Heinlein, Malte and Janka, Hans-Thomas",
    title = "{Diagnosing electron-neutrino lepton number crossings in core-collapse supernovae: A comparison of methods}",
    eprint = "2506.20723",
    archivePrefix = "arXiv",
    primaryClass = "astro-ph.HE",
    doi = "10.1103/gqd7-4ynz",
    journal = "Phys. Rev. D",
    volume = "112",
    number = "6",
    pages = "063004",
    year = "2025"
}

@article{Duan2010,
    author = "Duan, Huaiyu and Fuller, George M. and Qian, Yong-Zhong",
    title = "{Collective Neutrino Oscillations}",
    eprint = "1001.2799",
    archivePrefix = "arXiv",
    primaryClass = "hep-ph",
    reportNumber = "LA-UR-09-08309, INT-PUB-10-001",
    doi = "10.1146/annurev.nucl.012809.104524",
    journal = "Ann. Rev. Nucl. Part. Sci.",
    volume = "60",
    pages = "569--594",
    year = "2010"
}

@article{Ehring2023,
    author = "Ehring, Jakob and Abbar, Sajad and Janka, Hans-Thomas and Raffelt, Georg and Tamborra, Irene",
    title = "{Fast Neutrino Flavor Conversions Can Help and Hinder Neutrino-Driven Explosions}",
    eprint = "2305.11207",
    archivePrefix = "arXiv",
    primaryClass = "astro-ph.HE",
    doi = "10.1103/PhysRevLett.131.061401",
    journal = "Phys. Rev. Lett.",
    volume = "131",
    number = "6",
    pages = "061401",
    year = "2023"
}

@article{Eichler89,
	Adsurl = {http://adsabs.harvard.edu/abs/1989Natur.340..126E},
	Author = {{Eichler}, D. and {Livio}, M. and {Piran}, T. and {Schramm}, D.~N.},
	Doi = {10.1038/340126a0},
	Journal = {Nature},
	Month = jul,
	Pages = {126-128},
	Title = {{Nucleosynthesis, neutrino bursts and gamma-rays from coalescing neutron stars}},
	Volume = 340,
	Year = 1989}

@article{Endeve2026,
    author = "Endeve, Eirik and others",
    title = "{thornado+FLASH-X: A Hybrid Discontinuous Galerkin{\textendash}Implicit-explicit and Finite-volume Framework for Neutrino-radiation Hydrodynamics in Core-collapse Supernovae*}",
    eprint = "2601.00976",
    archivePrefix = "arXiv",
    primaryClass = "astro-ph.IM",
    doi = "10.3847/1538-4365/ae57ad",
    journal = "Astrophys. J. Suppl.",
    volume = "284",
    number = "2",
    pages = "40",
    year = "2026"
}

@article{Espino2024b,
       author = {{Espino}, Pedro Luis and {Hammond}, Peter and {Radice}, David and {Bernuzzi}, Sebastiano and {Gamba}, Rossella and {Zappa}, Francesco and {Micchi}, Lu{\'\i}s Felipe Longo and {Perego}, Albino},
        title = "{Neutrino Trapping and Out-of-Equilibrium Effects in Binary Neutron-Star Merger Remnants}",
      journal = {Phys. Rev. Lett.},
         year = 2024,
        month = may,
       volume = {132},
       number = {21},
          eid = {211001},
        pages = {211001},
          doi = {10.1103/PhysRevLett.132.211001},
archivePrefix = {arXiv},
       eprint = {2311.00031},
 primaryClass = {astro-ph.HE},
       adsurl = {https://ui.adsabs.harvard.edu/abs/2024PhRvL.132u1001E}
}

@ARTICLE{FermiLat2017,
   author = {{Fermi-LAT Collaboration}},
    title = "{Fermi-LAT observations of the LIGO/Virgo event GW170817}",
  journal = {ArXiv e-prints},
archivePrefix = "arXiv",
   eprint = {1710.05450},
 primaryClass = "astro-ph.HE",
     year = 2017,
    month = oct,
   adsurl = {http://adsabs.harvard.edu/abs/2017arXiv171005450F}
}

@article{Fernandez2022,
    author = "Fern{\'a}ndez, Rodrigo and Richers, Sherwood and Mulyk, Nicole and Fahlman, Steven",
    title = "{Fast flavor instability in hypermassive neutron star disk outflows}",
    eprint = "2207.10680",
    archivePrefix = "arXiv",
    primaryClass = "astro-ph.HE",
    doi = "10.1103/PhysRevD.106.103003",
    journal = "Phys. Rev. D",
    volume = "106",
    number = "10",
    pages = "103003",
    year = "2022"
}

@article{Fiorillo2024,
    author = "Fiorillo, Damiano F. G. and Raffelt, Georg G.",
    title = "{Theory of neutrino fast flavor evolution. Part I. Linear response theory and stability conditions.}",
    eprint = "2406.06708",
    archivePrefix = "arXiv",
    primaryClass = "hep-ph",
    doi = "10.1007/JHEP08(2024)225",
    journal = "JHEP",
    volume = "08",
    pages = "225",
    year = "2024"
}

@article{Fiorillo2024b,
    author = "Fiorillo, Damiano F. G. and Raffelt, Georg G.",
    title = "{Theory of neutrino fast flavor evolution. Part II. Solutions at the edge of instability}",
    eprint = "2409.17232",
    archivePrefix = "arXiv",
    primaryClass = "hep-ph",
    doi = "10.1007/JHEP12(2024)205",
    journal = "JHEP",
    volume = "12",
    pages = "205",
    year = "2024"
}

@article{Fischer2020b,
  title={Muonization of supernova matter},
  author={Fischer, Tobias and Guo, Gang and Mart{\'\i}nez-Pinedo, Gabriel and Liebend{\"o}rfer, Matthias and Mezzacappa, Anthony},
  journal={Physical Review D},
  volume={102},
  number={12},
  pages={123001},
  year={2020},
  publisher={APS}
}

@article{Foucart2015a,
	Adsurl = {http://adsabs.harvard.edu/abs/2015PhRvD..91l4021F},
	Archiveprefix = {arXiv},
	Author = {{Foucart}, F. and {O'Connor}, E. and {Roberts}, L. and {Duez}, M.~D. and {Haas}, R. and {Kidder}, L.~E. and {Ott}, C.~D. and {Pfeiffer}, H.~P. and {Scheel}, M.~A. and {Szilagyi}, B.},
	Doi = {10.1103/PhysRevD.91.124021},
	Eid = {124021},
	Eprint = {1502.04146},
	Journal = {Phys. Rev. D},
	Month = jun,
	Number = 12,
	Pages = {124021},
	Primaryclass = {astro-ph.HE},
	Title = {{Post-merger evolution of a neutron star-black hole binary with neutrino transport}},
	Volume = 91,
	Year = 2015}

@article{Foucart2023,
    author = "Foucart, Francois",
    title = "{Neutrino Transport in General Relativistic Neutron Star Merger Simulations}",
    eprint = "2209.02538",
    archivePrefix = "arXiv",
    primaryClass = "astro-ph.HE",
    doi = "10.1007/s41115-023-00016-y",
    journal = "Living Rev. Comput. Astrophys.",
    volume = "9",
    number = "1",
    pages = "1",
    year = "2023"
}

@article{Foucart2026,
    journal = {},

    author = "Foucart, Francois and Rath, Samantha and Davidson, Rowan and Cheong, Patrick Chi-Kit and Duez, Matthew D. and Kidder, Lawrence and Pfeiffer, Harald and Scheel, Mark",
    title = "{Impact of neutrino-electron scattering and an improved treatment of pair processes on binary neutron star mergers}",
    eprint = "2606.27425",
    archivePrefix = "arXiv",
    primaryClass = "astro-ph.HE",
    doi = "10.48550/arXiv.2606.27425",
    year = "2026"
}

@article{Froustey2023,
    author = "Froustey, Julien and Richers, Sherwood and Grohs, Evan and Flynn, Samuel D. and Foucart, Francois and Kneller, James P. and McLaughlin, Gail C.",
    title = "{Neutrino fast flavor oscillations with moments: Linear stability analysis and application to neutron star mergers}",
    eprint = "2311.11968",
    archivePrefix = "arXiv",
    primaryClass = "astro-ph.HE",
    reportNumber = "N3AS-23-032",
    doi = "10.1103/PhysRevD.109.043046",
    journal = "Phys. Rev. D",
    volume = "109",
    number = "4",
    pages = "043046",
    year = "2024"
}

@article{Froustey2025,
    author = "Froustey, Julien",
    title = "{Predicting the outcome of collisional neutrino flavor conversion}",
    eprint = "2505.16961",
    archivePrefix = "arXiv",
    primaryClass = "hep-ph",
    reportNumber = "N3AS-25-008",
    doi = "10.1103/syxg-sqfn",
    journal = "Phys. Rev. D",
    volume = "112",
    number = "2",
    pages = "023029",
    year = "2025"
}

@article{Froustey2026,
    author = "Froustey, Julien and Foucart, Francois and Hall, Christian and Kneller, James P. and Kundu, Debraj and Lin, Zidu and McLaughlin, Gail C. and Richers, Sherwood",
    title = "{Neutrino flavor instabilities in neutron star mergers with moment transport: Slow, fast, and collisional modes}",
    eprint = "2601.02461",
    archivePrefix = "arXiv",
    primaryClass = "astro-ph.HE",
    reportNumber = "N3AS-25-020",
    doi = "10.1103/htq2-d1t7",
    journal = "Phys. Rev. D",
    volume = "113",
    number = "6",
    pages = "063050",
    year = "2026"
}

@article{George2020,
	title={Fast neutrino flavor conversion, ejecta properties, and nucleosynthesis in newly-formed hypermassive remnants of neutron-star mergers},
	volume={102},
	ISSN={2470-0029},
	url={http://dx.doi.org/10.1103/PhysRevD.102.103015},
	DOI={10.1103/physrevd.102.103015},
	number={10},
	journal={Physical Review D},
	publisher={American Physical Society (APS)},
	author={George, Manu and Wu, Meng-Ru and Tamborra, Irene and Ardevol-Pulpillo, Ricard and Janka, Hans-Thomas},
	year={2020},
	month={Nov}
}

@article{Gieg2024,
    author = "Gieg, Henrique and Schianchi, Federico and Ujevic, Maximiliano and Dietrich, Tim",
    title = "{On the Role of Muons in Binary Neutron Star Mergers: First Simulations}",
    eprint = "2409.04420",
    journal = {arXiv e-prints},
    archivePrefix = "arXiv",
    primaryClass = "gr-qc",
    month = "9",
    year = "2024"
}

@ARTICLE{Gieg2026,
       author = {{Gieg}, Henrique and {Jaeger}, Ramon and {Ujevic}, Maximiliano and {Dietrich}, Tim},
        title = "{Consistent Treatment of Muons in Binary Neutron Star Mergers}",
      journal = {arXiv e-prints},
         year = 2026,
        month = apr,
          eid = {arXiv:2604.14225},
        pages = {arXiv:2604.14225},
          doi = {10.48550/arXiv.2604.14225},
archivePrefix = {arXiv},
       eprint = {2604.14225},
 primaryClass = {astro-ph.HE},
       adsurl = {https://ui.adsabs.harvard.edu/abs/2026arXiv260414225G}
}

@article{Glas2019,
    author = "Glas, Robert and Janka, H. -Thomas and Capozzi, Francesco and Sen, Manibrata and Dasgupta, Basudeb and Mirizzi, Alessandro and Sigl, Guenter",
    title = "{Fast Neutrino Flavor Instability in the Neutron-star Convection Layer of Three-dimensional Supernova Models}",
    eprint = "1912.00274",
    archivePrefix = "arXiv",
    primaryClass = "astro-ph.HE",
    reportNumber = "MPP-2019-239",
    doi = "10.1103/PhysRevD.101.063001",
    journal = "Phys. Rev. D",
    volume = "101",
    number = "6",
    pages = "063001",
    year = "2020"
}

@article{Grohs2022,
    author = "Grohs, Evan and Richers, Sherwood and Couch, Sean M. and Foucart, Francois and Kneller, James P. and McLaughlin, G. C.",
    title = "{Neutrino fast flavor instability in three dimensions for a neutron star merger}",
    eprint = "2207.02214",
    archivePrefix = "arXiv",
    primaryClass = "hep-ph",
    doi = "10.1016/j.physletb.2023.138210",
    journal = "Phys. Lett. B",
    volume = "846",
    pages = "138210",
    year = "2023"
}

@article{Guo2019,
    author = "Guo, Gang and Mart{\'\i}nez-Pinedo, Gabriel",
    title = "{Chiral effective field theory description of neutrino nucleon-nucleon Bremsstrahlung in supernova matter}",
    eprint = "1905.13634",
    archivePrefix = "arXiv",
    primaryClass = "astro-ph.HE",
    doi = "10.3847/1538-4357/ab536d",
    journal = "Astrophys. J.",
    volume = "887",
    pages = "58",
    year = "2019"
}

@article{Guo2020,
  title={Charged-current muonic reactions in core-collapse supernovae},
  author={Guo, Gang and Mart{\'\i}nez-Pinedo, Gabriel and Lohs, Andreas and Fischer, Tobias},
  journal={Physical Review D},
  volume={102},
  number={2},
  pages={023037},
  year={2020},
  publisher={APS}
}

@article{Hannestad1997,
    author = "Hannestad, Steen and Raffelt, Georg",
    title = "{Supernova neutrino opacity from nucleon-nucleon Bremsstrahlung and related processes}",
    eprint = "astro-ph/9711132",
    archivePrefix = "arXiv",
    doi = "10.1086/306303",
    journal = "Astrophys. J.",
    volume = "507",
    pages = "339--352",
    year = "1998"
}

@article{Harten83,
	Author = {A. Harten and P. D. Lax and B. van Leer},
	Doi = {10.1137/1025002},
	Journal = {SIAM Rev.},
	Key = {Harten83},
	Pages = {35},
	Title = {On Upstream Differencing and Godunov-Type Schemes for Hyperbolic Conservation Laws},
	Volume = {25},
	Year = {1983}}

@article{Hernandez2026,
    journal = {},

    author = "Hern{\'a}ndez, Jos{\'e} Luis and Manuel, Cristina and Tolos, Laura",
    title = "{Impact of muons on the bulk viscosity of neutron star matter metamodels}",
    eprint = "2603.23904",
    archivePrefix = "arXiv",
    primaryClass = "hep-ph",
    month = "3",
    year = "2026"
}

@article{Horowitz2002,
  title={Weak magnetism for antineutrinos in supernovae},
  author={Horowitz, CJ},
  journal={Physical Review D},
  volume={65},
  number={4},
  pages={043001},
  year={2002},
  publisher={APS}
}

@article{Horowitz2016,
    author = "Horowitz, C. J. and Caballero, O. L. and Lin, Zidu and O'Connor, Evan and Schwenk, A.",
    title = "{Neutrino-nucleon scattering in supernova matter from the virial expansion}",
    eprint = "1611.05140",
    archivePrefix = "arXiv",
    primaryClass = "nucl-th",
    doi = "10.1103/PhysRevC.95.025801",
    journal = "Phys. Rev. C",
    volume = "95",
    number = "2",
    pages = "025801",
    year = "2017"
}

@article{Hotokezaka2013,
	Adsurl = {http://adsabs.harvard.edu/abs/2013PhRvD..87b4001H},
	Archiveprefix = {arXiv},
	Author = {{Hotokezaka}, K. and {Kiuchi}, K. and {Kyutoku}, K. and {Okawa}, H. and {Sekiguchi}, Y.-i. and {Shibata}, M. and {Taniguchi}, K.},
	Doi = {10.1103/PhysRevD.87.024001},
	Eid = {024001},
	Eprint = {1212.0905},
	Journal = {Phys. Rev. D},
	Month = jan,
	Number = 2,
	Pages = {024001},
	Primaryclass = {astro-ph.HE},
	Title = {{Mass ejection from the merger of binary neutron stars}},
	Volume = 87,
	Year = 2013}

@article{Izaguirre2017,
  title = {Fast Pairwise Conversion of Supernova Neutrinos: A Dispersion Relation Approach},
  author = {Izaguirre, Ignacio and Raffelt, Georg and Tamborra, Irene},
  journal = {Phys. Rev. Lett.},
  volume = {118},
  issue = {2},
  pages = {021101},
  numpages = {6},
  year = {2017},
  month = {Jan},
  publisher = {American Physical Society},
  doi = {10.1103/PhysRevLett.118.021101},
  url = {https://link.aps.org/doi/10.1103/PhysRevLett.118.021101}
}

@article{Johns2021,
    author = "Johns, Lucas",
    title = "{Collisional Flavor Instabilities of Supernova Neutrinos}",
    eprint = "2104.11369",
    archivePrefix = "arXiv",
    primaryClass = "hep-ph",
    doi = "10.1103/PhysRevLett.130.191001",
    journal = "Phys. Rev. Lett.",
    volume = "130",
    number = "19",
    pages = "191001",
    year = "2023"
}

@article{Johns2022,
    author = "Johns, Lucas and Xiong, Zewei",
    title = "{Collisional instabilities of neutrinos and their interplay with fast flavor conversion in compact objects}",
    eprint = "2208.11059",
    archivePrefix = "arXiv",
    primaryClass = "hep-ph",
    doi = "10.1103/PhysRevD.106.103029",
    journal = "Phys. Rev. D",
    volume = "106",
    number = "10",
    pages = "103029",
    year = "2022"
}

@article{Johns2025,
    author = "Johns, Lucas and Richers, Sherwood and Wu, Meng-Ru",
    title = "{Neutrino Oscillations in Core-Collapse Supernovae and Neutron Star Mergers}",
    eprint = "2503.05959",
    archivePrefix = "arXiv",
    primaryClass = "astro-ph.HE",
    reportNumber = "LA-UR-25-21809",
    doi = "10.1146/annurev-nucl-121423-100853",
    journal = "Ann. Rev. Nucl. Part. Sci.",
    volume = "75",
    number = "1",
    pages = "399--423",
    year = "2025"
}

@article{Just2022,
    author = "Just, Oliver and Abbar, Sajad and Wu, Meng-Ru and Tamborra, Irene and Janka, Hans-Thomas and Capozzi, Francesco",
    title = "{Fast neutrino conversion in hydrodynamic simulations of neutrino-cooled accretion disks}",
    eprint = "2203.16559",
    archivePrefix = "arXiv",
    primaryClass = "astro-ph.HE",
    doi = "10.1103/PhysRevD.105.083024",
    journal = "Phys. Rev. D",
    volume = "105",
    number = "8",
    pages = "083024",
    year = "2022"
}

@ARTICLE{Kasen2017,
   author = {{Kasen}, D. and {Metzger}, B. and {Barnes}, J. and {Quataert}, E. and
	{Ramirez-Ruiz}, E.},
    title = "{Origin of the heavy elements in binary neutron-star mergers from a gravitational-wave event}",
  journal = {Nature},
archivePrefix = "arXiv",
   eprint = {1710.05463},
 primaryClass = "astro-ph.HE",
     year = 2017,
    month = nov,
   volume = 551,
    pages = {80-84},
      doi = {10.1038/nature24453},
   adsurl = {http://adsabs.harvard.edu/abs/2017Natur.551...80K}}

@article{Kato2023,
    author = "Kato, Chinami and Nagakura, Hiroki and Johns, Lucas",
    title = "{Collisional flavor swap with neutrino self-interactions}",
    eprint = "2309.02619",
    archivePrefix = "arXiv",
    primaryClass = "astro-ph.HE",
    doi = "10.1103/PhysRevD.109.103009",
    journal = "Phys. Rev. D",
    volume = "109",
    number = "10",
    pages = "103009",
    year = "2024"
}

@article{Kawaguchi2023b,
  title={Monte Carlo--based relativistic radiation hydrodynamics code with a higher-order scheme},
  author={Kawaguchi, Kyohei and Fujibayashi, Sho and Shibata, Masaru},
  journal={Physical Review D},
  volume={107},
  number={2},
  pages={023026},
  year={2023},
  publisher={APS}
}

@article{Kawaguchi2025,
    author = "Kawaguchi, Kyohei and Fujibayashi, Sho and Shibata, Masaru",
    title = "{Neutrino pair annihilation driven jets from black-hole torus systems}",
    eprint = "2506.01679",
    archivePrefix = "arXiv",
    primaryClass = "astro-ph.HE",
    doi = "10.1103/yygv-6268",
    journal = "Phys. Rev. D",
    volume = "112",
    number = "4",
    pages = "043001",
    year = "2025"
}

@article{Kiuchi2024,
  title={General relativistic magnetohydodynamics simulations for binary neutron star mergers},
  author={Kiuchi, Kenta},
  journal={arXiv preprint arXiv:2405.10081},
  year={2024}
}

@article{Kneller2024,
    author = "Kneller, James P. and Froustey, Julien and Grohs, Evan B. and Foucart, Francois and McLaughlin, Gail C. and Richers, Sherwood",
    title = "{Quantum closures for neutrino moment transport}",
    eprint = "2410.00719",
    archivePrefix = "arXiv",
    primaryClass = "hep-ph",
    reportNumber = "N3AS-24-035",
    doi = "10.1103/PhysRevD.111.063046",
    journal = "Phys. Rev. D",
    volume = "111",
    number = "6",
    pages = "063046",
    year = "2025"
}

@article{Kost2024,
    author = "Kost, Anson and Johns, Lucas and Duan, Huaiyu",
    title = "{Once-in-a-lifetime encounter models for neutrino media: From coherent oscillations to flavor equilibration}",
    eprint = "2402.05022",
    archivePrefix = "arXiv",
    primaryClass = "hep-ph",
    reportNumber = "INT-PUB-24-006, LA-UR-24-21029",
    doi = "10.1103/PhysRevD.109.103037",
    journal = "Phys. Rev. D",
    volume = "109",
    number = "10",
    pages = "103037",
    year = "2024"
}

@ARTICLE{Kuroda2016,
       author = {{Kuroda}, Takami and {Takiwaki}, Tomoya and {Kotake}, Kei},
        title = "{A New Multi-energy Neutrino Radiation-Hydrodynamics Code in Full General Relativity and Its Application to the Gravitational Collapse of Massive Stars}",
      journal = {Astrophys. J., Supp.},
         year = "2016",
        month = feb,
       volume = {222},
       number = {2},
          eid = {20},
        pages = {20},
          doi = {10.3847/0067-0049/222/2/20},
archivePrefix = {arXiv},
       eprint = {1501.06330},
 primaryClass = {astro-ph.HE},
       adsurl = {https://ui.adsabs.harvard.edu/abs/2016ApJS..222...20K}
}

@article{Laiu2021,
    author = "Laiu, M. Paul and Endeve, Eirik and Chu, Ran and Harris, J. Austin and Messer, O. E. Bronson",
    title = "{A DG-IMEX Method for Two-moment Neutrino Transport: Nonlinear Solvers for Neutrino{\textendash}Matter Coupling}",
    eprint = "2102.02186",
    archivePrefix = "arXiv",
    primaryClass = "astro-ph.HE",
    doi = "10.3847/1538-4365/abe2a8",
    journal = "Astrophys. J. Suppl.",
    volume = "253",
    number = "2",
    pages = "52",
    year = "2021"
}

@article{Laraib2025,
    author = "Laraib, Zoha and Richers, Sherwood",
    title = "{Many-body simulations of the fast flavor instability}",
    eprint = "2507.02040",
    archivePrefix = "arXiv",
    primaryClass = "astro-ph.HE",
    doi = "10.1103/d6w8-7j9s",
    journal = "Phys. Rev. D",
    volume = "112",
    number = "10",
    pages = "L101304",
    year = "2025"
}

@article{Lattimer74,
	Adsurl = {http://adsabs.harvard.edu/abs/1974ApJ...192L.145L},
	Author = {{Lattimer}, J.~M. and {Schramm}, D.~N.},
	Doi = {10.1086/181612},
	Journal = {Astrophys. J. Lett.},
	Month = sep,
	Pages = {L145-L147},
	Title = {{Black-hole-neutron-star collisions}},
	Volume = 192,
	Year = 1974}

@article{Li2021prl,
    author = "Li, Xinyu and Siegel, Daniel M.",
    title = "{Neutrino Fast Flavor Conversions in Neutron-Star Postmerger Accretion Disks}",
    eprint = "2103.02616",
    archivePrefix = "arXiv",
    primaryClass = "astro-ph.HE",
    doi = "10.1103/PhysRevLett.126.251101",
    journal = "Phys. Rev. Lett.",
    volume = "126",
    number = "25",
    pages = "251101",
    year = "2021"
}

@article{Lin2023,
  title = {Collision-induced flavor instability in dense neutrino gases with energy-dependent scattering},
  author = {Lin, Yu-Chia and Duan, Huaiyu},
  journal = {Phys. Rev. D},
  volume = {107},
  issue = {8},
  pages = {083034},
  numpages = {7},
  year = {2023},
  month = {Apr},
  publisher = {American Physical Society},
  doi = {10.1103/PhysRevD.107.083034},
  url = {https://link.aps.org/doi/10.1103/PhysRevD.107.083034}
}

@article{Liu2023,
  title = {Systematic study of the resonancelike structure in the collisional flavor instability of neutrinos},
  author = {Liu, Jiabao and Zaizen, Masamichi and Yamada, Shoichi},
  journal = {Phys. Rev. D},
  volume = {107},
  issue = {12},
  pages = {123011},
  numpages = {21},
  year = {2023},
  month = {Jun},
  publisher = {American Physical Society},
  doi = {10.1103/PhysRevD.107.123011},
  url = {https://link.aps.org/doi/10.1103/PhysRevD.107.123011}
}

@article{Liu2024,
    author = "Liu, Jiabao and Nagakura, Hiroki and Akaho, Ryuichiro and Ito, Akira and Zaizen, Masamichi and Furusawa, Shun and Yamada, Shoichi",
    title = "{Muon-induced collisional flavor instability in core-collapse supernova}",
    eprint = "2407.10604",
    archivePrefix = "arXiv",
    primaryClass = "hep-ph",
    doi = "10.1103/PhysRevD.110.043039",
    journal = "Phys. Rev. D",
    volume = "110",
    number = "4",
    pages = "043039",
    year = "2024"
}

@article{Liu2026,
    author = "Liu, Jiabao and Nagakura, Hiroki",
    title = "{Approximate energy-integration method for identifying collisional neutrino flavor instabilities}",
    eprint = "2604.01096",
    archivePrefix = "arXiv",
    primaryClass = "astro-ph.HE",
    doi = "10.1103/559p-9d8p",
    journal = "Phys. Rev. D",
    volume = "113",
    number = "12",
    pages = "123047",
    year = "2026"
}

@ARTICLE{Loffredo2023,
       author = {{Loffredo}, Eleonora and {Perego}, Albino and {Logoteta}, Domenico and {Branchesi}, Marica},
        title = "{Muons in the aftermath of neutron star mergers and their impact on trapped neutrinos}",
      journal = {Astronomy and Astrophysics},
         year = 2023,
        month = apr,
       volume = {672},
          eid = {A124},
        pages = {A124},
          doi = {10.1051/0004-6361/202244927},
archivePrefix = {arXiv},
       eprint = {2209.04458},
 primaryClass = {astro-ph.HE},
       adsurl = {https://ui.adsabs.harvard.edu/abs/2023A&A...672A.124L}
}

@article{Lund2025,
    author = "Lund, Kelsey A. and Mukhopadhyay, Payel and Miller, Jonah M. and McLaughlin, G. C.",
    title = "{Angle-dependent in Situ Fast Flavor Transformations in Post-neutron-star-merger Disks}",
    eprint = "2503.23727",
    archivePrefix = "arXiv",
    primaryClass = "astro-ph.HE",
    reportNumber = "LA-UR-25-23002, INT-PUB-25-005, N3AS-25-005",
    doi = "10.3847/2041-8213/add0a7",
    journal = "Astrophys. J. Lett.",
    volume = "985",
    number = "1",
    pages = "L9",
    year = "2025"
}

@article{Metzger:2010,
	Adsurl = {http://adsabs.harvard.edu/abs/2010MNRAS.406.2650M},
	Archiveprefix = {arXiv},
	Author = {{Metzger}, B.~D. and {Mart{\'{\i}}nez-Pinedo}, G. and {Darbha}, S. and {Quataert}, E. and {Arcones}, A. and {Kasen}, D. and {Thomas}, R. and {Nugent}, P. and {Panov}, I.~V. and {Zinner}, N.~T.},
	Doi = {10.1111/j.1365-2966.2010.16864.x},
	Eprint = {1001.5029},
	Journal = {Mon. Not. R. Astron. Soc.},
	Month = aug,
	Pages = {2650-2662},
	Primaryclass = {astro-ph.HE},
	Title = {{Electromagnetic counterparts of compact object mergers powered by the radioactive decay of r-process nuclei}},
	Volume = 406,
	Year = 2010}

@ARTICLE{Metzger2017,
   author = {{Metzger}, B.~D.},
    title = "{Kilonovae}",
  journal = {Living Reviews in Relativity},
archivePrefix = "arXiv",
   eprint = {1610.09381},
 primaryClass = "astro-ph.HE",
     year = 2017,
    month = may,
   volume = 20,
      eid = {3},
    pages = {3},
      doi = {10.1007/s41114-017-0006-z},
   adsurl = {http://adsabs.harvard.edu/abs/2017LRR....20....3M}
}

@ARTICLE{Mezzacappa1993b,
       author = {{Mezzacappa}, Anthony and {Bruenn}, Stephen W.},
        title = "{A Numerical Method for Solving the Neutrino Boltzmann Equation Coupled to Spherically Symmetric Stellar Core Collapse}",
      journal = {Astrophys. J.},
         year = 1993,
        month = mar,
       volume = {405},
        pages = {669},
          doi = {10.1086/172395},
       adsurl = {https://ui.adsabs.harvard.edu/abs/1993ApJ...405..669M}
}

@inproceedings{Mezzacappa2026,
    author = "Mezzacappa, Anthony",
    title = "{Core Collapse Supernova Modeling: The Next Ten Years}",
    eprint = "2604.24970",
    archivePrefix = "arXiv",
    primaryClass = "astro-ph.HE",
    month = "4",
    year = "2026"
}

@ARTICLE{Miller2019a,
       author = {{Miller}, Jonah M. and {Ryan}, Ben. R. and {Dolence}, Joshua C.},
        title = "{{\ensuremath{\nu}} bhlight: Radiation GRMHD for Neutrino-driven Accretion Flows}",
      journal = {Astrophys. J., Supp.},
         year = "2019",
        month = apr,
       volume = {241},
       number = {2},
          eid = {30},
        pages = {30},
          doi = {10.3847/1538-4365/ab09fc},
archivePrefix = {arXiv},
       eprint = {1903.09273},
 primaryClass = {astro-ph.IM},
       adsurl = {https://ui.adsabs.harvard.edu/abs/2019ApJS..241...30M}
}

@article{Morinaga2019,
    author = "Morinaga, Taiki and Nagakura, Hiroki and Kato, Chinami and Yamada, Shoichi",
    title = "{Fast neutrino-flavor conversion in the preshock region of core-collapse supernovae}",
    eprint = "1909.13131",
    archivePrefix = "arXiv",
    primaryClass = "astro-ph.HE",
    doi = "10.1103/PhysRevResearch.2.012046",
    journal = "Phys. Rev. Res.",
    volume = "2",
    number = "1",
    pages = "012046",
    year = "2020"
}

@ARTICLE{Most2019b,
       author = {{Most}, E.~R. and {Papenfort}, L. Jens and {Rezzolla}, L.},
        title = "{Beyond second-order convergence in simulations of magnetized binary neutron stars with realistic microphysics}",
      journal = {Mon. Not. R. Astron. Soc.},
         year = 2019,
        month = dec,
       volume = {490},
       number = {3},
        pages = {3588-3600},
          doi = {10.1093/mnras/stz2809},
archivePrefix = {arXiv},
       eprint = {1907.10328},
 primaryClass = {astro-ph.HE},
       adsurl = {https://ui.adsabs.harvard.edu/abs/2019MNRAS.490.3588M}
}

@article{MuLan2011,
    author = "Webber, D. M. and others",
    collaboration = "MuLan",
    title = "{Measurement of the Positive Muon Lifetime and Determination of the Fermi Constant to Part-per-Million Precision}",
    eprint = "1010.0991",
    archivePrefix = "arXiv",
    primaryClass = "hep-ex",
    doi = "10.1103/PhysRevLett.106.079901",
    journal = "Phys. Rev. Lett.",
    volume = "106",
    pages = "041803",
    year = "2011"
}

@article{Musolino2023,
       author = {{Musolino}, Carlo and {Rezzolla}, Luciano},
        title = "{A practical guide to a moment approach for neutrino transport in numerical relativity}",
      journal = {Mon. Not. R. Astron. Soc.},
         year = 2024,
        month = mar,
       volume = {528},
       number = {4},
        pages = {5952-5971},
          doi = {10.1093/mnras/stae224},
archivePrefix = {arXiv},
       eprint = {2304.09168},
 primaryClass = {gr-qc},
       adsurl = {https://ui.adsabs.harvard.edu/abs/2024MNRAS.528.5952M}
}

@article{Nagakura2023c,
    author = "Nagakura, Hiroki and Johns, Lucas and Zaizen, Masamichi",
    title = "{Bhatnagar-Gross-Krook subgrid model for neutrino quantum kinetics}",
    eprint = "2312.16285",
    archivePrefix = "arXiv",
    primaryClass = "astro-ph.HE",
    reportNumber = "LA-UR-23-34188",
    doi = "10.1103/PhysRevD.109.083013",
    journal = "Phys. Rev. D",
    volume = "109",
    number = "8",
    pages = "083013",
    year = "2024"
}

@ARTICLE{Nagakura2025,
       author = {{Nagakura}, Hiroki and {Sumiyoshi}, Kohsuke and {Fujibayashi}, Sho and {Sekiguchi}, Yuichiro and {Shibata}, Masaru},
        title = "{Neutrino flavor instabilities in a binary neutron star merger remnant: Roles of a long-lived hypermassive neutron star}",
      journal = {arXiv e-prints},
         year = 2025,
        month = apr,
          eid = {arXiv:2504.20143},
        pages = {arXiv:2504.20143},
          doi = {10.48550/arXiv.2504.20143},
archivePrefix = {arXiv},
       eprint = {2504.20143},
 primaryClass = {astro-ph.HE},
       adsurl = {https://ui.adsabs.harvard.edu/abs/2025arXiv250420143N}
}

@article{Ng2024a,
       author = {{Ng}, Harry Ho-Yin and {Cheong}, Patrick Chi-Kit and {Lam}, Alan Tsz-Lok and {Li}, Tjonnie Guang Feng},
        title = "{General-relativistic Radiation Transport Scheme in Gmunu. II. Implementation of Novel Microphysical Library for Neutrino Radiation{\textemdash}Weakhub}",
      journal = {Astrophys. J., Supp.},
         year = 2024,
        month = may,
       volume = {272},
       number = {1},
          eid = {9},
        pages = {9},
          doi = {10.3847/1538-4365/ad2fbd},
archivePrefix = {arXiv},
       eprint = {2309.03526},
 primaryClass = {astro-ph.HE},
       adsurl = {https://ui.adsabs.harvard.edu/abs/2024ApJS..272....9N}
}

@ARTICLE{Ng2024b,
       author = {{Ng}, Harry Ho-Yin and {Jiang}, Jin-Liang and {Musolino}, Carlo and {Ecker}, Christian and {Tootle}, Samuel D. and {Rezzolla}, Luciano},
        title = "{Hybrid approach to long-term binary neutron-star simulations}",
      journal = {Phys. Rev. D},
         year = 2024,
        month = mar,
       volume = {109},
       number = {6},
          eid = {064061},
        pages = {064061},
          doi = {10.1103/PhysRevD.109.064061},
archivePrefix = {arXiv},
       eprint = {2312.11358},
 primaryClass = {gr-qc},
       adsurl = {https://ui.adsabs.harvard.edu/abs/2024PhRvD.109f4061N}
}

@article{Ng2024c,
    author = "Ng, Harry Ho-Yin and Musolino, Carlo and Tootle, Samuel D. and Rezzolla, Luciano",
    title = "{Accurate Muonic Interactions in Neutron Star Mergers and Impact on Heavy-element Nucleosynthesis}",
    eprint = "2411.19178",
    archivePrefix = "arXiv",
    primaryClass = "astro-ph.HE",
    doi = "10.3847/2041-8213/add324",
    journal = "Astrophys. J. Lett.",
    volume = "985",
    number = "2",
    pages = "L36",
    year = "2025"
}

@article{Ng2026b,
    journal = {},
    author = "Ng, Harry Ho-Yin and Most, Elias R.",
    title = "{Spinning down neutron-star merger remnants with the Tayler-Spruit dynamo: Global simulations reveal the formation of massive disks and neutron-rich ejecta}",
    eprint = "2607.28556",
    archivePrefix = "arXiv",
    primaryClass = "astro-ph.HE",
    month = "7",
    year = "2026"
}

@ARTICLE{Oconnor2015,
       author = {{O'Connor}, Evan},
        title = "{An Open-source Neutrino Radiation Hydrodynamics Code for Core-collapse Supernovae}",
      journal = {Astrophys. J., Supp.},
         year = "2015",
        month = aug,
       volume = {219},
       number = {2},
          eid = {24},
        pages = {24},
          doi = {10.1088/0067-0049/219/2/24},
archivePrefix = {arXiv},
       eprint = {1411.7058},
 primaryClass = {astro-ph.HE},
       adsurl = {https://ui.adsabs.harvard.edu/abs/2015ApJS..219...24O}
}

@article{Offermans2026,
    journal = {},

    author = "Offermans, Arthur and Ng, Harry Ho-Yin and Cheong, Patrick Chi-Kit and Mezzacappa, Anthony and Li, Tjonnie Guang Feng",
    title = "{A Multidimensional General-Relativistic Boltzmann Solver for Neutrino Transport: Implementation, Discretization and Optimization}",
    eprint = "2609.31611",
    archivePrefix = "arXiv",
    primaryClass = "astro-ph.HE",
    month = "9",
    year = "2026"
}

@ARTICLE{Olsen2025,
       author = {{Olsen}, Tom and {Rezzolla}, Luciano},
        title = "{General-relativistic lattice-Boltzmann method for radiation transport}",
      journal = {Mon. Not. R. Astron. Soc.},
         year = 2025,
        month = aug,
       volume = {541},
       number = {2},
        pages = {1305-1328},
          doi = {10.1093/mnras/staf992},
archivePrefix = {arXiv},
       eprint = {2502.17552},
 primaryClass = {gr-qc},
       adsurl = {https://ui.adsabs.harvard.edu/abs/2025MNRAS.541.1305O}
}

@article{Padilla-Gay2025,
    author = "Padilla-Gay, Ian and Chen, Heng-Hao and Abbar, Sajad and Wu, Meng-Ru and Xiong, Zewei",
    title = "{Flavor equilibration of supernova neutrinos: Exploring the dynamics of slow modes}",
    eprint = "2505.11588",
    archivePrefix = "arXiv",
    primaryClass = "astro-ph.HE",
    reportNumber = "SLAC-PUB-250421",
    doi = "10.1103/jg14-8p4l",
    journal = "Phys. Rev. D",
    volume = "112",
    number = "4",
    pages = "043039",
    year = "2025"
}

@article{Pajkos2024,
    author = "Pajkos, Michael A. and Most, Elias R.",
    title = "{The Influence of Muons, Pions, and Trapped Neutrinos on Neutron Star Mergers}",
    journal = {arXiv e-prints},
    eprint = "2409.09147",
    archivePrefix = "arXiv",
    primaryClass = "astro-ph.HE",
    month = "9",
    year = "2024"
}

@article{Pantaleone1992,
    author = "Pantaleone, James T.",
    title = "{Neutrino oscillations at high densities}",
    reportNumber = "DOE-ER-40561-056, INT-92-07-01",
    doi = "10.1016/0370-2693(92)91887-F",
    journal = "Phys. Lett. B",
    volume = "287",
    pages = "128--132",
    year = "1992"
}

@book{Press1992,
        title={Numerical Recipes in Fortran: The Art of Scientific Computing},
        author={Press, WH and Teukolsky, SA and Vetterling, WT and Flannery, BP},
        publisher={Cambridge University Press},
        year={1992}}

@ARTICLE{Qiu2025,
       author = {{Qiu}, Yi and {Radice}, David and {Richers}, Sherwood and {Bhattacharyya}, Maitraya},
        title = "{Neutrino Flavor Transformation in Neutron Star Mergers}",
      journal = {arXiv e-prints},
         year = 2025,
        month = mar,
          eid = {arXiv:2503.11758},
        pages = {arXiv:2503.11758},
          doi = {10.48550/arXiv.2503.11758},
archivePrefix = {arXiv},
       eprint = {2503.11758},
 primaryClass = {astro-ph.HE},
       adsurl = {https://ui.adsabs.harvard.edu/abs/2025arXiv250311758Q}
}

@article{Qiu2025b,
    author = "Qiu, Yi and Radice, David and Richers, Sherwood and Guercilena, Federico Maria and Perego, Albino and Bhattacharyya, Maitraya",
    title = "{Impact of neutrino flavor conversions on neutron star merger dynamics, ejecta, nucleosynthesis, and multimessenger signals}",
    eprint = "2510.15028",
    archivePrefix = "arXiv",
    primaryClass = "astro-ph.HE",
    doi = "10.1103/qckq-78gt",
    journal = "Phys. Rev. D",
    volume = "112",
    number = "12",
    pages = "123039",
    year = "2025"
}

@article{Radice2013,
	Adsurl = {http://adsabs.harvard.edu/abs/2013JCoPh.242..648R},
	Archiveprefix = {arXiv},
	Author = {{Radice}, D. and {Abdikamalov}, E. and {Rezzolla}, L. and {Ott}, C.~D.},
	Doi = {10.1016/j.jcp.2013.01.048},
	Eprint = {1209.1634},
	Journal = {Journal of Computational Physics},
	Month = jun,
	Pages = {648-669},
	Primaryclass = {astro-ph.HE},
	Title = {{A new spherical harmonics scheme for multi-dimensional radiation transport I. Static matter configurations}},
	Volume = 242,
	Year = 2013}

@ARTICLE{Radice2020,
       author = {{Radice}, David},
        title = "{Binary Neutron Star Merger Simulations with a Calibrated Turbulence Model}",
      journal = {Symmetry},
         year = 2020,
        month = jul,
       volume = {12},
       number = {8},
        pages = {1249},
          doi = {10.3390/sym12081249},
archivePrefix = {arXiv},
       eprint = {2005.09002},
 primaryClass = {astro-ph.HE},
       adsurl = {https://ui.adsabs.harvard.edu/abs/2020Symm...12.1249R}
}

@article{Radice2022,
    author = "Radice, David and Bernuzzi, Sebastiano and Perego, Albino and Haas, Roland",
    title = "{A new moment-based general-relativistic neutrino-radiation transport code: Methods and first applications to neutron star mergers}",
    eprint = "2111.14858",
    archivePrefix = "arXiv",
    primaryClass = "astro-ph.HE",
    doi = "10.1093/mnras/stac589",
    journal = "Mon. Not. Roy. Astron. Soc.",
    volume = "512",
    number = "1",
    pages = "1499--1521",
    year = "2022"
}

@article{Rath2026,
    journal = {},

    author = "Rath, Samantha and Foucart, Francois and Kidder, Lawrence E. and Pfeiffer, Harald P. and Scheel, Mark A.",
    title = "{Assessing the Relative Importance of Neutrino Matter Interaction Channels in Post-Merger Remnant of Binary Neutron Stars}",
    eprint = "2605.29187",
    archivePrefix = "arXiv",
    primaryClass = "astro-ph.HE",
    month = "5",
    year = "2026"
}

@article{Rezzolla:2011,
	Adsurl = {http://adsabs.harvard.edu/abs/2011ApJ...732L...6R},
	Archiveprefix = {arXiv},
	Author = {{Rezzolla}, L. and {Giacomazzo}, B. and {Baiotti}, L. and {Granot}, J. and {Kouveliotou}, C. and {Aloy}, M.~A.},
	Doi = {10.1088/2041-8205/732/1/L6},
	Eid = {L6},
	Eprint = {1101.4298},
	Journal = {Astrophys. J. Letters},
	Month = may,
	Pages = {L6},
	Primaryclass = {astro-ph.HE},
	Title = {{The Missing Link: Merging Neutron Stars Naturally Produce Jet-like Structures and Can Power Short Gamma-ray Bursts}},
	Volume = 732,
	Year = 2011}

@article{Richers2019,
    author = "Richers, Sherwood A. and McLaughlin, Gail C. and Kneller, James P. and Vlasenko, Alexey",
    title = "{Neutrino Quantum Kinetics in Compact Objects}",
    eprint = "1903.00022",
    archivePrefix = "arXiv",
    primaryClass = "astro-ph.HE",
    doi = "10.1103/PhysRevD.99.123014",
    journal = "Phys. Rev. D",
    volume = "99",
    number = "12",
    pages = "123014",
    year = "2019",
    note = "[Erratum: Phys.Rev.D 109, 129902 (2024)]"
}

@ARTICLE{Ruiz2021rev,
       author = {{Ruiz}, Milton and {Shapiro}, Stuart L. and {Tsokaros}, Antonios},
        title = "{Multimessenger Binary Mergers Containing Neutron Stars: Gravitational Waves, Jets, and {\ensuremath{\gamma}}-Ray Bursts}",
      journal = {Frontiers in Astronomy and Space Sciences},
         year = 2021,
        month = apr,
       volume = {8},
          eid = {39},
        pages = {39},
          doi = {10.3389/fspas.2021.656907},
archivePrefix = {arXiv},
       eprint = {2102.03366},
 primaryClass = {astro-ph.HE},
       adsurl = {https://ui.adsabs.harvard.edu/abs/2021FrASS...8...39R}
}

@article{Sawyer2005,
  title={Speed-up of neutrino transformations in a supernova environment},
  author={Sawyer, Raymond F},
  journal={Physical Review D—Particles, Fields, Gravitation, and Cosmology},
  volume={72},
  number={4},
  pages={045003},
  year={2005},
  publisher={APS}
}

@article{Sawyer2016,
  title = {Neutrino Cloud Instabilities Just above the Neutrino Sphere of a Supernova},
  author = {Sawyer, R. F.},
  journal = {Phys. Rev. Lett.},
  volume = {116},
  issue = {8},
  pages = {081101},
  numpages = {5},
  year = {2016},
  month = {Feb},
  publisher = {American Physical Society},
  doi = {10.1103/PhysRevLett.116.081101},
  url = {https://link.aps.org/doi/10.1103/PhysRevLett.116.081101}
}

@article{Schianchi2023,
    author = "Schianchi, Federico and Gieg, Henrique and Nedora, Vsevolod and Neuweiler, Anna and Ujevic, Maximiliano and Bulla, Mattia and Dietrich, Tim",
    title = "{M1 neutrino transport within the numerical-relativistic code BAM with application to low mass binary neutron star mergers}",
    eprint = "2307.04572",
    archivePrefix = "arXiv",
    primaryClass = "gr-qc",
    doi = "10.1103/PhysRevD.109.044012",
    journal = "Phys. Rev. D",
    volume = "109",
    number = "4",
    pages = "044012",
    year = "2024"
}

@article{Shalgar2025,
    author = "Shalgar, Shashank and Tamborra, Irene",
    title = "{Neutrino quantum kinetics in three flavors}",
    eprint = "2503.03835",
    archivePrefix = "arXiv",
    primaryClass = "astro-ph.HE",
    doi = "10.1088/1475-7516/2025/12/026",
    journal = "JCAP",
    volume = "12",
    pages = "026",
    year = "2025"
}

@article{ShibataRev19,
author = {Shibata, Masaru and Hotokezaka, Kenta},
title = {Merger and Mass Ejection of Neutron Star Binaries},
journal = {Annual Review of Nuclear and Particle Science},
volume = {69},
number = {1},
pages = {41-64},
year = {2019},
doi = {10.1146/annurev-nucl-101918-023625},

URL = {
        https://doi.org/10.1146/annurev-nucl-101918-023625

},
eprint = {
        https://doi.org/10.1146/annurev-nucl-101918-023625

}
}

@article{Sigl1993,
    author = "Sigl, G. and Raffelt, G.",
    title = "{General kinetic description of relativistic mixed neutrinos}",
    reportNumber = "MPI-PH-92-112",
    doi = "10.1016/0550-3213(93)90175-O",
    journal = "Nucl. Phys. B",
    volume = "406",
    pages = "423--451",
    year = "1993"
}

@ARTICLE{Skinner2019,
       author = {{Skinner}, M. Aaron and {Dolence}, Joshua C. and {Burrows}, Adam and
         {Radice}, David and {Vartanyan}, David},
        title = "{FORNAX: A Flexible Code for Multiphysics Astrophysical Simulations}",
      journal = {Astrophys. J., Supp.},
         year = 2019,
        month = mar,
       volume = {241},
       number = {1},
          eid = {7},
        pages = {7},
          doi = {10.3847/1538-4365/ab007f},
archivePrefix = {arXiv},
       eprint = {1806.07390},
 primaryClass = {astro-ph.IM},
       adsurl = {https://ui.adsabs.harvard.edu/abs/2019ApJS..241....7S}
}

@article{Steiner2013,
	Adsurl = {http://adsabs.harvard.edu/abs/2013ApJ...774...17S},
	Archiveprefix = {arXiv},
	Author = {{Steiner}, A.~W. and {Hempel}, M. and {Fischer}, T.},
	Doi = {10.1088/0004-637X/774/1/17},
	Eid = {17},
	Eprint = {1207.2184},
	Journal = {Astrophys. J.},
	Month = sep,
	Pages = {17},
	Primaryclass = {astro-ph.SR},
	Title = {{Core-collapse Supernova Equations of State Based on Neutron Star Observations}},
	Volume = 774,
	Year = 2013}

@article{Sugiura2022,
    author = "Sugiura, Ken'ichi and Furusawa, Shun and Sumiyoshi, Kohsuke and Yamada, Shoichi",
    title = "{Leptonic and semi-leptonic neutrino interactions with muons in proto-neutron star cooling}",
    eprint = "2211.06944",
    archivePrefix = "arXiv",
    primaryClass = "astro-ph.HE",
    doi = "10.1093/ptep/ptac118",
    journal = "PTEP",
    volume = "2022",
    number = "11",
    pages = "113E01",
    year = "2022"
}

@article{Sumiyoshi2012,
	Adsurl = {http://adsabs.harvard.edu/abs/2012ApJS..199...17S},
	Archiveprefix = {arXiv},
	Author = {{Sumiyoshi}, K. and {Yamada}, S.},
	Doi = {10.1088/0067-0049/199/1/17},
	Eid = {17},
	Eprint = {1201.2244},
	Journal = {Astrophys. J., Supp.},
	Month = mar,
	Pages = {17},
	Primaryclass = {astro-ph.HE},
	Title = {{Neutrino Transfer in Three Dimensions for Core-collapse Supernovae. I. Static Configurations}},
	Volume = 199,
	Year = 2012}

@article{Sumiyoshi2020,
    author = "Sumiyoshi, Kohsuke and Fujibayashi, Sho and Sekiguchi, Yuichiro and Shibata, Masaru",
    title = "{Properties of Neutrino Transfer in a Deformed Remnant of a Neutron Star Merger}",
    eprint = "2010.10865",
    archivePrefix = "arXiv",
    primaryClass = "astro-ph.HE",
    doi = "10.3847/1538-4357/abce63",
    journal = "Astrophys. J.",
    volume = "907",
    number = "2",
    pages = "92",
    year = "2021"
}

@article{Takahashi2026,
    journal = {},

    author = "Takahashi, Shota and Nagakura, Hiroki and Zaizen, Masamichi and Kato, Chinami and Liu, Jiabao",
    title = "{Dynamic Competition of Fast and Collisional Neutrino Flavor Instabilities with Collisional Damping in Spatially Inhomogeneous Systems}",
    eprint = "2605.10435",
    archivePrefix = "arXiv",
    primaryClass = "astro-ph.HE",
    doi = "10.1103/7c32-rfjb",
    month = "5",
    year = "2026"
}

@article{Tamborra2020,
    author = "Tamborra, Irene and Shalgar, Shashank",
    title = "{New Developments in Flavor Evolution of a Dense Neutrino Gas}",
    eprint = "2011.01948",
    archivePrefix = "arXiv",
    primaryClass = "astro-ph.HE",
    doi = "10.1146/annurev-nucl-102920-050505",
    journal = "Ann. Rev. Nucl. Part. Sci.",
    volume = "71",
    pages = "165--188",
    year = "2021"
}

@ARTICLE{Tanaka2020_opacity,
       author = {{Tanaka}, Masaomi and {Kato}, Daiji and {Gaigalas}, Gediminas and {Kawaguchi}, Kyohei},
        title = "{Systematic opacity calculations for kilonovae}",
      journal = {Mon. Not. R. Astron. Soc.},
         year = 2020,
        month = aug,
       volume = {496},
       number = {2},
        pages = {1369-1392},
          doi = {10.1093/mnras/staa1576},
archivePrefix = {arXiv},
       eprint = {1906.08914},
 primaryClass = {astro-ph.HE},
       adsurl = {https://ui.adsabs.harvard.edu/abs/2020MNRAS.496.1369T}
}

@article{Typel:2009sy,
    author = "Typel, S. and Ropke, G. and Klahn, T. and Blaschke, D. and Wolter, H. H.",
    title = "{Composition and thermodynamics of nuclear matter with light clusters}",
    eprint = "0908.2344",
    archivePrefix = "arXiv",
    primaryClass = "nucl-th",
    doi = "10.1103/PhysRevC.81.015803",
    journal = "Phys. Rev. C",
    volume = "81",
    pages = "015803",
    year = "2010"
}

@article{Urquilla2026,
    journal = {},

    author = "Urquilla, Erick and Shankar, Swapnil and Kundu, Debraj and Froustey, Julien and Richers, Sherwood and Miller, Jonah M. and McLaughlin, Gail C. and Kneller, James P. and Foucart, Francois",
    title = "{Neutrino transport and flavor instabilities in a post-merger disk}",
    eprint = "2604.06404",
    archivePrefix = "arXiv",
    primaryClass = "astro-ph.HE",
    month = "4",
    year = "2026"
}

@article{Wang2025,
    author = "Wang, Tianshu and Nagakura, Hiroki and Johns, Lucas and Burrows, Adam",
    title = "{Effect of the collisional flavor instability on core-collapse supernova models}",
    eprint = "2507.01100",
    archivePrefix = "arXiv",
    primaryClass = "astro-ph.HE",
    doi = "10.1103/z3qh-nj18",
    journal = "Phys. Rev. D",
    volume = "112",
    number = "6",
    pages = "063039",
    year = "2025"
}

@article{Wang2025a,
    journal = {},

    author = "Wang, Tianshu and Burrows, Adam",
    title = "{The Effect of the Fast-Flavor Instability on Core-Collapse Supernova Models}",
    eprint = "2503.04896",
    archivePrefix = "arXiv",
    primaryClass = "astro-ph.HE",
    doi = "10.3847/1538-4357/add889",
    month = "3",
    year = "2025"
}

@ARTICLE{Weih2020c,
       author = {{Weih}, L.~R. and {Gabbana}, A. and {Simeoni}, D. and {Rezzolla}, L. and {Succi}, S. and {Tripiccione}, R.},
        title = "{Beyond moments: relativistic lattice Boltzmann methods for radiative transport in computational astrophysics}",
      journal = {Mon. Not. R. Astron. Soc.},
         year = 2020,
        month = nov,
       volume = {498},
       number = {3},
        pages = {3374-3394},
          doi = {10.1093/mnras/staa2575},
archivePrefix = {arXiv},
       eprint = {2007.05718},
 primaryClass = {physics.comp-ph},
       adsurl = {https://ui.adsabs.harvard.edu/abs/2020MNRAS.498.3374W}
}

@article{Wu2017b,
    author = "Wu, Meng-Ru and Tamborra, Irene",
    title = "{Fast neutrino conversions: Ubiquitous in compact binary merger remnants}",
    eprint = "1701.06580",
    archivePrefix = "arXiv",
    primaryClass = "astro-ph.HE",
    doi = "10.1103/PhysRevD.95.103007",
    journal = "Phys. Rev. D",
    volume = "95",
    number = "10",
    pages = "103007",
    year = "2017"
}

@article{Xiong2022,
    author = "Xiong, Zewei and Wu, Meng-Ru and Mart{\'\i}nez-Pinedo, Gabriel and Fischer, Tobias and George, Manu and Lin, Chun-Yu and Johns, Lucas",
    title = "{Evolution of collisional neutrino flavor instabilities in spherically symmetric supernova models}",
    eprint = "2210.08254",
    archivePrefix = "arXiv",
    primaryClass = "astro-ph.HE",
    doi = "10.1103/PhysRevD.107.083016",
    journal = "Phys. Rev. D",
    volume = "107",
    number = "8",
    pages = "083016",
    year = "2023"
}

@article{Xiong2023b,
    author = "Xiong, Zewei and Johns, Lucas and Wu, Meng-Ru and Duan, Huaiyu",
    title = "{Collisional flavor instability in dense neutrino gases}",
    eprint = "2212.03750",
    archivePrefix = "arXiv",
    primaryClass = "hep-ph",
    doi = "10.1103/PhysRevD.108.083002",
    journal = "Phys. Rev. D",
    volume = "108",
    number = "8",
    pages = "083002",
    year = "2023"
}

@article{Xiong2024,
    author = "Xiong, Zewei and Wu, Meng-Ru and George, Manu and Lin, Chun-Yu",
    title = "{Robust Integration of Fast Flavor Conversions in Classical Neutrino Transport}",
    eprint = "2403.17269",
    archivePrefix = "arXiv",
    primaryClass = "astro-ph.HE",
    doi = "10.1103/PhysRevLett.134.051003",
    journal = "Phys. Rev. Lett.",
    volume = "134",
    number = "5",
    pages = "051003",
    year = "2025"
}

@article{Xiong2026,
    journal = {},

    author = "Xiong, Zewei and Wu, Meng-Ru",
    title = "{Neutrino quantum kinetics for fast flavor conversion in a time-dependent environment}",
    eprint = "2608.07773",
    archivePrefix = "arXiv",
    primaryClass = "astro-ph.HE",
    month = "8",
    year = "2026"
}

@article{Zaizen2021,
    author = "Zaizen, Masamichi and Morinaga, Taiki",
    title = "{Nonlinear evolution of fast neutrino flavor conversion in the preshock region of core-collapse supernovae}",
    eprint = "2104.10532",
    archivePrefix = "arXiv",
    primaryClass = "hep-ph",
    doi = "10.1103/PhysRevD.104.083035",
    journal = "Phys. Rev. D",
    volume = "104",
    number = "8",
    pages = "083035",
    year = "2021"
}

@article{Zaizen2025,
    author = "Zaizen, Masamichi",
    title = "{Spectral diversity in collisional neutrino-flavor conversion: Flavor equipartition or swap}",
    eprint = "2502.09260",
    archivePrefix = "arXiv",
    primaryClass = "hep-ph",
    doi = "10.1103/PhysRevD.111.103029",
    journal = "Phys. Rev. D",
    volume = "111",
    number = "10",
    pages = "103029",
    year = "2025"
}

@article{Zaizen2026,
    journal = {},

    author = "Zaizen, Masamichi and Nagakura, Hiroki",
    title = "{Fast Neutrino-Flavor Conversion with Attenuation and Global Lepton Gradient}",
    eprint = "2604.13617",
    archivePrefix = "arXiv",
    primaryClass = "astro-ph.HE",
    month = "4",
    year = "2026"
}

@article{Zappa2023,
       author = {{Zappa}, Francesco and {Bernuzzi}, Sebastiano and {Radice}, David and {Perego}, Albino},
        title = "{Binary neutron star merger simulations with neutrino transport and turbulent viscosity: impact of different schemes and grid resolution}",
      journal = {Mon. Not. R. Astron. Soc.},
         year = 2023,
        month = mar,
       volume = {520},
       number = {1},
        pages = {1481-1503},
          doi = {10.1093/mnras/stad107},
archivePrefix = {arXiv},
       eprint = {2210.11491},
 primaryClass = {astro-ph.HE},
       adsurl = {https://ui.adsabs.harvard.edu/abs/2023MNRAS.520.1481Z}
}

\end{document}